\documentclass[12pt]{article}
\pdfoutput=1
\usepackage{hyperref}
\usepackage{color}
\usepackage{epsfig, palatino}
\usepackage{epic}
\usepackage{dsfont}
\usepackage{colonequals}
\usepackage{mathrsfs}
\usepackage{ae} 
\usepackage[T1]{fontenc}
\usepackage[ansinew]{inputenc}
\usepackage{amsmath}
\usepackage{amssymb}
\usepackage{graphicx}
\usepackage[normalem]{ulem}
\usepackage{xcolor}
\definecolor{darkblue}{cmyk}{0.9,0.9,0,0}
\usepackage{cite}
\usepackage{varioref}
\usepackage{makeidx}
\usepackage[english]{babel}
\usepackage{simplewick}
\usepackage{array}
\usepackage{subcaption}
\usepackage{multirow}

\usepackage[font={small}]{caption}

\newcommand{\orange}{\color [rgb]{1,.5,0}}

\newcommand{\comment}[1]{}

\newcommand{\begBvR}[1]{\begin{#1}} 
\newcommand{\beq}{\begBvR{equation}}
\newcommand{\eeq}{\end{equation}}

\newcommand{\eeqq}{\end{equation*}}

\newcommand\eeqaa{\end{eqnarray*}}

\newcommand\eeqa{\end{array}}

\newcommand{\eea}{\end{eqnarray}}

\newcommand{\ud}{\mathrm d}

\newcommand{\neqa}{\nonumber\end{eqnarray}} 
\newcommand{\la}[1]{\label{#1}}

\renewcommand{\d}{\partial}

\newcommand{\<}{{\langle}}
\renewcommand{\>}{{\rangle}}

\newcommand{\re}{\relax{\rm I\kern-.18em R}}

\renewcommand{\sp}{p\hspace{-.40em}/}

\definecolor{darkgreen}{rgb}{0.0, 0.45, 0.0}
\definecolor{mathematicablue}{RGB}{94,130,182}

\def\XXint#1#2#3{{\setbox0=\hbox{$#1{#2#3}{\int}$}
\vcenter{\hbox{$#2#3$}}\kern-.5\wd0}}

\def\su2{{SU(2)}}

\def\a{{\alpha}}

\def\[{\left[}
\def\]{\right]}

\def\s{\sigma}
\def\a{\alpha}

\def\({\left(}
\def\){\right)}
\def\[{\left[}
\def\]{\right]}

\def\<{\langle}
\def\>{\rangle}

\def\i2{\frac{i}{2}}

\def\spi{\relax{\rm \pi\kern-0.5em /}}
\def\sA{\relax{\rm A\kern-0.5em /}}
\def\sp{\relax{\rm p\kern-0.5em /}}
\def\sd{\relax{\rm \d\kern-0.5em /}}
\def\sk{\relax{\rm k\kern-0.5em /}}
\def\sn{\relax{\rm n\kern-0.5em /}}
\def\sl{\relax{\rm l\kern-0.5em /}}
\def\sP{\relax{\rm P\kern-0.7em /}}
\def\sBethe{\relax{\rm \Bethe\kern-0.5em /}}

\def\2F1{\,_2{\rm F}_1}

\newtheorem{lemm}{Lemma}

\makeindex

\newcommand\blfootnote[1]{%
  \begingroup
  \renewcommand\thefootnote{}\footnote{\hspace{-6mm}#1}%
  \addtocounter{footnote}{-1}%
  \endgroup
}

\begin{document}

\thispagestyle{empty}

\renewcommand{\thefootnote}{\fnsymbol{footnote}}
\setcounter{page}{1}
\setcounter{footnote}{0}
\setcounter{figure}{0}

\vspace{-0.4in}

\begin{center}
$$$$
{\Large\textbf{\mathversion{bold}
The S-matrix Bootstrap IV: \\ Multiple Amplitudes
}\par}
\vspace{1.0cm}

\textrm{Alexandre Homrich$^\text{\tiny 1,\tiny 2}$, Jo\~ao Penedones$^\text{\tiny 3}$, Jonathan Toledo$^\text{\tiny 3}$, Balt C. van Rees$^\text{\tiny 4}$, Pedro Vieira$^\text{\tiny 1,\tiny 2}$}
\blfootnote{\tt  \#@gmail.com\&/@\{alexandre.homrich,jpenedones,jonathan.campbell.toledo,baltvanrees,pedrogvieira\}}
\\ \vspace{1.2cm}
\footnotesize{\textit{
$^\text{\tiny 1}$Perimeter Institute for Theoretical Physics,
Waterloo, Ontario N2L 2Y5, Canada\\
$^\text{\tiny 2}$ICTP South American Institute for Fundamental Research, IFT-UNESP, S\~ao Paulo, SP Brazil 01440-070  
$^\text{\tiny 3}$Fields and Strings Laboratory, 
Institute of Physics, \'Ecole Polytechnique F\'ed\'erale de Lausanne (EPFL),
 CH-1015 Lausanne,
Switzerland\\
$^\text{\tiny 4}$Centre for Particle Theory, Department of Mathematical Sciences, Durham University, Lower Mountjoy, Stockton Road, Durham, England, DH1 3LE\\
}  
\vspace{4mm}
}
\end{center}

\par\vspace{1.5cm}


\vspace{2mm}
\begin{abstract}
We explore the space of consistent three-particle couplings in $\mathbb Z_2$-symmetric two-dimensional QFTs using two first-principles approaches. Our first approach relies solely on unitarity, analyticity and crossing symmetry of the two-to-two scattering amplitudes and extends the techniques of \cite{Paper2} to a multi-amplitude setup. Our second approach is based on placing QFTs in AdS to get upper bounds on couplings  with the numerical conformal bootstrap, and is a multi-correlator version of \cite{Paper1}. The space of allowed couplings that we carve out is rich in features, some of which we can link to amplitudes in integrable theories with a $\mathbb Z_2$ symmetry, e.g., the three-state Potts and tricritical Ising field theories. Along a specific line our maximal coupling agrees with that of a new exact S-matrix that corresponds to an elliptic deformation of the supersymmetric Sine-Gordon model which preserves unitarity and solves the Yang-Baxter equation.
\end{abstract}

\noindent

\setcounter{page}{1}
\renewcommand{\thefootnote}{\arabic{footnote}}
\setcounter{footnote}{0}

\setcounter{tocdepth}{2}

 \def\nref#1{{(\ref{#1})}}

\newpage

\tableofcontents

\parskip 5pt plus 1pt   \jot = 1.5ex

\newpage
\section{Introduction} \la{intro}

The bootstrap of the two-to-two S-matrix of the lightest particle in a relativistic unitarity quantum field theory was recently revived in \cite{Paper1,Paper2,Paper3} and extended to particles with flavour in \cite{Martin,Lucia,Miguel,Andrea}. These works can be seen as gapped counterparts of the conformal bootstrap explorations in \cite{CFTNoFlavour,Caracciolo:2009bx} and \cite{CFTFlavour} (without and with flavour respectively). Here we initiate the bootstrap analysis of S-matrix elements involving different external particles in $\mathbb{Z}_2$ symmetric theories. This multiple amplitude study again mimics a similar recent development in the conformal bootstrap, namely the multiple correlator analysis of the Ising model which famously gave rise to the CFT islands in \cite{CFTMixed}.\footnote{While walking hand in hand as illustrated in the previous paragraph, the S-matrix and the conformal bootstrap also have significant differences. In the CFT bootstrap we exclude theories; once excluded, a theory can never be accepted; with better computers we exclude more. In the S-matrix bootstrap of \cite{Paper2,Paper3} we include theories; by constructing explicit solutions to crossing and unitarity some parameters are shown to be allowed; with better computers we include more. (In the S-matrix bootstrap when we impose new physical conditions we will exclude some of the previously found S-matrices though so the process is not as monotonic as in the CFT bootstrap.) The two bootstraps are thus two faces of a same coin, the Yin and the Yang, the darkness and the light, the chaos and the order. For a recent review of the conformal bootstrap state of the art see \cite{CFTreview}. \la{firstF}}

We will consider two-dimensional QFTs with exactly two stable particles of masses $m_1$ and $m_2$. We will assume the theory to be parity and time-reversal invariant and both particles to be parity even. For simplicity we will also postulate the existence of a ${\mathbb Z}_2$ symmetry, under which the first particle is \emph{odd} and the second particle is \emph{even}.\footnote{In two dimensions theories with fermions and scalars are naturally $\mathbb{Z}_2$ symmetric theories so the setup here applies as well to any theories with scalars and fermions, not necessarily supersymmetric.} This means that the nonzero three-particle couplings are $g_{112}$ and $g_{222}$, which can be defined non-perturbatively in terms of the residues of a pole in a suitable S-matrix element. In the first part of this paper we will analyze all the two-to-two S-matrices of particles 1 and 2 and use crossing symmetry, analyticity and unitarity to explore the space of possible points in the (non-dimensionalized) $(g_{112},g_{222})$ plane as a function of $m_2/m_1$ -- see figure \ref{Panels} on page \pageref{Panels} to get an idea. In order to avoid singularities or Coleman-Thun poles  \cite{CT}, which complicate the analytic structure of the scattering amplitudes, we will restrict ourselves to
\beq
m_2 \le \sqrt{2} m_1 \,. \la{massRange}
\eeq
Note that we allow $m_2 < m_1$ also.

\begin{figure}[t]
\center \includegraphics[scale=.57]{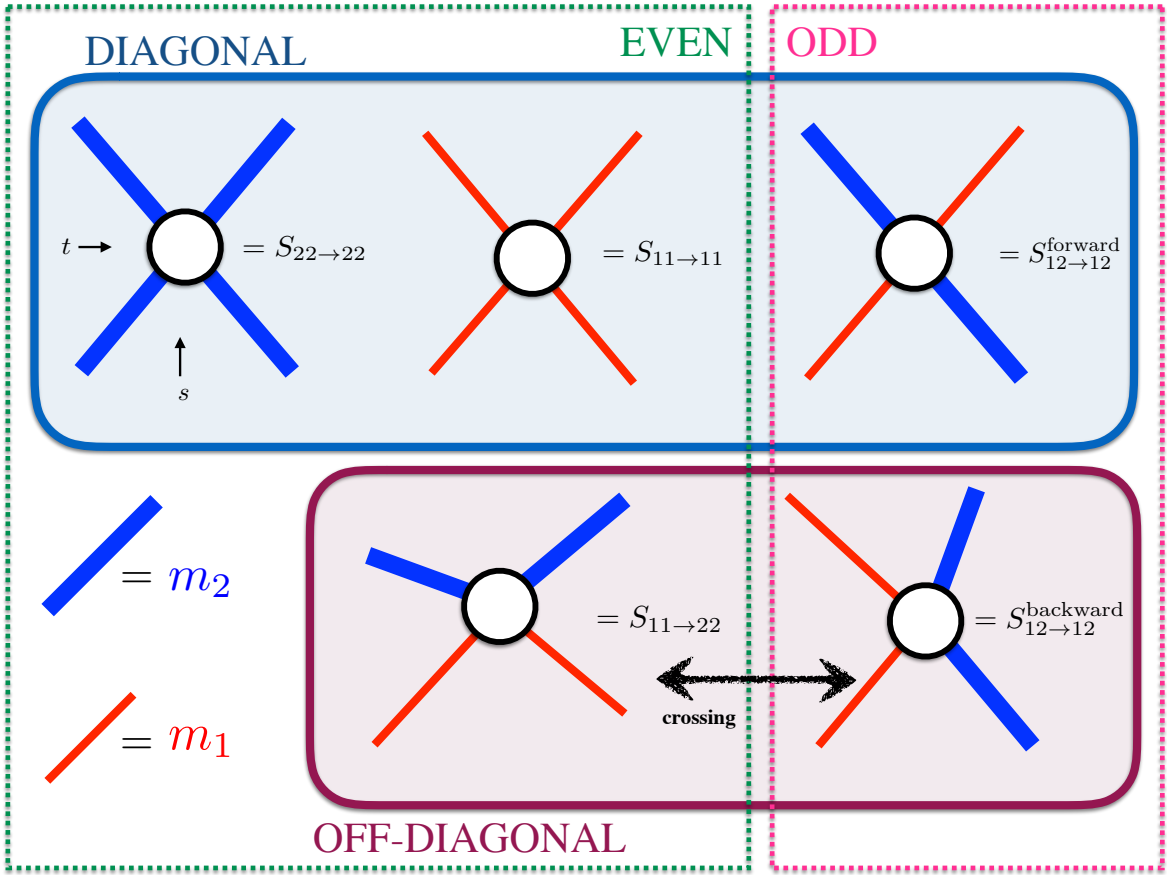}
\caption{Diagonal processes are those where the  incoming and outgoing particles have the same momenta as illustrated in the first row; they are all crossing invariant. 
The non-diagonal processes in the second row are those for which the final momenta are not the same as the initial momenta. Swapping space and time interchanges the odd and even off-diagonal processes so these off-diagonal processes play a crucial role in connecting these two sectors of different $\mathbb{Z}_2$ charge. 
} \label{AllProcesses} 
\end{figure}

Under the stated assumptions there are five different physical two-to-two scattering processes as shown in figure \ref{AllProcesses}. These can be grouped either according to the nature of their intermediate states, which can be $\mathbb Z_2$ odd or even, or according to whether they are `diagonal' or not. To wit, for a diagonal process the incoming and outgoing momenta are the same whereas for an off-diagonal process they are different.\footnote{As explained further below, in two dimensions the scattering angle can take only two values by kinematical restrictions; the outgoing momenta are essentially `locked' in terms of the incoming momenta. Unlike in higher dimensions, there is therefore no (analytic) function interpolating between forward and backward scattering.} As is also indicated in figure \ref{AllProcesses}, we call the $12\to12$ diagonal process `forward' scattering, and the $12 \to 12$ off-diagonal process `backward'.

In section \ref{Multiple} we will state in detail the conditions of unitarity, analyticity, and crossing symmetry that these five processes must obey. To guide ideas let us mention two conspicuous facts. First, we note that crossing symmetry flips the $s$ and $t$ axes on the diagram. This relates the two off-diagonal processes and thereby reduces the number of independent amplitudes (\emph{i.e.} functions of the Mandelstam invariant) to four. Of course, it also imposes a non-trivial constraint on the amplitudes for the diagonal processes. Second, we observe that particle 1 can appear as an intermediate state in all the odd processes and gives rise to a pole in these amplitudes with residue proportional to $g_{112}^2$, whereas particle 2 gives rise to poles in all the even amplitudes with residues proportional to $g_{112}^2$, $g_{112} g_{222}$ or $g_{222}^2$, depending on the process. These poles can be thought of as our definition of the corresponding couplings.\footnote{The astute reader will have noticed that this defines the couplings only up to an overall sign flip, leading to an obvious reflection symmetry in some of our plots.}

\begin{figure}[t!]
\center \includegraphics[scale=.8]{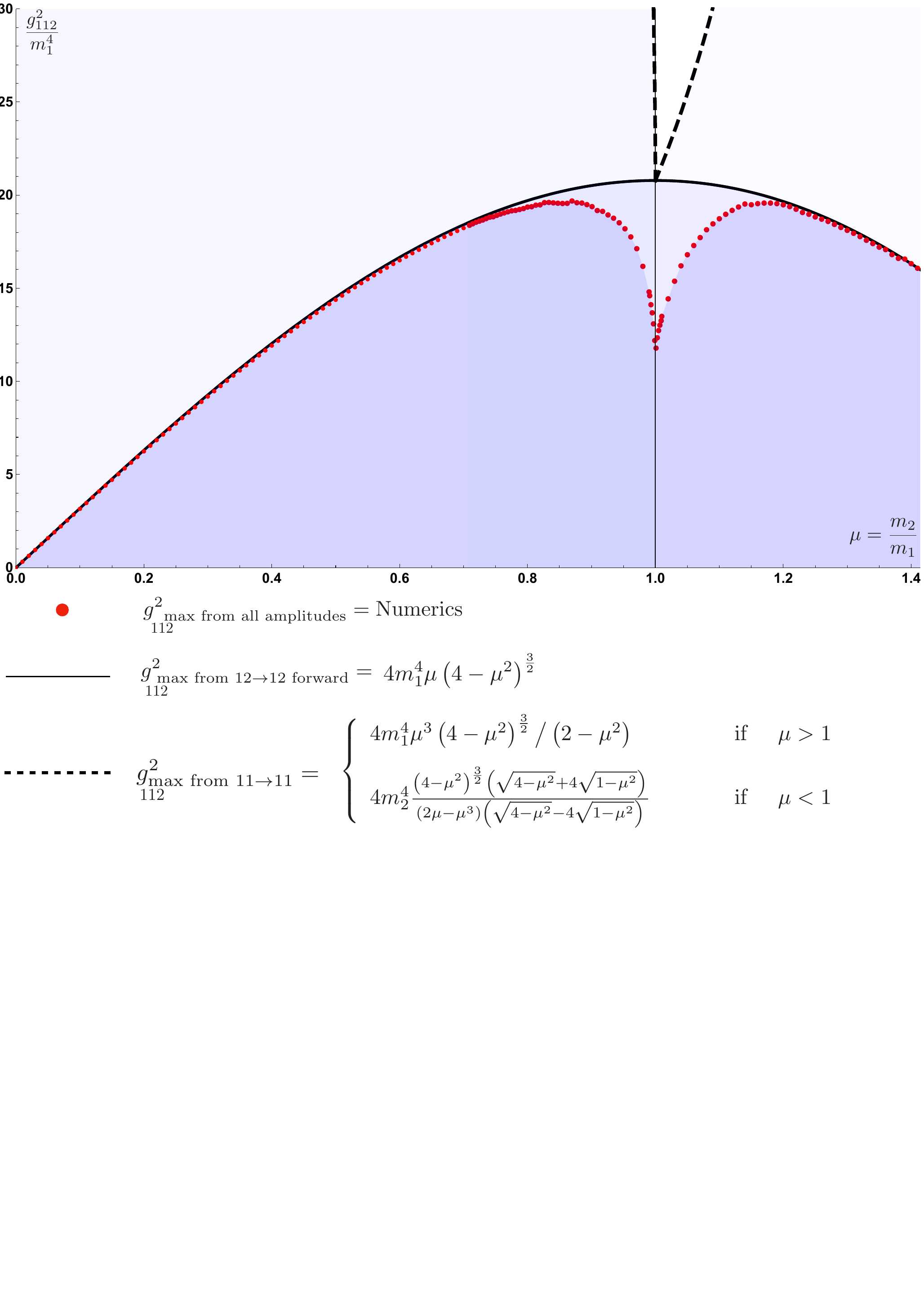}
\vspace{-9cm}
\caption{Upper bounds on the cubic coupling $g_{112}^2$ as a function of $\mu\equiv m_2/m_1$.
\emph{Dashed line:} Analytic bound based on the scattering of the lightest odd particle, from \cite{Paper2}. 
\emph{Solid line:} Analytic bound arising from the forward (or transmission) scattering of the odd particle against the even particle; it is a much stronger bound. 
\emph{Red dots:} The numeric bound obtained from \textit{all} two-to-two 
processes as discussed in the main text. The shaded regions represent the allowed regions which nicely shrink as we include more constraints. Any relativistic, unitary, $\mathbb{Z}_2$ invariant theory theory with two stable particles (one odd with mass $m_1$ and one even with mass $m_2$) must lie inside the darkest blue region.} \label{g112}
\end{figure}

\begin{figure}[t!]
\center \includegraphics[scale=.8]{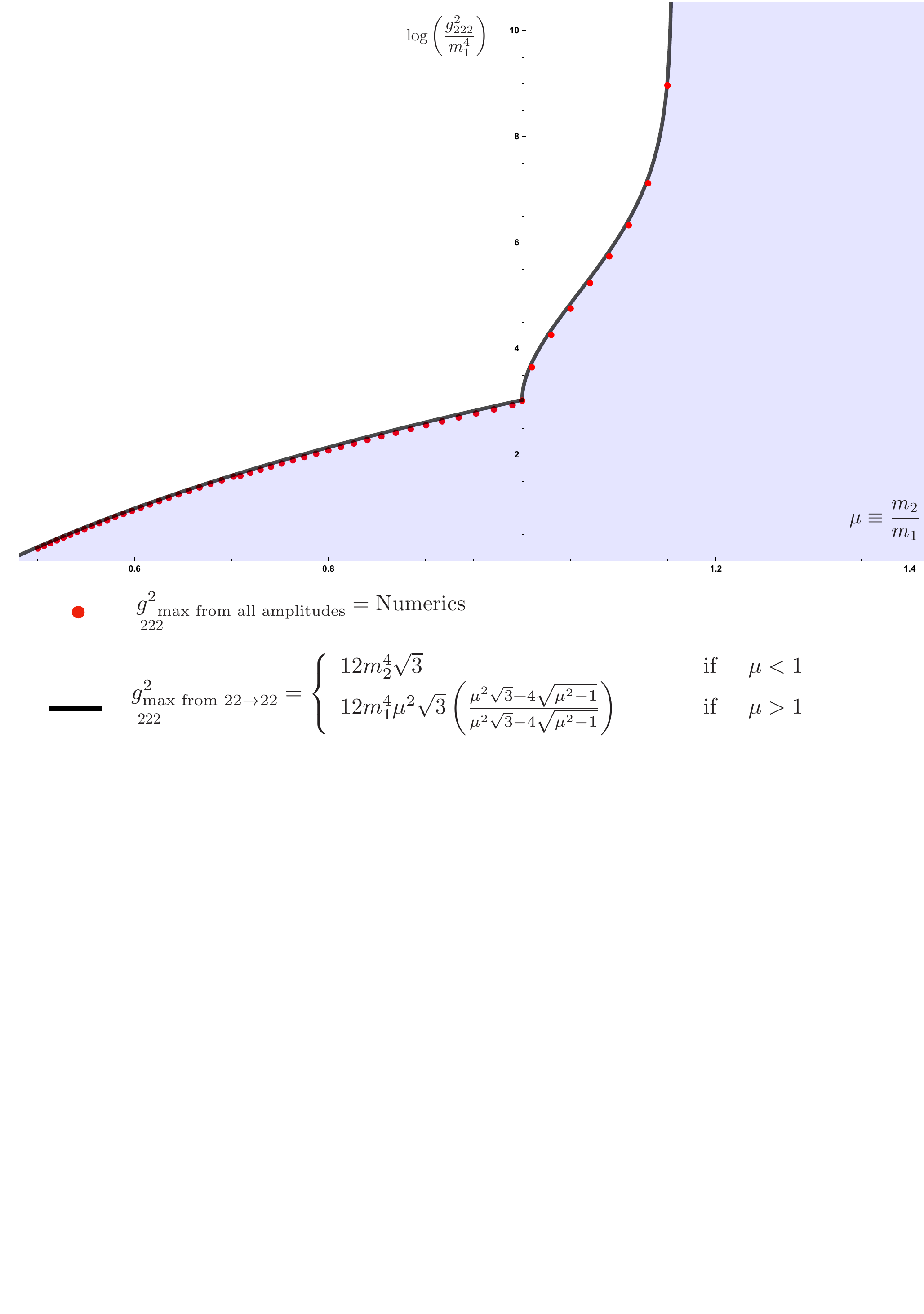}
\vspace{-10cm}
\caption{Upper bounds on the cubic coupling $g_{222}$ as a function of $\mu\equiv m_2/m_1$.
\emph{Solid line:} Analytic bound based on the scattering of the lightest even particle, from \cite{Paper2}. 
\emph{Red dots:} The numeric bound obtained from \textit{all} two-to-two 
processes as discussed in the main text. The shaded region represent the allowed region. When the even particle is the lightest, we can solve analytically for the maximal coupling, even considering the full set of amplitudes. When the odd particle is the lightest, the coupling can be bigger, diverging when singularities of the amplitudes corresponding to physical processes collide. This happens at $m_2/m_1 = 2/\sqrt{3}$. After this mass ratio the upper bound disappears.
}\label{g222}
\end{figure}

\subsection{Quick comparison with single-correlator bounds}
As a warm up exercise let us first discuss the three diagonal processes in isolation and explain how the methods discussed in \cite{Paper2,Paper3,Creutz,nobodycanfindSymanzik} already lead to some constraints on the couplings.

The analyticity and crossing symmetry of the diagonal processes in the Mandelstam $s$ plane is pretty straightforward. For example, the odd process has a two-particle $s$-channel cut starting at $s = (m_1 + m_2)^2$ and a pole at $m_1^2$ with residue proportional to $g_{112}^2$, plus the crossed $t$-channel singularities obtained by swapping $s \to 2m_1^2 + 2m_2^2 -s$. The even processes $S_{11\to11}$ and $S_{22\to22}$ have their two-particle $s$-channel cuts starting at $\text{min} (4m_1^2,4m_2^2)$ and a pole with residue $g_{112}^2$ or $g_{222}^2$, again plus the crossed $t$-channel singularities obtained by swapping $s \to 4m_1^2 -s$ or $s \to 4m_2^2 - s$. As for unitarity, notice that the discontinuity across the cut is always positive, but it is bounded from \emph{above} only for physical $s$, which means $s > 4m_1^2$ for $S_{11\to11}$ and $s > 4m_2^2$ for $S_{22\to 22}$. Therefore only for the \emph{lightest} of the two particles is the discontinuity everywhere bounded from above, whereas for the other particle the discontinuity can be arbitrarily large (but not negative) in the interval between $\text{min}(4m_1^2,4m_2^2)$ and $\text{max} (4m_1^2,4m_2^2)$.

We can bound the couplings as follows. First let us bound $g_{112}^2$ by using the maximum modulus principle for $S^{\text{forward}}_{12\to 12}$ following \cite{Paper2,Paper3,Creutz,nobodycanfindSymanzik}. We define
\beq
f_{12\to 12}(s)\equiv  S^{\text{forward}}_{12\to 12}(s) / \frac{h_{12}(s)+h_{12}(m_1^2)}{h_{12}(s)-h_{12}(m_1^2)}\,,
\eeq
with $h_{ab}(s) \equiv \sqrt{(s-(m_a-m_b)^2)( (m_a+m_b)^2-s)}$. The function $f_{12\to12}$ is free of singularities (since we divided out by functions with poles at the pole location of the amplitudes) and is bounded at the $s$-- and $t$-- channel cuts (since the functions we divided by are phases at those cuts and the amplitude is bounded). Therefore $f_{12\to 12}(s) $ must  have absolute value smaller or equal to $1$ everywhere, and in particular at $m_2^2$ and $m_1^2$ where we can simply read off the maximally allowed couplings in these amplitudes. This leads to a universal upper bound on $g_{112}^2$, which is the solid line in figure \ref{g112}.

The exact same analysis can be used for the elastic amplitude for the lightest of the two particles. If we denote this by $\ell$, so $m_\ell= \min(m_1,m_2)$, then the maximum modulus principle for 
\beq
f_{\ell\ell\to \ell\ell}(s) \equiv  S_{\ell\ell\to \ell\ell}(s) / \frac{h_{\ell\ell}(s)+h_{\ell\ell}(m_2^2)}{h_{\ell\ell}(s)-h_{\ell\ell}(m_2^2)}
\eeq
gives a bound on the coupling appearing on $S_{\ell \ell \to \ell \ell}$, which is $g_{\ell \ell 2}^2$. This is the dashed line for $m_2 > m_1$ in figure \ref{g112} and the solid line for $m_2 < m_1$ in figure \ref{g222}.

Finally we can use the techniques of \cite{Paper2} to also derive a bound on $g_{222}$ from the amplitude $S_{22\to 22}$ even when $m_2$ is not the lightest particle. In this case there is a cut which is not bounded directly by unitarity as depicted in figure \ref{screening2222}. As we derive in appendix \ref{maxg222}, the amplitude with maximal $g_{222}^2$ is given by
\beq
S_{22\to 22}(s) =  -\frac{h_{22}(s)+h_{22}(m_2^2)}{h_{22}(s)-h_{22}(m_2^2)} \times \frac{h_{22}(s)+h_{22}(4m_1^2)}{h_{22}(s)-h_{22}(4m_1^2)}
\label{S2222opt}
\eeq
The corresponding bound on $g_{222}^2$ is plotted as the solid line in figure \ref{g222} for $m_2 > m_1$. As the figure shows, the bound actually disappears for $m_2 \geq \frac{2}{\sqrt{3}} m_1$, which is due the t-channel pole colliding with the s-channel cut in the $22\to22$ process at this mass ratio.  This is the simplest instance of a phenomenon we call \textit{screening}. It is detailed in figure \ref{screening2222} and we will encounter it again below. In the same way we could obtain a bound on $g^2_{112}$ from the $11\to11$ process even when $m_1$ is the heaviest particle. This bound corresponds to the dashed line in figure \ref{g112} for $\mu<1$, and is always less restrictive than the bound from $S_{12\to12}^{\text{forward}}$.

\begin{figure}[t]
\center \includegraphics[width=0.8\textwidth]{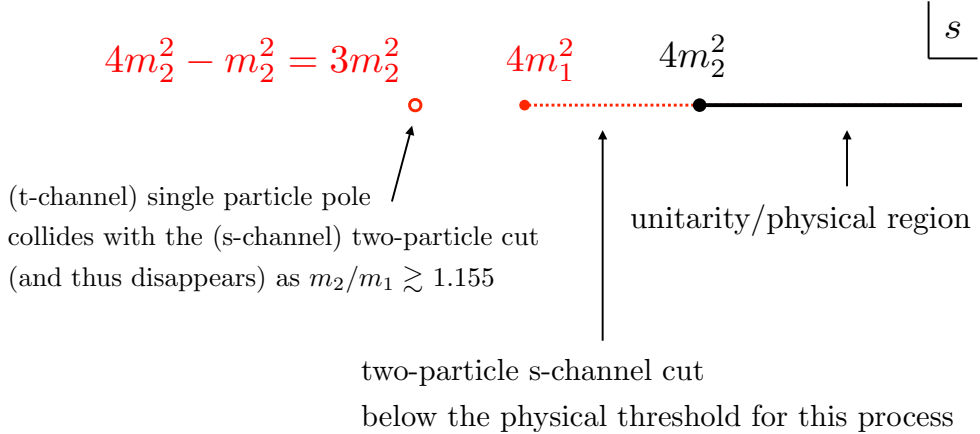}
\vspace{-3.5cm}
\caption{
Analytic structure of the $S_{22\to 22}$ amplitude (for clarity we do not show the left cut and s-channel pole following from crossing symmetry $S_{22\to 22}(s)=S_{22\to 22}(4m_2^2-s)$). 
If $m_2$ is not the lightest particle, there is a new feature in the $S_{22\to 22}$ amplitude: a two particle cut starting at $s=4m_1^2$ corresponding to the contribution of two particles $m_1$. This cut appears before the cut for two particles $m_2$ at physical energies $s \ge 4m_2^2$ where regular unitarity is imposed and the amplitude needs to be bounded. 
As $m_2$ grows beyond $ 2/\sqrt{3} m_1$ the t-channel pole corresponding to the exchange of particle $m_2$ enters the new cut (by crossing symmetry the $s$-channel pole enters the $t$-channel cut) so we ``lose'' this pole. Beyond this point we can no longer bound $g_{222}$ since it does not appear in any other diagonal amplitude.
 This is indeed what we observe in the numerics as illustrated in figure \ref{g222}. 
 Note that before the bound on $g_{222}$ disappears it diverges. This divergence, arising from the collision of the $t$-channel pole with an $s$-channel cut is analogous 
 to the divergences in bounds on couplings when $s$-- and $t$-- channel poles collide as already observed in \cite{Paper2}; the dashed line in figure \ref{g112} which was taken from \cite{Paper2} diverges at $m_2=\sqrt{2} m_1$ for exactly this reason. 
}\label{screening2222}
\end{figure}

This concludes our discussion of the single-amplitude results. As a preview for the more detailed numerical results presented below, we already marked in figures \ref{g112} and \ref{g222} in red dots our best numerical values of the coupling obtained from a simultaneous analysis of the full set of two-to-two amplitudes depicted in figure \ref{AllProcesses}. Figure \ref{g112} displays a clear improvement over the quick single-amplitude analysis for $m_1/\sqrt{2} <m_2<\sqrt{2} m_1$, with an intriguing kink at $m_2=m_1$. It would be fascinating to find if this kink corresponds to a physical theory. On the other hand, in figure \ref{g222} we see no improvement over the single-amplitude results. In fact, in section \ref{Numerics} we will prove that the maximal value of $g_{222}$ in the multi-amplitude analysis saturates the single-amplitude analytic bounds just derived. In the same section we will show a more complete picture by considering the entire $(g_{112},g_{222})$ plane.

\subsection{QFT in AdS}
As shown in \cite{Paper1}, there exists a completely orthogonal approach towards the problem of determining the maximal couplings in QFT. Rather than working from the S-matrix, which required analyticity assumptions that in general dimension $D$ are not very well understood, the idea is to consider QFTs on an AdS background. The boundary correlators of such a QFT, which are defined in a similar way as in the AdS/CFT correspondence, behave much like conformal correlation functions in one lower dimension $d = D -1$. By applying numerical conformal bootstrap methods of \cite{CFTNoFlavour} one can put a universal upper bound on the three-point couplings of QFTs in AdS. One can then extrapolate this bound to the flat-space limit (by sending all scaling dimensions to infinity), resulting in putative bounds for flat-space QFTs. In \cite{Paper1,Paper2} this was shown to work extremely well for two-dimensional QFTs: a precise match was found between the \emph{single-correlator} analysis using the conformal bootstrap, and the \emph{single-amplitude} analysis that we partially reviewed above.

In this work we continue these explorations. As discussed further in section \ref{sec:qftinads}, the $\mathbb Z_2$ symmetric setup that we consider is easily translated to a multi-correlator conformal bootstrap problem for QFTs in AdS. In most cases we again find a very good match, and in particular we are able to recover the coupling of the 3-state Potts field theory from the conformal crossing equations. For large-ish mass ratios, however, we will see that the multi-correlator bootstrap appears to be less powerful than even the single-amplitude bootstrap.

\subsection{Outline}
The $\mathbb{Z}_2$ symmetric S-matrix bootstrap is fully spelled out in section \ref{Multiple} and analysed numerically in section \ref{Numerics} leading to various bounds on the allowed coupling space for various mass ratios as illustrated in figure \ref{Panels}. In section \ref{Theories} we discuss integrable $\mathbb{Z}_2$ symmetric theories with $m_2=m_1$ and how some of them nicely show up at the boundary of the allowed S-matrix space found in the numerical bootstrap. These include a massive deformation of the $3$-state Potts model, the super-symmetric Sine-Gordon model and a  SUSY breaking integrable elliptic deformation of the super-symmetric Sine-Gordon which seems to be novel as far as we know. Section \ref{sec:qftinads} contains the results from the QFT in AdS analysis. Various appendices complement the main text with further extensions. (For example, the special role of the Tricritical Ising model as a kink in the space of S-matrices is discussed in appendix \ref{KinkIsing}.)

\section{Multiple amplitudes} \la{Multiple}

\subsection{Kinematics of the various \texorpdfstring{$\mathbb{Z}_2$}{Z2} preserving processes}
\begin{figure}[t]
\center \includegraphics[scale=.55]{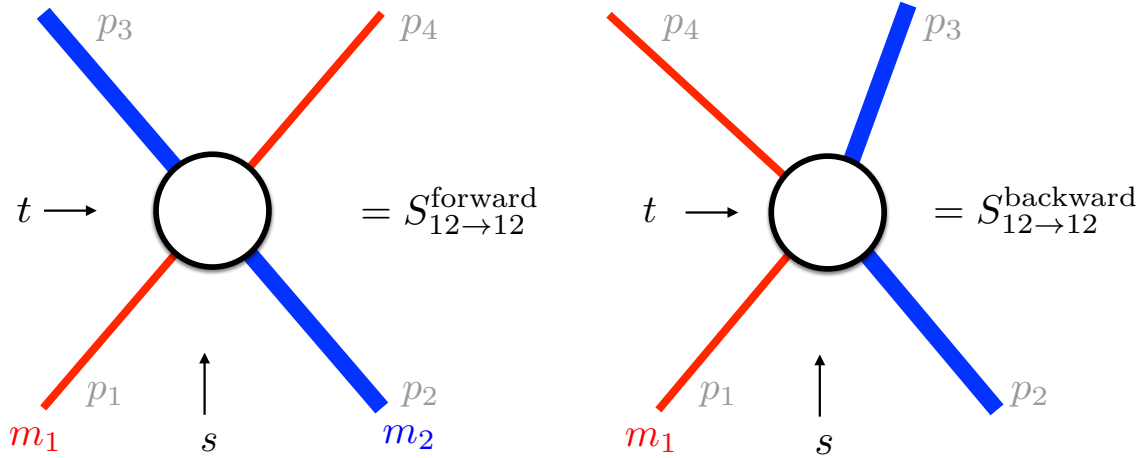}\vspace{-4.8cm}
\caption{In two dimensions when we scatter two particles $m_a$ and $m_b$ from the infinite past with $m_a$ to the left of $m_b$ we can end up, in the infinite future with $m_a$ to the right of $m_b$ or vice-versa. If the particles are distinguishable these are two genuinely different processes denoted as the forward or backward process. (They are sometimes also called the transmission and reflection processes.) 
In higher dimensions, these two scenarios are  limiting values of the a single amplitude when the scattering angle tends to $\theta=0$ or $\theta=\pi$, but in two dimensions there is no scattering angle and these processes are described by independent functions. 
As we exchange time and space, i.e. as we analytically continue these processes by swapping $t$ and $s$ we see that the forward process is mapped to itself while the backward process as seen from its crossed channel describes the $m_a m_a \to m_b m_b$ event. This translates into equations (\ref{C1}) and (\ref{C2}) in the main text. 
} \label{BackForward}
\end{figure}

There are \textit{six} two-to-two processes involving particles $m_1$ (odd) and $m_2$ (even) in a two dimensional $\mathbb{Z}_2$ symmetric theory.
We also assume time-reversal and parity symmetry.
Four of those six are even processes where we scatter either $11$ or $22$ into either $11$ or $22$. Of those four, two are trivially related by time-reversal, 
\beq
M_{22 \to 11} =  M_{11 \to 22}
\eeq
so we can ignore one of them (say $22 \to 11$) in what follows.  The remaining two processes are $\mathbb{Z}_2$ odd processes where we scatter the odd particle against the even particle obtaining those same two particles in the future. As explained in the introduction this process splits into two possibilities which we call the forward and the backward component, see figures \ref{AllProcesses} and \ref{BackForward}.  

In two dimensions, any process depends uniquely on the center of mass energy or equivalently on the Mandelstam invariant 
\beq
s = (p_1 + p_2)^2 \,.
\eeq
This in particular means that the other two Mandelstam invariants
\beq
t = (p_1 - p_3)^2\,, \qquad u = (p_1 - p_4)^2   \,.
\eeq
are completely determined in terms of $s$. It is important to find the precise relation because crossing symmetry permutes the three Mandelstam invariants and therefore leads to symmetries of the amplitudes $M(s)$ that we need to impose. In a process\footnote{In the convention $p_1^2=m_a^2$, $p_2^2=m_b^2$, $p_3^2=m_c^2$ and $p_4^2=m_d^2$. } involving $m_a m_b \to m_c m_d$ 
\beq
0 = 8\left|
\begin{array}{ccc}
p_1\cdot p_1 & p_1\cdot p_2 & p_1\cdot p_3 \\
p_2\cdot p_1 & p_2\cdot p_2 & p_2\cdot p_3 \\
p_3\cdot p_1 & p_3\cdot p_2 & p_3\cdot p_3 
\end{array}
 \right| = \left|
\begin{array}{ccc}
2m_a^2 & s-m_a^2-m_b^2 & -t + m_a^2 + m_c^2 \\
 s-m_a^2-m_b^2 & 2m_b^2 & -u+m_b^2+m_c^2\\
 -t + m_a^2 + m_c^2 & -u+m_b^2+m_c^2 & 2m_c^2
\end{array}
 \right|  
\eeq
The first equal sign is the two dimensional constraint: in two dimensions $p_3$ is always a linear combination of the two-vectors $p_1$ and $p_2$ and hence the determinant vanishes. In the second equal sign we used the on-shell conditions and momentum conservation. For example $2 p_2\cdot p_3=-(p_2-p_3)^2+p_2^2+p_3^2=-(p_1-p_4)^2+p_2^2+p_3^2 = -u+m_b^2+d_c^2$ and so on. Evaluated explicitly and combined with the previous linear constraint on the Mandelstam invariants, 
this can be cast  in a nice symmetric form:
\begin{eqnarray}
0=s t u +s
    (m_a^2+m_b^2)(m_c^2 +m_d^2 )+t  (m_a^2+m_c^2)(m_b^2 +m_d^2 )+u  (m_a^2+m_d^2)(m_b^2 +m_c^2 ) +C \la{constraint}
\end{eqnarray}
where $C= -\frac{1}{6} ( \sum m_i^2)^3-\frac{1}{2}  ( \sum m_i^4) ( \sum m_i^2)+\frac{2}{3}  \sum m_i^6$. 

Let us now specialize to the $\mathbb{Z}_2$ preserving cases mentioned above.
For the simplest processes corresponding to all equal masses (i.e. for $11\to 11$ and $22\to 22$) the condition dramatically simplifies into $stu=0$ which leads to $u=0$ or $t=0$ or $s=0$. In fact, we can not set $s=0$ since by definition we assume $s$ to be constructed from two incoming particles and setting $u=0$ or $t=0$ is the same up to a simple relabelling of the final particles which we can always do for indistinguishable particles. Hence without loss of generality we can set $u=0$ recovering the famous result that elastic scattering of identical particles in two dimensions has zero momentum transfer.  

Next we have the processes involving two particles of mass $m_1$ and two particles of mass $m_2$. Here it matters whether the two particles of the same mass are both incoming or if one is incoming and the other is outgoing. Let us start first with the second case so that we can set $m_a=m_d=m_1$ and $m_b=m_c=m_2$ in agreement with the conventions of figure \ref{BackForward}. Then we obtain a nice factorization of the constraint (\ref{constraint}) into 
\beq
0=u \left(\left(m_1^2 - m_2^2\right)^2 - \left(2 m_2^2  +  2 m_1^2 + s\right) s + s u\right)
\eeq
with two clear solutions: $u=0$ corresponding to forward scattering and $u= 2 m_1^2 + 2 m_2^2 - \left(m_1^2 - m_2^2\right)^2/s + s$ corresponding to the more complicated backward scattering. Note that in  forward scattering the final momenta are equal to the initial momenta but this is  not the case in   backward scattering where the momentum transfer is non-zero as   highlighted in figure \ref{BackForward}. 

Lastly we have the   even process $11\to 22$ where $m_a=m_b=m_1$ and $m_c=m_d=m_2$ which of course corresponds to a simple relabelling of the previous constraint in which $s \leftrightarrow u$ and thus leads, after discarding the $s=0$ solution, to 
\beq
0=\left(m_1^2 - m_2^2\right)^2 - \left(2 m_2^2  +  2 m_1^2 + u\right) u + s u \label{1122kin}
\eeq 
whose solutions are $u = \frac{1}{2} (2 m_1^2 + 2 m_2^2  \pm \sqrt{(4 m_1^2 - s)} \sqrt{(4 m_2^2 - s)} -s)$. In fact, these two solutions are equivalent up to relabelling of the two outgoing particles. Of course, the $s \leftrightarrow u$ relation between $11\to 22$ and backward $12 \to 12$ scattering is just crossing symmetry.

All in all we understood that all amplitudes can be thought of as functions of $s$ 
with the other Mandelstam invariants given by
\begin{alignat}{2}
&M_{11 \to 11}(s):  \qquad &&t = 4 m_1^2  - s, \qquad u =0, \la{S1}\\
&M_{22 \to 22}(s):  \qquad &&t = 4 m_2^2 - s, \qquad u =0, \\
&M_{12 \to 12}^{\text{forward}}(s):  \qquad &&t = 2 m_1^2 + 2m_2^2 - s, \qquad u =0, \\
&M_{12 \to 12}^{\text{backward}}(s):  \qquad &&u+t =  2 m_1^2 + 2 m_2^2  - s, \qquad t = \frac{(m_1^2 - m_2^2)^2}{s}, \la{S4}\\
&M_{11 \to 22}(s): \qquad &&u-t =  \sqrt{(4 m_1^2 - s)} \sqrt{(4 m_2^2 - s)} , \qquad u+t= 2 m_1^2 + 2 m_2^2  - s  \la{S5} \,.
\end{alignat}
The above equations allow us to state the crossing symmetry equations which we will impose in the sequel. They are:
\begin{alignat}{2}
&M_{11 \to 11}(4 m_1^2-s)=M_{11 \to 11}(s) \,,\la{C0}\\
&M_{22 \to 22}(4m_2^2-s)=M_{22 \to 22}(s) \la{C01}\,,\\
&M_{12 \to 12}^{\text{forward}}(2 m_1^2 + 2m_2^2 - s)=M_{12 \to 12}^{\text{forward}}(s) \la{C1} \,,\\
&M_{11 \to 22}(2 m_1^2 + 2 m_2^2  -\frac{(m_1^2 - m_2^2)^2}{s} - s)=M_{12 \to 12}^{\text{backward}}(s) \,. \la{C2}
\end{alignat}
Note in particular that the last crossing relation plays quite an important role: it connects the even and the odd sectors. 

For more on how the above discussion can be related to a similar analysis in higher dimensions see appendix \ref{triangles}.

\subsection{Analyticity, Unitarity and Extended Unitarity}
\label{analyticity}

The central hypothesis for the S-matrix bootstrap is that the scattering amplitudes are analytic for arbitrary complex values of $s$ up to so-called Landau singularities \cite{Landau} corresponding to on-shell intermediate processes. For the amplitudes and mass range discussed in this paper, these singularities in the physical sheet correspond to the possibility of the full $a \to b$ scattering process to factorise into two scatterings, first $a \to c$ and then $c \to b$.  Each on-shell state $c$ of the theory will produce a singularity in the $a \to b$ process for $s$ equal to the center of mass energy squared of the state $c$. This singularity will then proliferate according to its image under crossing transformations, see e.g. (\ref{C0}--\ref{C2}). The discontinuities
around these singularities are governed by the generalized unitarity equations \cite{Landau},
\beq	
M_{12 \to 34}(s + i \epsilon) - M_{12 \to 34}(s - i \epsilon) = 2 \text{Im} M_{12 \to 34}(s + i \epsilon)  = \sum_c  \int \text{d}\Pi_c  M_{12 \to c}^* M_{c \to 34}\,
\label{extended}
\eeq
(where the first equality assumes time reversal invariance.) Equation (\ref{extended}) is very powerful and reduces to a number of familiar examples in special cases: 

\begin{itemize}
\item The contribution from one particle intermediate states corresponds to nothing but the usual bound-state poles: there the phase space integral reduces to the energy momentum delta function and the product of amplitudes to the physical three-point couplings, combining to the bound-state pole discontinuity $-2 \pi i \delta(s - m_k^2)  g_{12k} g_{34k}$. 
\item For real values of $s$ for which there are no on-shell states, (\ref{extended}) reduces to the reality condition $\text{Im} M_{12 \to 34} = 0$. 
\item If we are at physical energies, $s > \max\{(m_1+m_2)^2, (m_3+m_4)^2\}$, then (\ref{extended}) is just the physical unitarity condition $\< \textbf{3} \textbf{4} | S^\dagger S -  \mathds{1} | \textbf{1} \textbf{2} \> = 0 $. 
\item All of the above are very well known. Indeed, for the lightest two particle states in a given channel, there is nothing more to (\ref{extended}) than bound state poles, real analyticity and unitarity. For heavier external states, however, (\ref{extended}) extends the unitarity relation to the unphysical energy region $s < \max\{(m_1+m_2)^2, (m_3+m_4)^2\}$ by keeping the quadratic terms in the unitarity equation that correspond to physical intermediate states of energy $\sqrt{s}$. This is what is called \textit{extended} unitarity. 

In our $\mathbb{Z}_2$ symmetric setup 
and for $\sqrt{s}<\min(3m_2,2m_1+m_2)$,\footnote{The bound corresponds to the first  $\mathbb{Z}_2$ even three particle state. } we find
\begin{align}
2 \text{Im} M_{11 \to 11} &= \frac{|M_{11 \to 11}|^2}{2\sqrt{s(s-4m_1^2)}}
\theta(s-4m_1^2) + \frac{|M_{11 \to 22}|^2}{2\sqrt{s(s-4m_2^2)}} \theta(s-4m_2^2),\label{ex1111}\\
2 \text{Im} M_{11 \to 22} &= \frac{M_{11 \to 22} M^*_{11 \to 11}}{2\sqrt{s(s-4m_1^2)}}\theta(s-4m_1^2) + \frac{M^*_{11 \to 22} M_{22\to22}}{2\sqrt{s(s-4m_2^2)}} \theta(s-4m_2^2),\label{ex1122}\\ 
2 \text{Im} M_{22 \to 22} &= \frac{|M_{11 \to 22}|^2}{2\sqrt{s(s-4m_1^2)}} \theta(s-4m_1^2)+ \frac{|M_{22 \to 22}|^2}{2\sqrt{s(s-4m_2^2)}} \theta(s-4m_2^2) \label{ex2222},
\end{align}
and for $\sqrt{s}<\min(3m_1,2m_2+m_1)$,
\begin{align}
2 \text{Im} M^{\text{Forward}}_{12 \to 12} &= \frac{|M^{\text{Forward}}_{12 \to 12}|^2 + |M^{\text{Backward}}_{12 \to 12}|^2}{2\sqrt{(s - (m_1 - m_2)^2)(s-(m_1 + m_2)^2)}}\theta(s-(m_1+m_2)^2),\label{ex12e}\\
2 \text{Im} M^{\text{Backward}}_{12 \to 12} &= \frac{M^{*\text{Forward}}_{12 \to 12} M^{\text{Backward}}_{12 \to 12} + M^{*\text{Backward}}_{12 \to 12} M^{\text{Forward}}_{12 \to 12}}{2\sqrt{(s - (m_1 - m_2)^2)(s-(m_1 + m_2)^2)}}\theta(s-(m_1+m_2)^2), \label{ex12h}
\end{align}
where the denominators come from the phase space factors and $\theta$ is the Heaviside step function.  For example, if $m_2>m_1$ then equation (\ref{ex2222}) for $s>4m_2^2$ is just unitarity for the $22 \to 22$ process, but for $4m_1^2<s<4m_2^2$ it is a ``new" constraint over the $|\textbf{1} \textbf{1}\>$ production cut.
\end{itemize}

Of course, the scattering amplitudes also have cuts and poles corresponding to crossed intermediate processes. The discontinuities around those singularities are governed by the generalised unitarity equations for the crossed scattering, together with the crossing equations (\ref{C0}--\ref{C2}).

For energies above the three particle  threshold, new terms corresponding to three-particle intermediate states should be introduced in the r.h.s.\ of equations (\ref{ex1111}-\ref{ex12h}) 
It is useful, however, to keep only the contributions from two-particle intermediate states and replace the full set of equations (\ref{ex1111}-\ref{ex12h}) by a positive semidefinite constraint on the amplitudes.  For the $\mathbb{Z}_2$ even sector, by dropping the contributions from  intermediate states with three or more particles in (\ref{extended}), we can write in matrix form
\beq
2\text{Im} \mathbb{M} \succeq \mathbb{M}^\dagger \rho^2 \mathbb{M}, \qquad \mathbb{M} = \bordermatrix{~ & ~ & \cr
         ~ & M_{11\to11} &  M_{11\to22}\cr
         ~ & M_{11\to22}  & M_{22\to22}\cr}, \qquad 
         \rho = \bordermatrix{~ & ~ & \cr
         ~ & \rho_{11} & 0\cr
         ~ & 0  & \rho_{22}\cr}, 
         \label{matrixextended}
\eeq
where  $\rho^2_{ab} = \frac{\theta\left(s-\left(m_a + m_b\right)^2\right)}{2  \sqrt{s-\left(m_a + m_b\right)^2}\sqrt{s-\left(m_a - m_b\right)^2}}$ takes into account the phase space volume. Note that (\ref{matrixextended}) is saturated for $\sqrt{s}<\min(3m_2,2m_1+m_2)$. As discussed in section \ref{Numerics}, for the numerical implementation we impose (\ref{matrixextended}) even before multiparticle thresholds, leaving for the computer to achieve saturation where (\ref{ex1111}-\ref{ex2222}) applies. A similar discussion holds for the $\mathbb{Z}_2$ odd sector. 

In appendix \ref{ap:unitarity}, we provide a direct derivation of \eqref{matrixextended} for $\sqrt{s}>2\max(m_1,m_2)$. This derivation  elucidates the physical meaning of the matrix $\mathbb{M}$ and its relation to transition probabilitues between initial and final states.

\section{Numerics} \la{Numerics}

\subsection{Implementation} 
\begin{figure}[t]
\center \includegraphics[width=\textwidth]{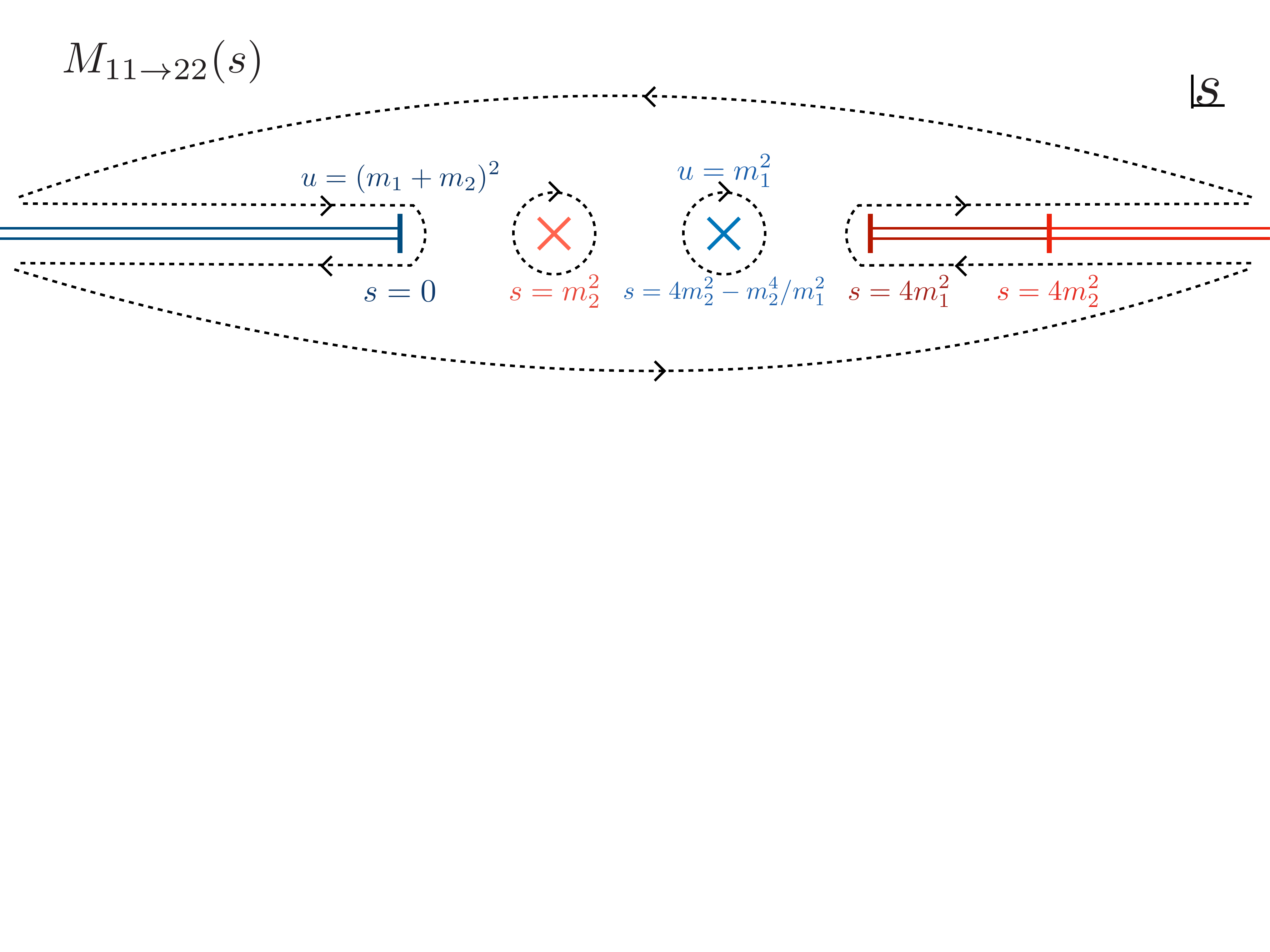}\vspace{-7cm}
\caption{Analytic structure of $M_{11 \to 22}(s)$. According to the kinematics in equation (\ref{S4}), as we move in the $s$ plane, we hit poles and two-particle (as well as multiparticle) thresholds in the $s$ and $u$ channel, but not in the $t$-channel. This is a consequence of our arbitrary definition of $t$ and $u$ (in the language of appendix \ref{triangles}, this comes from choosing for each $s$ a single point in the two-valued hyperbolas of figure \ref{triangle2}). To derive the dispersion (\ref{disp1122}) we start by assuming the amplitude approaches a constant at infinity (but see discussion in the main text) and write the identity $M_{11\to22}(s) = M_\infty +  \frac{1}{2 \pi i} \int_\gamma \frac{M_{11\to22}(s^*) -M_\infty}{s^*-s} \mathrm{d}s^*$, where $\gamma$ is the dotted contour above. We can then neglect the arcs at infinity. The contribution from the arcs around the red singularities correspond to the $s$-channel pole and $s^*$ integral in (\ref{disp1122}). After changing the integration variable in the remaining terms to $u^*(s^*)$ according to equation (\ref{S4}), we find the kernel transformation $\int_0^\infty \frac{\mathrm{d}s^*}{s^*-s} \to \int_{(m_1 + m_2)^2}^\infty (\frac{1}{u^*-u} + \frac{1}{u^* -t})\mathrm{d}u^* + C_{11\to22}$, where we could relabel $u^* \to t^*$ in the second term. Then, after absorbing $M_\infty$ into the constant $C_{11\to22}$ and using the crossing relation (\ref{C2}) and the discontinuity formula (\ref{extended}) for the pole terms, we obtain the dispersion relation (\ref{disp1122}).} \label{dispersion} 
\end{figure}

\la{implementation}
As discussed in section \ref{analyticity}, the $\mathbb{Z}_2$ symmetric scattering amplitudes in the mass range (\ref{massRange}) are analytic functions in the physical sheet of the the kinematical variable $s$ up to poles corresponding to bound states. This sheet is defined by continuing the amplitudes away from physical kinematics respecting the $i \epsilon$ prescription and has as its boundaries cuts corresponding to two and higher particle production thresholds which may happen in the $s$, $t$ and $u$ channels. These can be summarised by expressing the amplitudes through dispersion relations, as illustrated in figure \ref{dispersion}. For the case $m_1<m_2$, we obtain
\begin{align}
M_{11\to11}(s) =&\text{ } C_{11\to11} -\frac{g^2_{112}}{s-m^2_2} - \frac{g^2_{112}}{t(s)-m^2_2} + \frac{1}{\pi}\int_{4m^2_1}^{\infty}\frac{\text{Im}M_{11\to11}(s^*)}{s^*-s} \mathrm{d}s^* \label{disp1111}  \\ & +  \frac{1}{\pi}\int_{4m^2_1}^{\infty}\frac{\text{Im}M_{11\to11}(t^*)}{t^*-t(s)} \mathrm{d}t^* \nonumber, \\
M_{22\to22}(s) = &\text{ } C_{22\to22} -\frac{g^2_{222}}{s-m^2_2} - \frac{g^2_{222}}{t(s)-m^2_2} + \frac{1}{\pi}\int_{4m^2_1}^{\infty}\frac{\text{Im}M_{22\to22}(s^*)}{s^*-s} \mathrm{d}s^* \label{disp2222}\\ &+ \frac{1}{\pi}\int_{4m^2_1}^{\infty}\frac{\text{Im}M_{22\to22}(t^*)}{t^*-t(s)} \mathrm{d}t^* \nonumber, \\
M_{12\to12}^\text{Forward}(s) = &\text{ } C_{12\to12} -\frac{g^2_{112}}{s-m^2_1} - \frac{g^2_{112}}{t(s)-m^2_1} + \frac{1}{\pi}\int_{(m_1 + m_2)^2}^{\infty}\frac{\text{Im}M_{12\to12}(s^*)}{s^*-s} \mathrm{d}s^* \label{disp1212}\\ &+  \frac{1}{\pi}\int_{(m_1 + m_2)^2}^{\infty}\frac{\text{Im}M_{12\to12}(t^*)}{t^*-t(s)} \mathrm{d}t^*, \nonumber\\
M_{11\to22}(s) =&\text{ } C_{11\to22} -\frac{g_{112} g_{222}}{s-m^2_2} - \frac{g^2_{112}}{t(s)-m^2_1} - \frac{g^2_{112}}{u(s)-m^2_1} + \frac{1}{\pi}\int_{4m^2_1}^{\infty}\frac{\text{Im}M_{11\to22}(s^*)}{s^*-s} \mathrm{d}s^* \label{disp1122}\\ &+  \frac{1}{\pi}\int_{(m1 + m2)^2}^{\infty}\frac{\text{Im}M^{Backward}_{12\to12}(t^*)}{t^*-t(s)} \mathrm{d}t^*+  \frac{1}{\pi}\int_{(m1 + m2)^2}^{\infty}\frac{\text{Im}M^{Backward}_{12\to12}(u^*)}{u^*-u(s)} \mathrm{d}u^* \nonumber,
\end{align}
with $C_{a\to b}$ constant. Equations for the $m_1 > m_2$ case are obtained by replacing \linebreak $4m_1^2$ $\to$ $4m^2_2$ in the lower limits of the integrals. Recall that these are the only independent amplitudes, since $M_{12\to12}^{\text{Backward}}(s) = M_{11 \to 22}(2 m_1^2 + 2 m_2^2  -\frac{(m_1^2 - m_2^2)^2}{s} - s)$. 

In deriving this relations, see figure \ref{dispersion}, we assumed that the scattering amplitudes have no essential singularities at infinity, and in fact approach a constant in this limit, i.e. the S-matrix becomes free. This latter assumption is not crucial nor required: it can be lifted by introducing subtractions as discussed in \cite{Paper2} and the numerical problem of maximising the couplings is not sensitive to this. This is to be expected physically, since the low energy physics of bound state poles should not be much sensitive to the behaviour of the amplitudes at high energies. 

To obtain a concrete numerical implementation to the problem, we proceed as follows. First, we define a dispersion grid $\{x_1,...,x_M\}$ along the integration domains in (\ref{disp1111}-\ref{disp1122}). We then approximate the discontinuities $\text{Im}M_{a \to b}(x^*)$ by splines $\sigma_{a\to b}(s)$ \footnote{If $m_1 < m_2$, extended unitarity, equations (\ref{ex1111}-\ref{ex2222}), allows for $M_{22\to22}$ to diverge as $1/\sqrt{s-4m^2_1}$ close to the $4m^2_1$ threshold. Due to this, between the first two grid points, we approximate $\text{Im}M_{22\to 22} \propto 1/\sqrt{s-4m^2_1}$. If $m_1 > m_2$ we should replace $1 \leftrightarrow 2$ in this discussion.} linear in between the grid points up to a cutoff point $x_M$, after which we assume the discontinuities decay as $\text{Im}M_{a\to b}(x^*) \sim 1/x^*$.\footnote{This is similar to the numerical implementation in \cite{Paper2}. We could have parametrised our amplitudes using the $\rho$ variables defined in \cite{Paper3}. These variables provide a cleaner framework for the numerics but, in practice, we find that convergence with the $\rho$ variables is much slower than with the use of discretized dispersion relations.} With this approximation we can analytically perform the integrals in (\ref{disp1111}-\ref{disp1122}) obtaining, in the case $m_1<m_2$ and for $M_{11\to11}$, as an example,
\beq
M_{11\to11}(s) \approx  C_{11\to11} -\frac{g^2_{112}}{s-m^2_2} - \frac{g^2_{112}}{t(s)-m^2_2} + \sum^{M}_{i=1} \sigma_{11\to11}(x_i) (K_i(s) + K_i(t(s))),
\eeq
where the functions $K_i$ are defined in the appendix A of \cite{Paper2}. Next, we impose (\ref{matrixextended}) along a fine grid over $s>\min\{4m_1^2 ,4m_2^2\}$ (we impose analogous constraints over analogous ranges for the $\mathbb{Z}_2$ odd channels). Note that we leave for the computer to achieve saturation of (\ref{matrixextended}) before the three-particles thresholds. As shown in appendix \ref{semidefiniteequivalence}, equation (\ref{matrixextended})is equivalent to the semidefiniteness constraint
\beq
\bordermatrix{~ & ~ & \cr
         ~ & \mathbb{I} & \rho  \mathbb{M} \cr
         ~ & (\rho \mathbb{M})^\dagger & 2\text{Im } \mathbb{M} \cr} \succeq 0\,,
         \label{matrixextendedlinear}
\eeq
and a similar rewriting can be done for the $\mathbb{Z}_2$ odd sector. If we fix $\alpha = \frac{g_{222}}{g_{112}}$, as well as the masses, then the matrix in the l.h.s.\ of (\ref{matrixextendedlinear}) is linear on the variables $\{C_{a \to b}, g^2_{112}, \sigma_{a \to b}(x_i)\}$. The problem of maximising $g_{112}^2$ in this space of variables under the positive semidefinite constraint (\ref{matrixextended}) (and equivalent for the $\mathbb{Z}_2$ odd sector) is therefore a semidefinite program and can be solved with, say, \texttt{SDPB} \cite{SDPB}. Details on the numerical implementation, such as parameter settings are available upon request.

\subsection{Results for any \texorpdfstring{$m_2/m_1$}{m2/m1}} \la{results}

\begin{figure}[t]
\center \includegraphics[width=\textwidth]{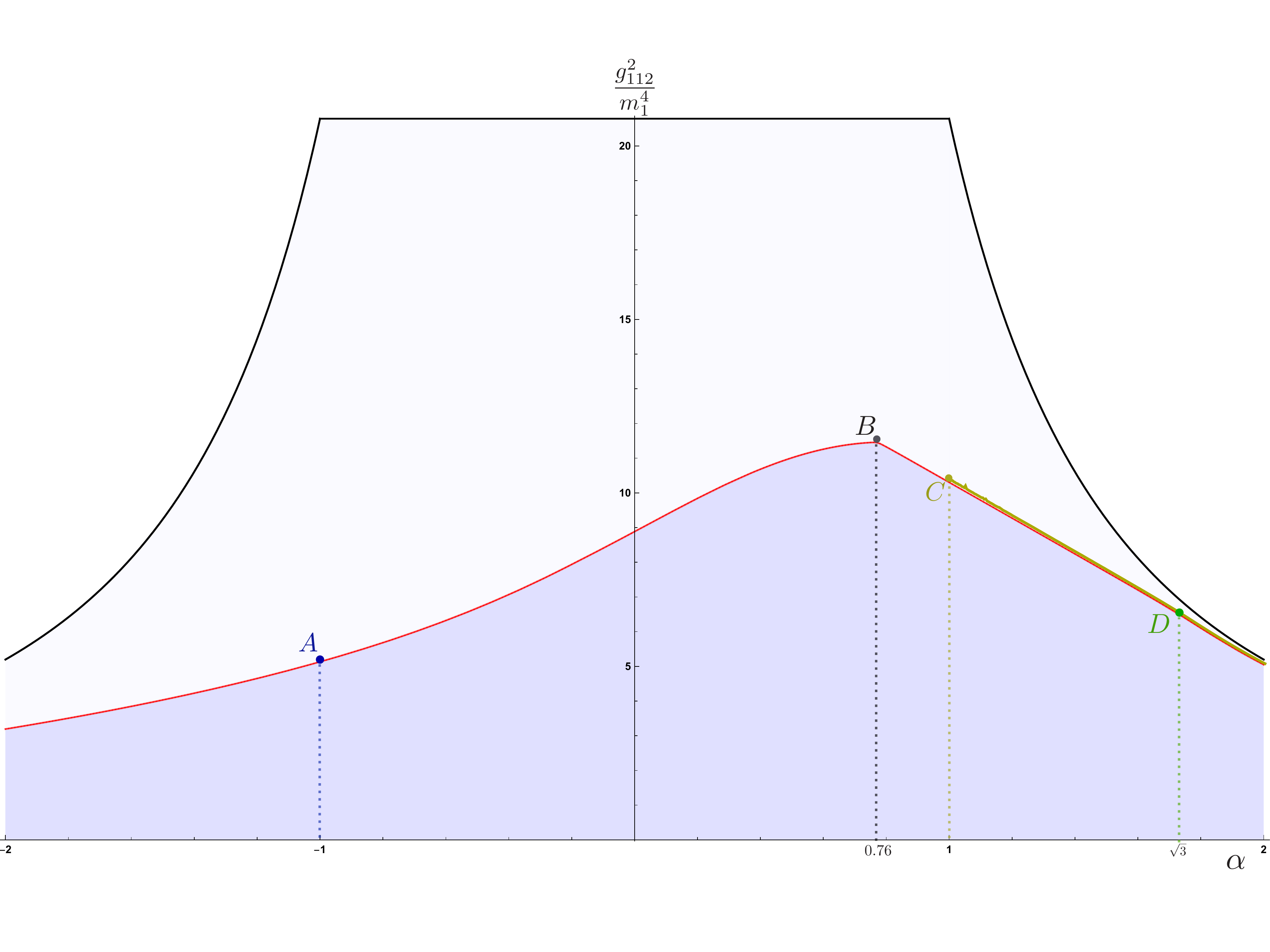}
\caption{Maximum coupling $g_{112}$ as a function of $\alpha =  g_{222}/g_{112}$ for a $\mathbb{Z}_2$ symmetric theory with an odd and an even particle both with the same mass. Solid black: bounds from single amplitude analytics. Red: bounds from multiple amplitudes numerics. The interesting points $A,B,C,D$ are discussed in more detail in the next section. Multiplying the $\alpha$ axis by $g_{112}$ we convert this plot into a plot of the allowed coupling space $(g_{112},g_{222})$, see figure \ref{gvgm1m2}.} \label{equalmass} 
\end{figure}

\begin{figure}[t!]
\centering
\vspace*{-1.8cm}
\begin{subfigure}{0.5\textwidth}
\center \includegraphics[width=\textwidth]{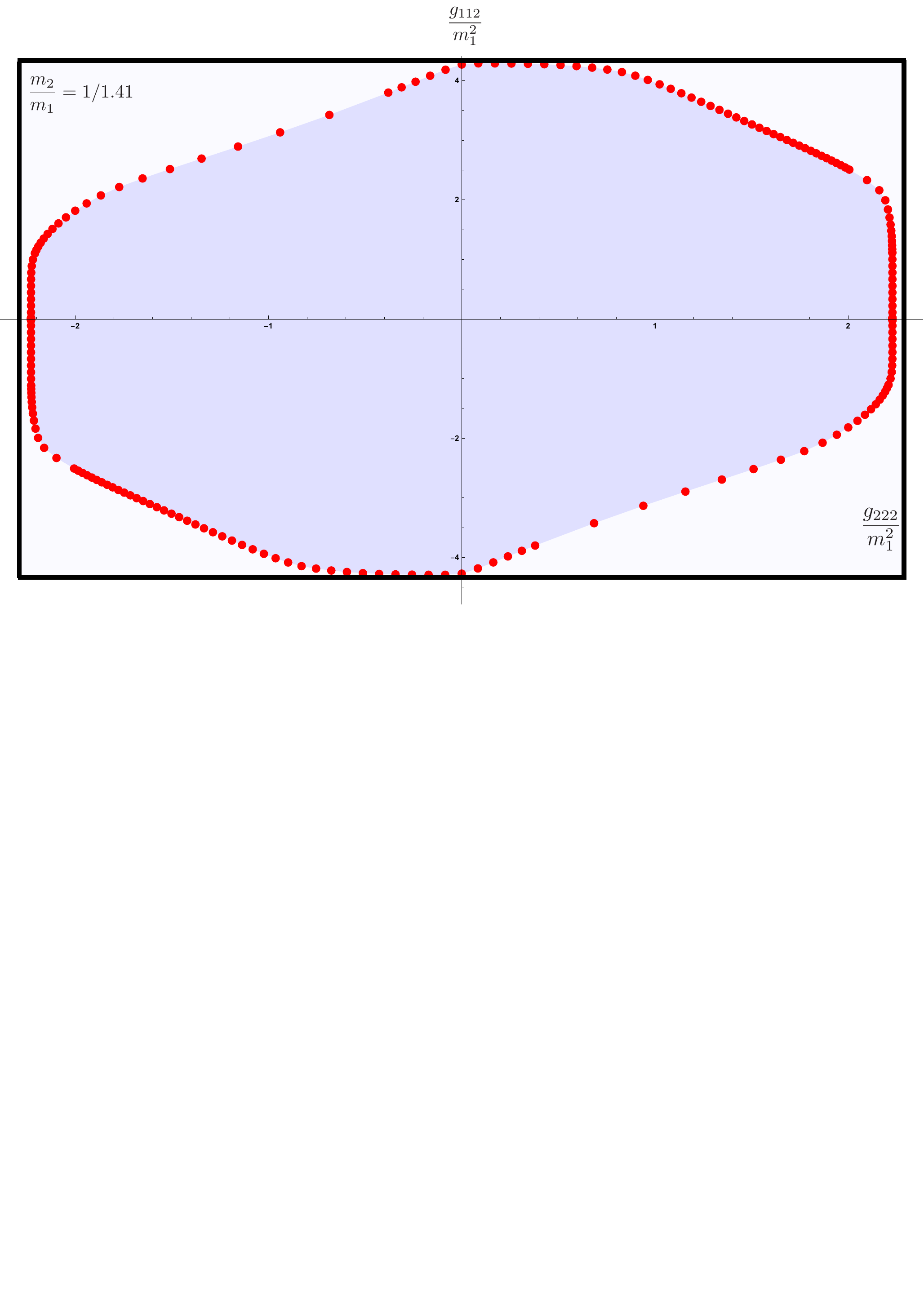} \label{pl4}
\vspace{-6.8cm}
\end{subfigure}\hspace*{\fill}
\begin{subfigure}{0.5\textwidth}
\center \includegraphics[width=\textwidth]{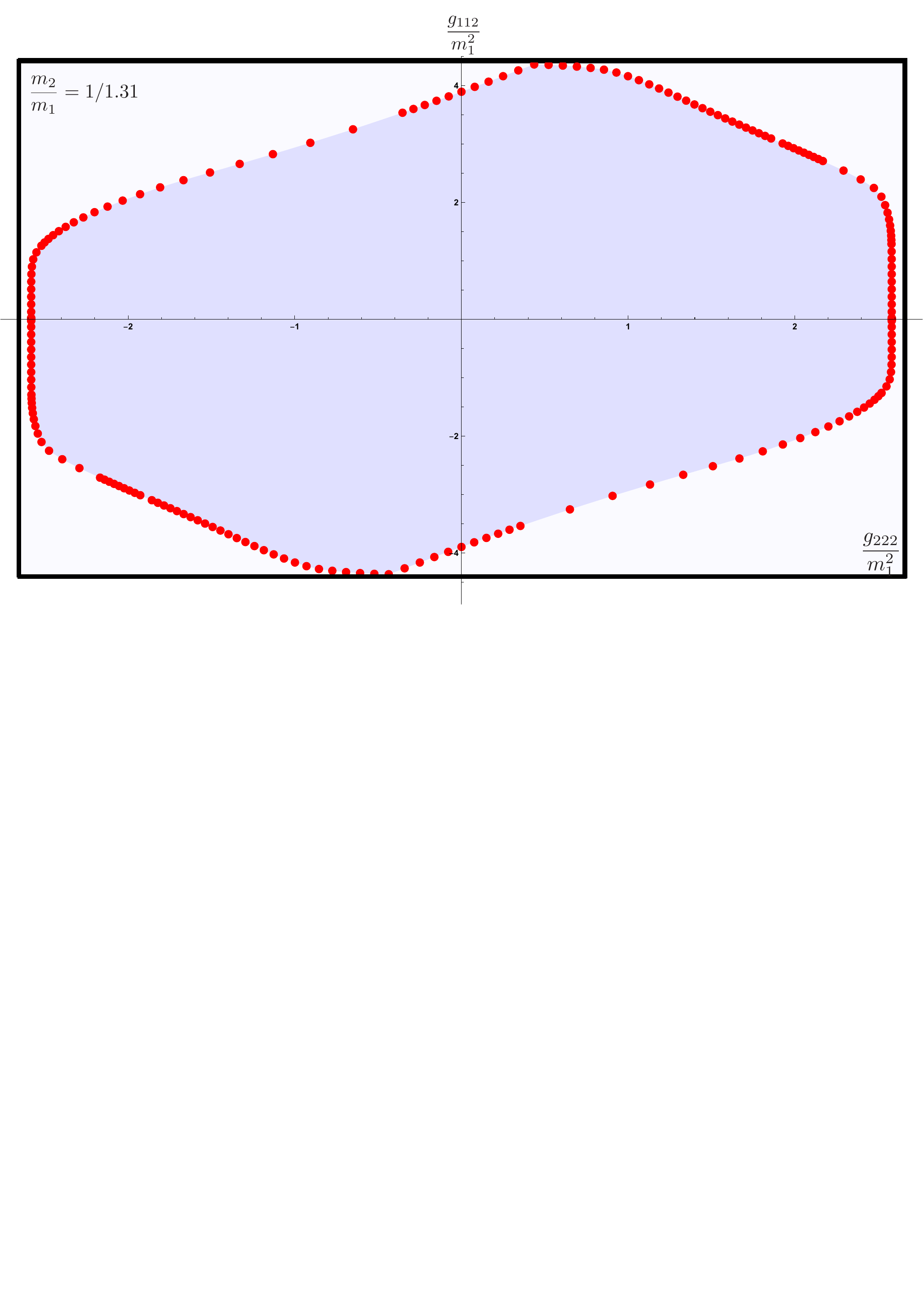} \label{pl3}
\vspace{-6.8cm}
\end{subfigure}
\medskip
\begin{subfigure}{0.5\textwidth}
\center \includegraphics[width=\textwidth]{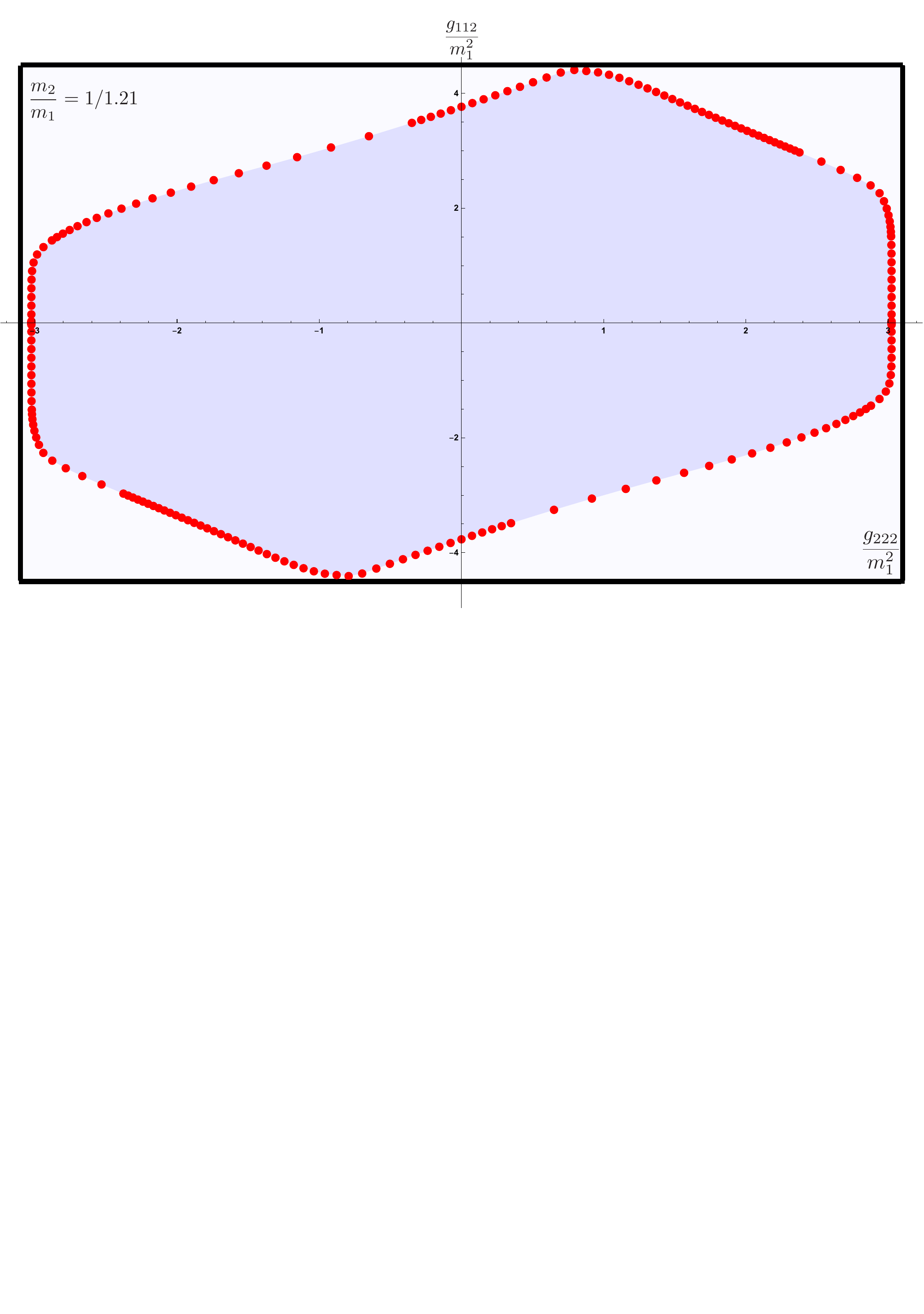} \label{pl2}
\vspace{-6.8cm}
\end{subfigure}\hspace*{\fill}
\begin{subfigure}{0.5\textwidth}
\center \includegraphics[width=\textwidth]{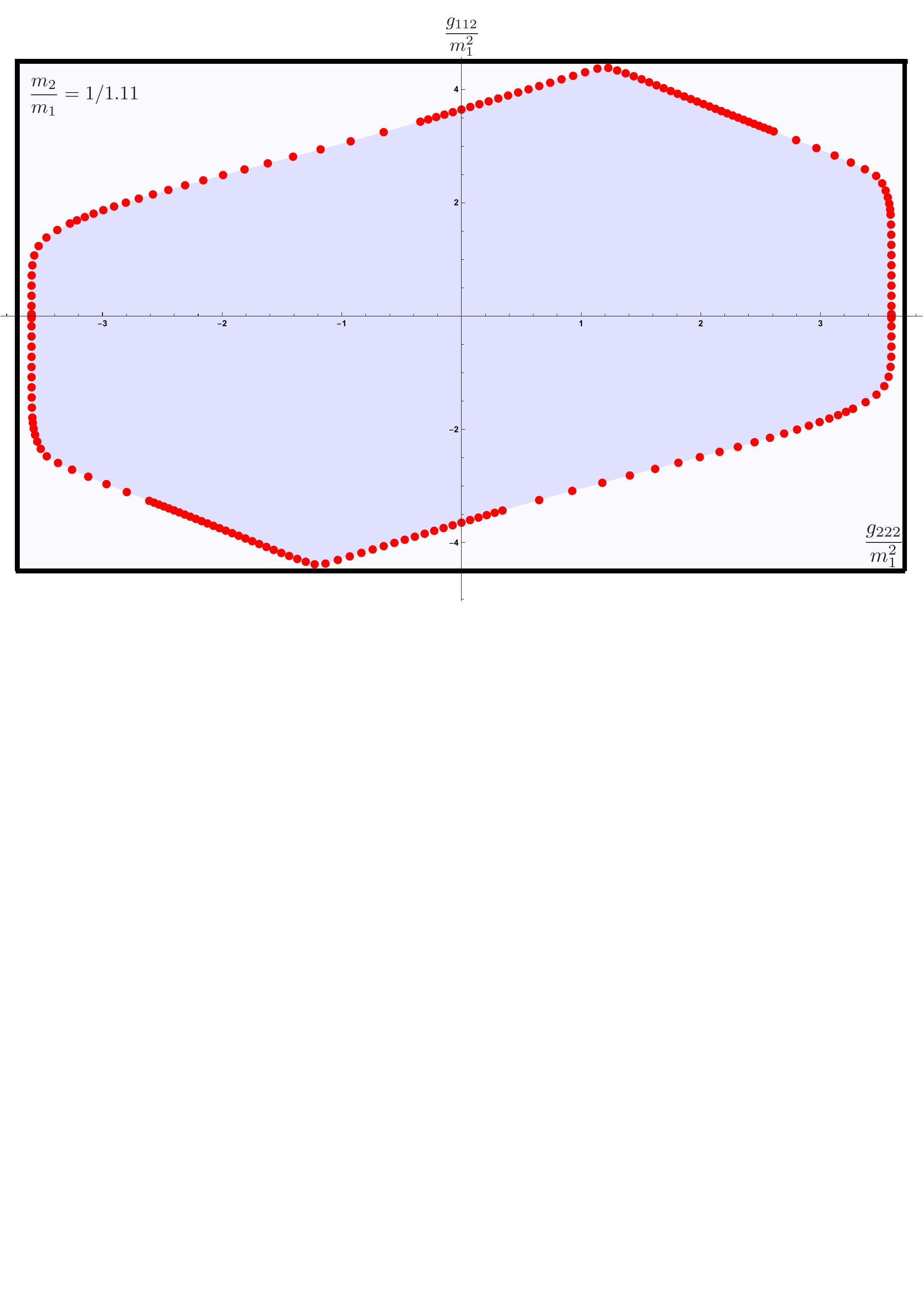} \label{pl1}
\vspace{-6.8cm}
\end{subfigure}
\begin{subfigure}{0.5\textwidth}
\center \includegraphics[width=\textwidth]{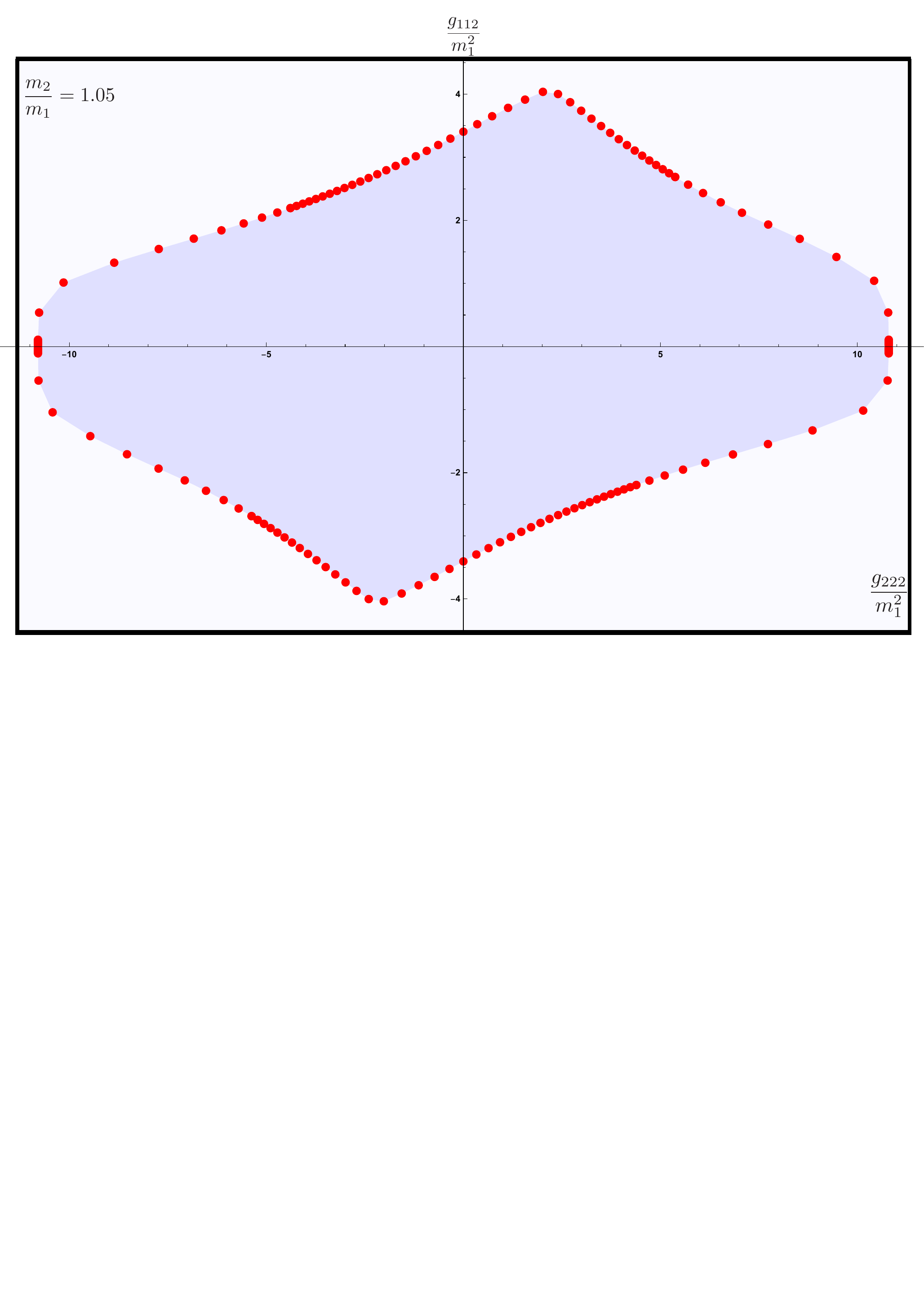} \label{ph4}
\vspace{-6.5cm}
\end{subfigure}\hspace*{\fill}
\begin{subfigure}{0.5\textwidth}
\center \includegraphics[width=\textwidth]{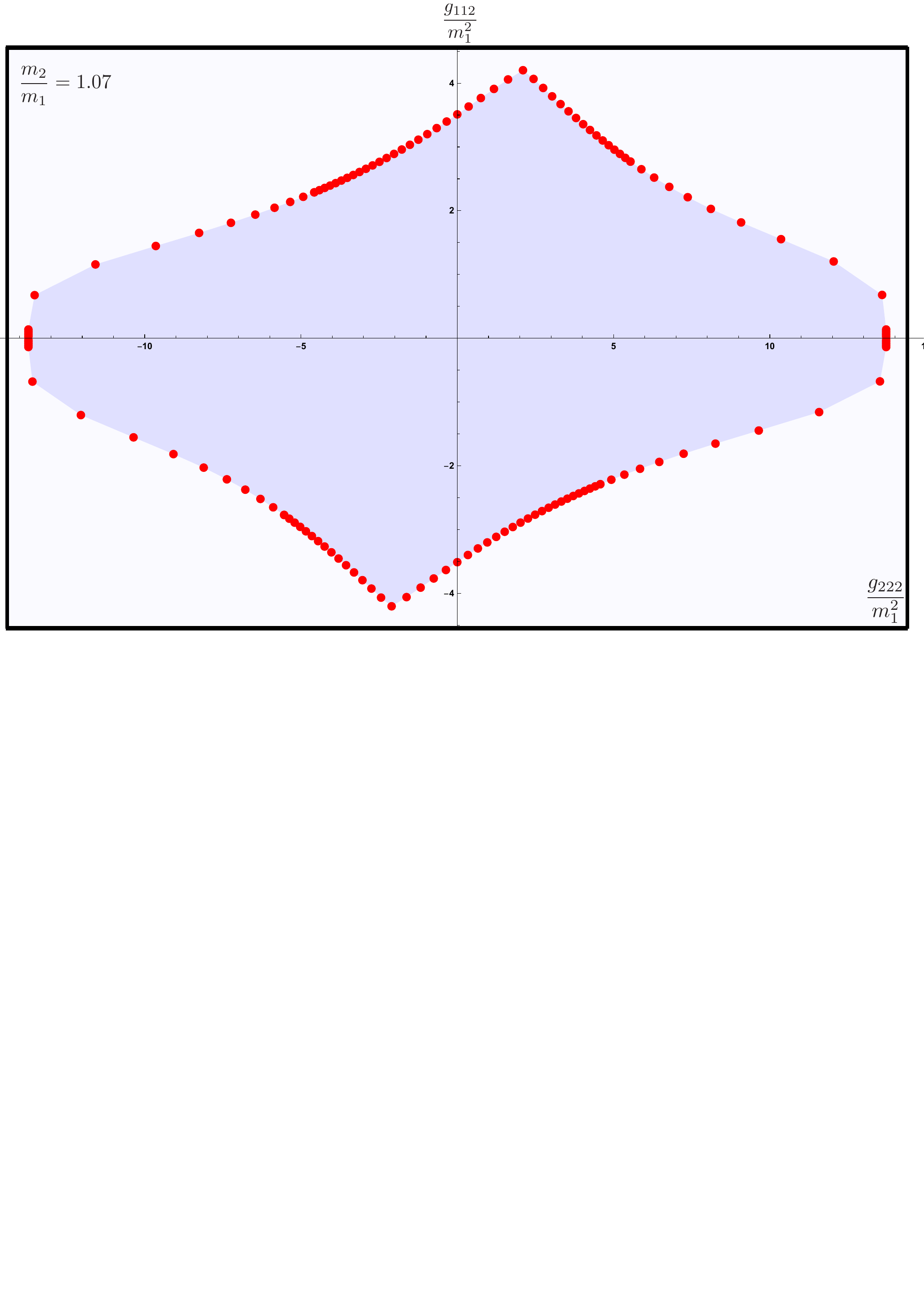} \label{ph3}
\vspace{-6.5cm}
\end{subfigure}
\medskip
\begin{subfigure}{0.5\textwidth}
\center \includegraphics[width=\textwidth]{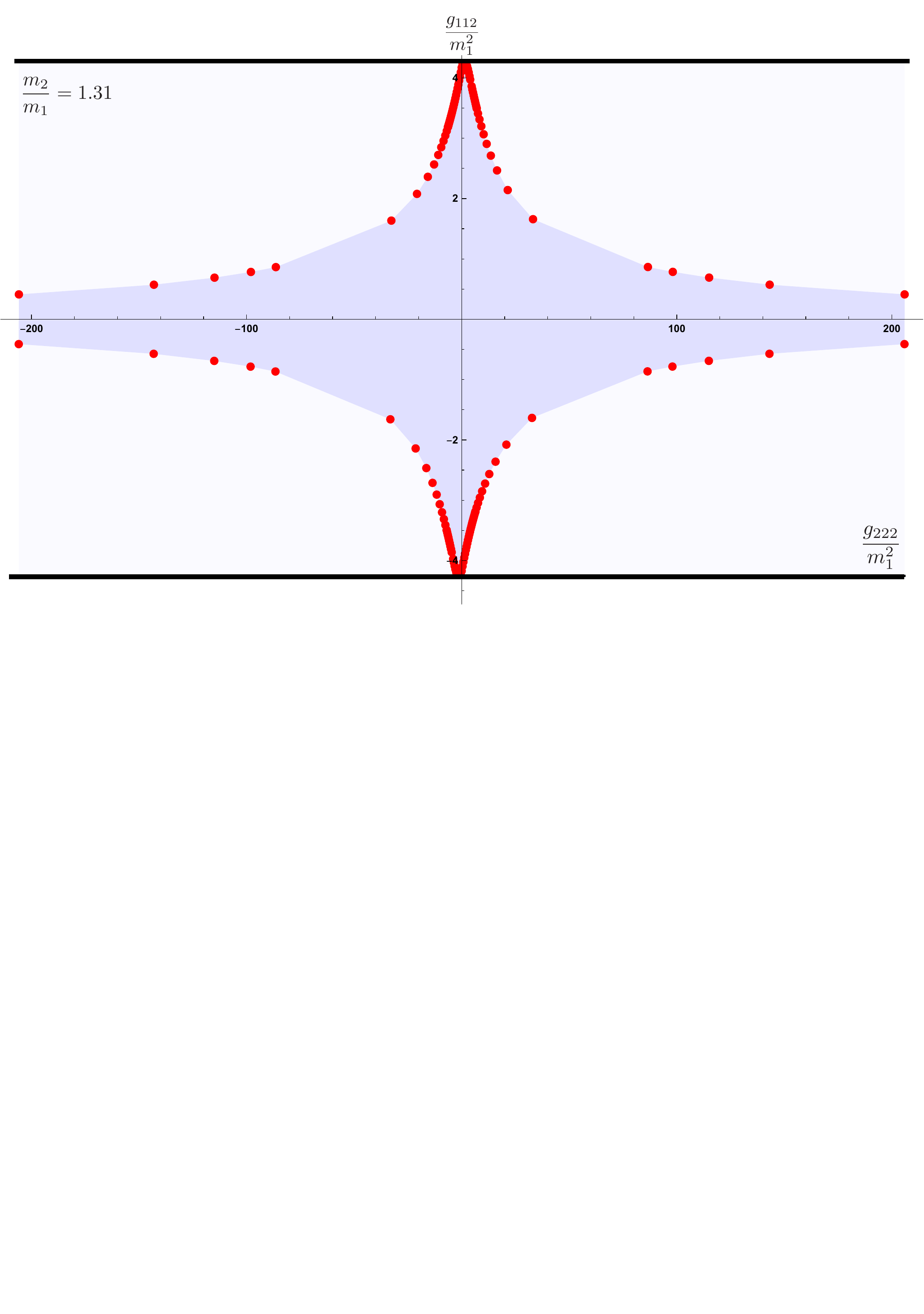} \label{ph2}
\vspace{-6.8cm}
\end{subfigure}\hspace*{\fill}
\begin{subfigure}{0.5\textwidth}
\center \includegraphics[width=\textwidth]{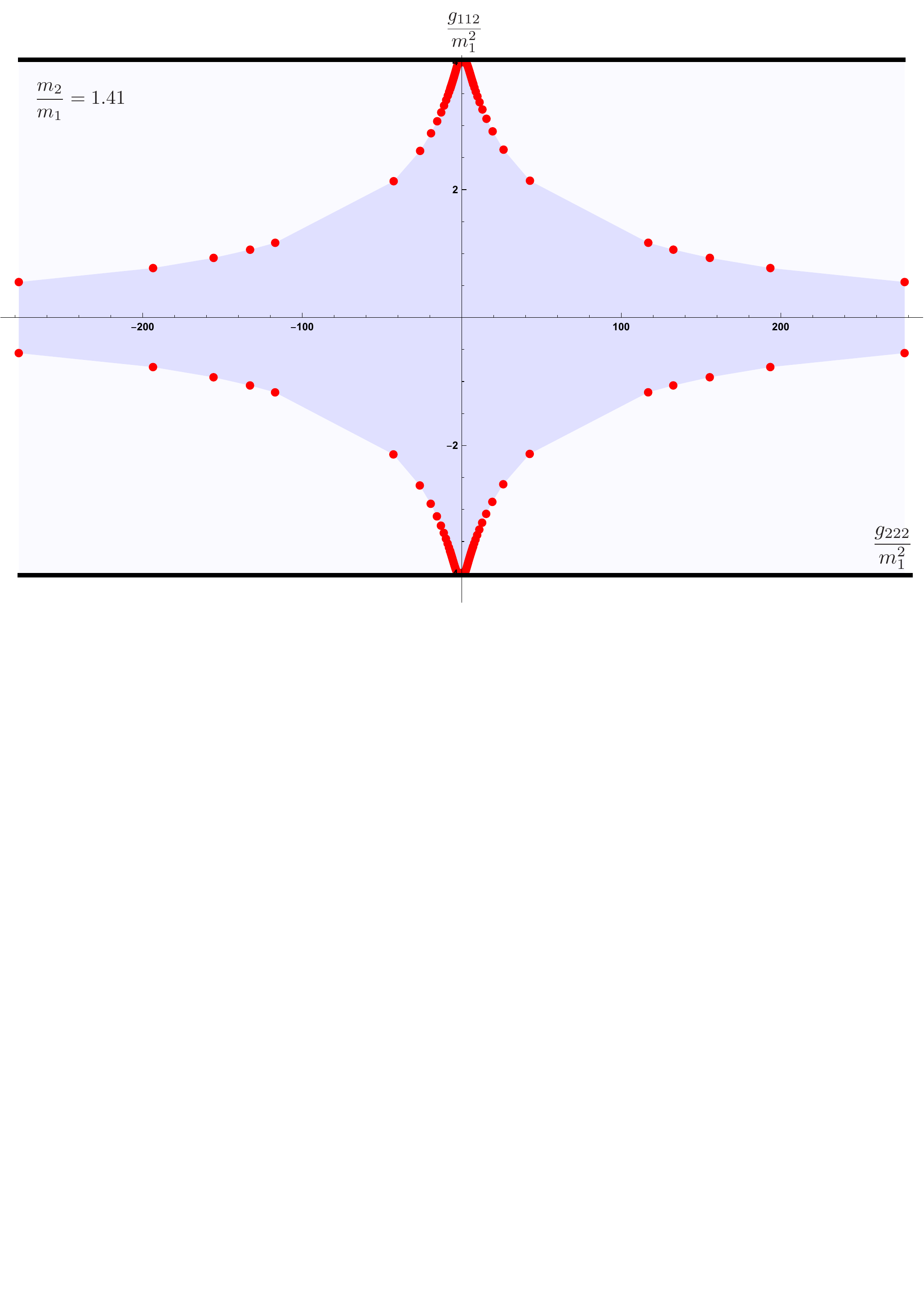} \label{ph1}
\vspace{-6.8cm}
\end{subfigure}
\vspace{-0.4cm}
\caption{Space of allowed couplings for fixed mass ratios. \emph{Horizontal and vertical solid lines:} Analytic bounds based on diagonal scattering derived in section \ref{intro}. 
\emph{Red dots:} The numeric bound obtained from \textit{all} two-to-two  
processes. Features of the panels discussed in the main text. } \label{Panels}
\end{figure}

For each mass ratio $m_2/m_1$ and for each coupling ratio $\alpha\equiv g_{222}/g_{112}$ we can now look for the maximum value of $g_{112}$. By varying all parameters we obtain a nice 3D plot which is presented in appendix \ref{3dplotAppendix}; by contrast, in this section we will restrict ourselves to showing only 2D plots that each correspond to a fixed value of $m_2/m_1$. For example, at equal masses $m_2/m_1=1$ we have figure \ref{equalmass} which shows the upper bound as a function of $\alpha$. Although holding $\alpha$ fixed is convenient for the numerics (as explained above), it is sometimes more useful to visualize the allowed space of couplings $(g_{112},g_{222})$ instead. To do this we simply multiply the $\alpha$ axis in the numerics by $g_{112}$, and in this way we can represent the same $m_2/m_1 = 1$ data as in figure \ref{gvgm1m2}. Applying the same mapping to other mass ratios in the range $m_1/\sqrt{2}<m_2<\sqrt{2}m_1$ we furthermore obtain the panels shown in figure \ref{Panels}. (As explained below, the results for $m_2<m_1/\sqrt{2}$ are somehow trivial due to screening.) For the most part, the numerical bounds in these figures significantly improve the bounds single amplitude bounds derived in the introduction which set the box sizes.

The most remarkable feature of  figures \ref{equalmass}  and  \ref{Panels} is the existence of a pronounced maximum of $g_{112}$, which is attained for a non-trivial value of the ratio $\alpha=g_{222}/g_{112}$. In particular, for equal masses this maximum (point B in figure  \ref{equalmass}) is a clear kink in the boundary of the allowed region.
It would be remarkable if there is a physical theory sitting close to this kink. As shown in figure \ref{YB}, such a theory should not be integrable.

Sometimes the numerical red dots in figures \ref{Panels} approach the solid black lines. When this happens the full numerical bounds saturate the analytically derived diagonal bounds. We see that this happens for very small $g_{112}$\footnote{The fact that the numerical points do not exactly touch the vertical lines in panels (a)-(d) in figure \ref{Panels} when $g_{112} \simeq 0$ is due to numerical convergence. In that region it would be more sensible to ask for the computer to maximise $g_{222}$ instead of $g_{112}$. This would lead to numerical saturation of the vertical lines.} and when we approach the boundaries of the mass range $m_1/\sqrt{2}<m_2<\sqrt{2}$ (for some small values of $\alpha$). This is not surprising: when $g_{112} \to 0$ we decouple the odd and even particles. Since there would be no poles in any amplitude but in $M_{22\to22}$, the bound would reduce to the single amplitude bound coming from the $22\to22$ process and yielding \begin{align}
g^2_{222}|_\text{max} &= 12\sqrt{3} m^4_2 \qquad \text{for } m_2<m_1 \text{ or}\\ g^2_{222}|_\text{max} &= 12m^4_1 \mu^2 \sqrt{3} \left(\frac{\mu^2 \sqrt{3} + 4\sqrt{\mu^2 -1}}{\mu^2 \sqrt{3} - 4\sqrt{\mu^2 -1}}\right) \qquad \text{ for } \sqrt{2} > \mu \equiv m_2/m_1>1.\end{align} This explains the analytic bound in figure \ref{g222}.
In the second case, when we approach the boundary of the mass range, we expect screening to be very important since the extended unitarity region becomes quite large. The poles in the $M_{11\to22}$ component can now be almost perfectly screened, see also appendix \ref{screeningAp2}, allowing for the diagonal amplitude bounds on $g_{112}$ to be saturated. We omitted panels for $m_2<m_1/\sqrt{2}$ since in this range we can have perfect screening for any value of $g_{112}/g_{222}$, so that the multiple amplitudes bounds in the $(g_{112},g_{222})$ plane coincide with the rectangular frame derived from diagonal processes.

Note also that there are no vertical walls in the last row of panels in figure \ref{Panels} since for $m_2>\frac{2}{\sqrt{3}}m_1$ there are no longer analytic bounds on $g_{222}$ from the $22 \to 22$ amplitude. As the extended unitarity region in $22 \to 22$ becomes bigger, it becomes increasingly more effective at screening the pole, until at $m_2/m_1 = 2/\sqrt{3}$ the s (t) channel $22\to11$ production threshold collide with the t (s) channel pole as discussed in figure \ref{screening2222}. After this mass ratio, the discontinuity across the cut can completely screen the bound state pole implying that its residue can be arbitrary. This is indeed nicely backed up by our numerics as seen in the last two panels in figure \ref{Panels} where we see that $g_{222}$ becomes unbounded at this mass range.

Finally, we can also look for the maximum value of either coupling ($g_{112}$ or $g_{222}$) leaving the other coupling arbitrary. In other words, how tall ($g_{112}$) and wide ($g_{222}$) are the darker allowed regions in (\ref{Panels}) where the allowed coupling live. Once plotted for various mass ratios, this gives figures \ref{g112} and \ref{g222} in the introduction.

Each optimal S-matrix at the boundary of the allowed coupling space is numerically seen to \emph{saturate} the extended unitarity equations (\ref{ex1111}-\ref{ex12h}). This means that the scattering of two particles of type $1$ or $2$ can never lead to multi-particle production. Processes such as $11 \to 222$ are forbidden. When dealing with 2D S-matrices, in particular extremal examples saturating unitarity such as the ones stemming from this numerical computation, we are commanded to look for integrable field theories. For $m_2 \neq m_1$, these are only possible if the inelastic amplitudes $M_{11\to22} = M^\text{Backward}_{12\to12} =  0$ but no $S$-matrices we found satisfy this condition\footnote{This is not an accident, we knew this to be the case apriori since this could only happen if the bound state poles in these amplitudes collided and cancelled or if some extra Landau poles were present. This is not a possibility in the mass range (\ref{massRange}).}  so the boundary S-matrices we find for $m_2\neq m_1$ can at most be \textit{close} to those describing good physical theories.
This still leaves the possibility of finding interesting physical theories along the equal mass line $m_2 = m_1$.\footnote{Actually, this line is a one-dimensional kink in the maximal coupling surface described in detail in appendix \ref{3dplotAppendix}.} 

\subsection{(Surprises at) the \texorpdfstring{$m_2=m_1$}{m2=m1} line} \la{Theories}

\begin{figure}[t!]
\center \includegraphics[width=\textwidth]{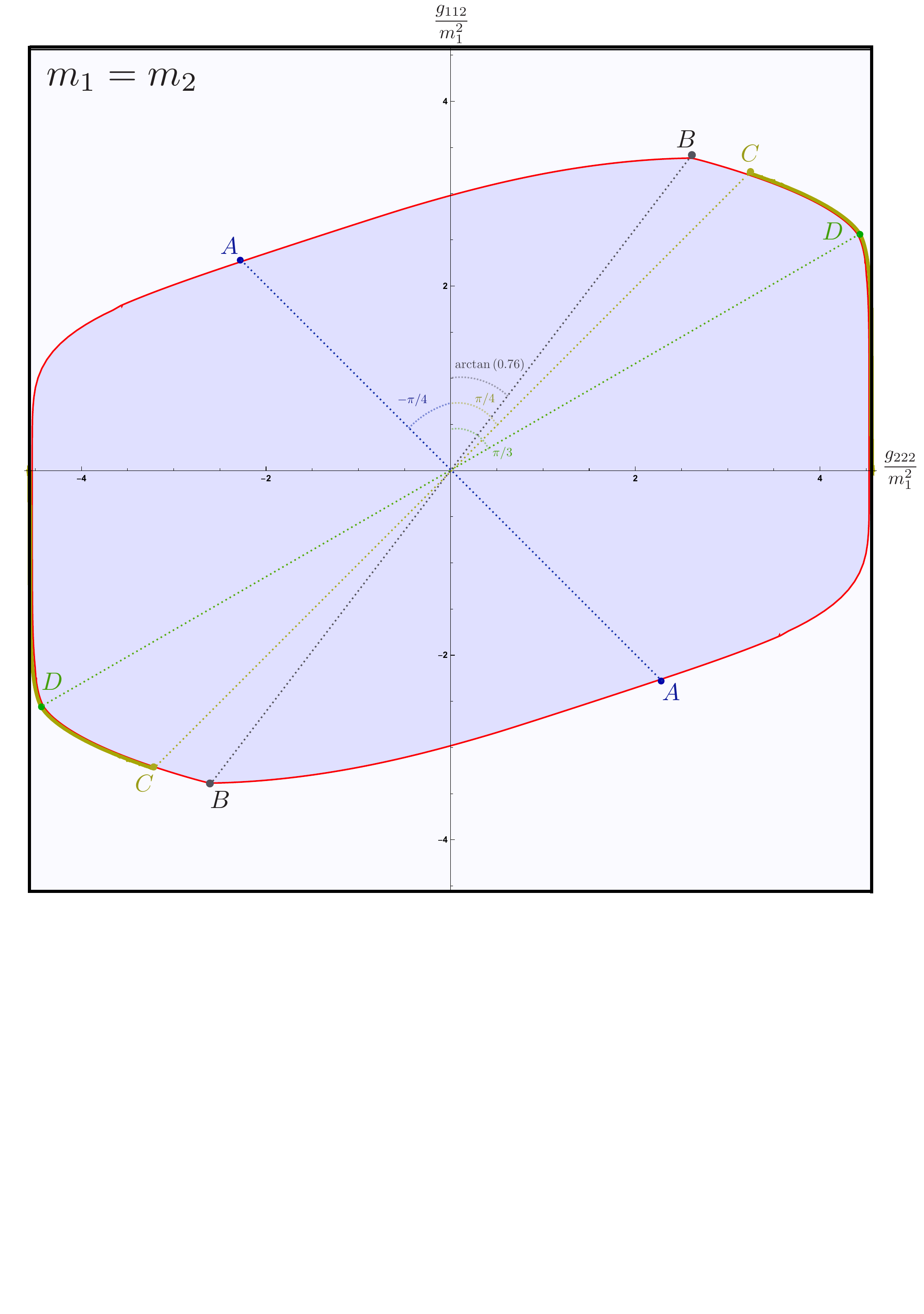}
\vspace{-7cm}
\caption{Allowed space of couplings for equal masses. A: Potts massive field theory, B: Maximum coupling $g_{112}$, $C$: Beginning of elliptic deformation line, $D$: Supersymmetric sine-Gordon (along the elliptic deformation line).} \label{gvgm1m2} 
\end{figure}

Indeed, nice surprises are to be found in the $m_2=m_1$ line depicted in the two equivalent figure \ref{gvgm1m2} (depicting the space of allowed couplings) and figure \ref{equalmass} (for the maximum coupling $g_{112}$ as a function of the coupling ratio $\alpha\equiv g_{222}/g_{112}$). Although equivalent these two figures highlight different aspects of this interesting line so it is worth having both in mind. 

\begin{figure}[t!]
\center \includegraphics[width=\textwidth]{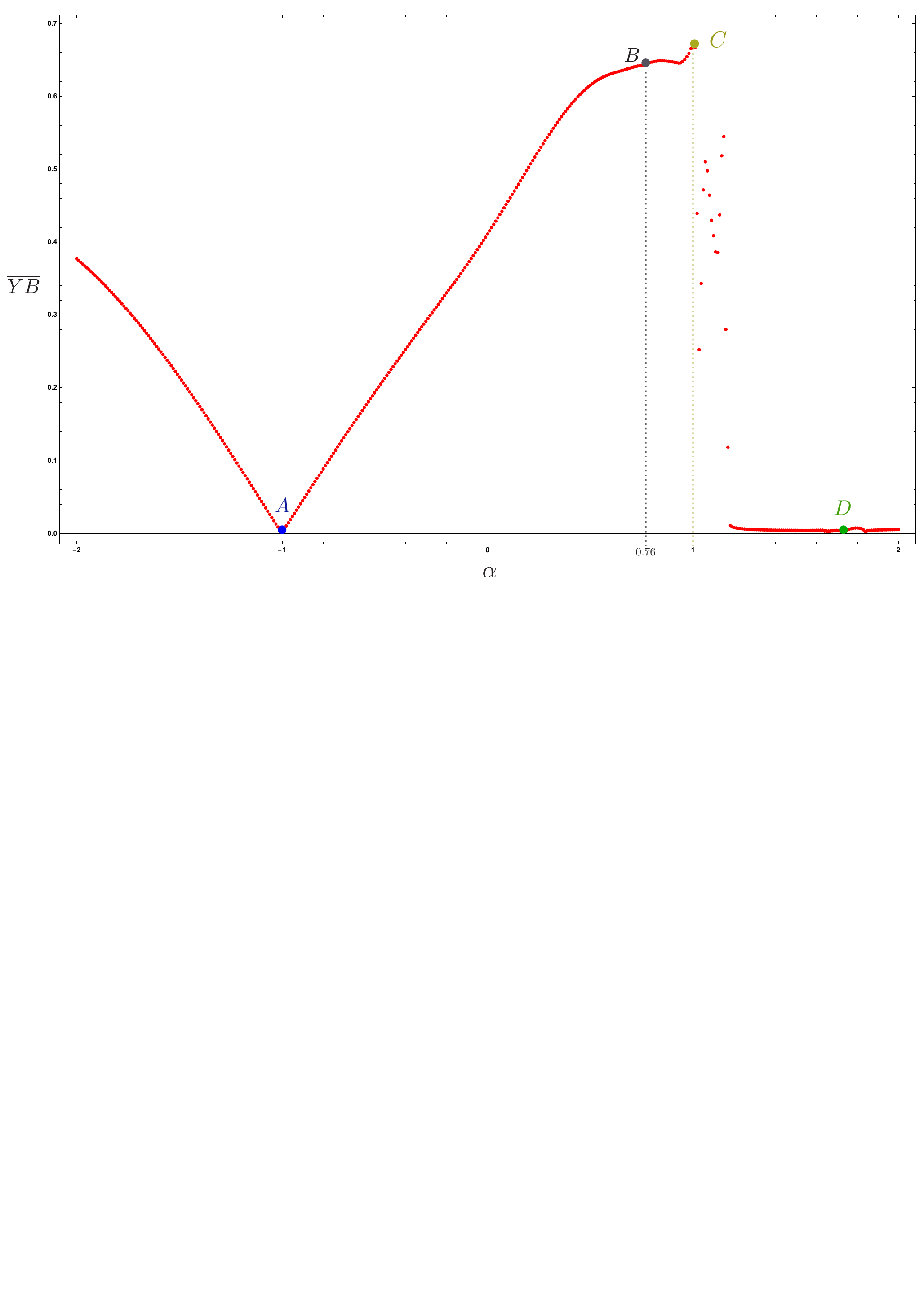}
\vspace{-13cm}
\caption{Because of $\mathbb{Z}_2$ symmetry, some Yang-Baxter equations are automatically satisfied. $\overline{YB}$ is an average over the non-trivial Yang-Baxter equations, with external rapidities at physical values. The qualitative features of the plot do not depend on averaging over the equations nor on the values of the external rapidities, taken here to be $\theta_1 = 1/2$, $\theta_2 = 0$, $\theta_3 = -1/3$.  Once again, A: Potts massive field theory, B: Maximum coupling $g_{112}$, $C$: Beginning of elliptic deformation line, $D$: Supersymmetric sine-Gordon. } \label{YB} 
\end{figure}

As concluded in the last section, the line $m_2=m_1$ is where our hope lies if we are to match the S-matrices we obtained numerically with physical integrable theories. This necessary condition is \textit{not} sufficient. For an extremal S-matrix to correspond to an integrable theory it should also obey the factorization conditions encoded in the so-called Yang-Baxter equations \cite{YangDeltaFunction,ZOn}. In figure \ref{YB} we see how our extremal S-matrices fail to satisfy these conditions as we move along the allowed coupling region (by sweeping $\alpha$).

Before unveiling which analytic S-matrices we successfully identified along the $m_1=m_2$ line let us go over these numerics in some detail: 
We observe that for large negative $\alpha$ Yang-Baxter is violated until we reach $\alpha=-1$ (i.e. when the couplings are equal up to a sign, $g_{222}=-g_{112}$) at which point Yang-Baxter is beautifully satisfied. This point is isolated; immediately to the right of  $\alpha=-1$ Yang-Baxter fails again.  It is curious to note that this special point -- our first candidate for a physical integrable theory -- marked with an $A$ in figures \ref{gvgm1m2} and \ref{equalmass} looks absolutely  innocent there, without any apparent kink features. As we increase $\alpha$ further into positive values we reach point $B$ for $\alpha \simeq 0.76$ where the coupling $g_{112}$ is maximal. As seen in figures \ref{gvgm1m2} and \ref{equalmass} this point is a nice kink. Since it does not obey Yang-Baxter, however, this can hardly correspond to a physical theory. As we increase $\alpha$ further we reach $\alpha= +1$ marked with a $C$ in figures \ref{gvgm1m2} and \ref{equalmass} where again something interesting happens. At that point something goes unstable as far as testing of Yang-Baxter goes indicated by the shower of points in figure \ref{YB} from $\alpha=1$ until somewhere around $\alpha \simeq 1.2$. Once this mess settles we observe a nice line where Yang-Baxter is satisfied throughout! A particularly nice point along that line is point $D$ located at $\alpha=\sqrt{3}$. That point actually is the one furthest from the origin in figure \ref{gvgm1m2}, in other words, it maximizes $g_{112}^2+g_{222}^2$ and as described below it corresponds to a nice known physical theory. 

Now we unveil what we found about these points. In short (setting $m_1=1$ here):
\begin{itemize}
\item Point $A$ is a massive deformation of the three state Potts Model. 

Here $g_{112}^\text{max}=-g^\text{max}_{222}= \sqrt{3\sqrt{3}} \simeq 2.28 $. 
\item Point $B$ is yet to be identified. We do not know the analytic form of the corresponding S-matrix; Since it does not obey YB it can at most be close to a physical theory. 

Here $g_{112}^\text{max}\simeq 1/0.76\, g^\text{max}_{222} \simeq 3.38 $.
\item Point $D$ is (an analytic continuation of the lightest breather S-matrix of) the super-symmetric Sine-Gordon model. 

Here $g_{112}^\text{max}=-g^\text{max}_{222}\simeq 2.56 $.
\item There is a line going from point $C$ at $\alpha=1$ all the way to $\alpha= \infty$ where the optimal S-matrix is given by an elliptic deformation of the super-symmetric Sine-Gordon. We are unaware of a physical theory with this S-matrix. Point $C$ is the tip of the elliptic deformation where it becomes hyperbolic. 

Here $g_{112}^\text{max}=-g^\text{max}_{222}= \sqrt{6\sqrt{3}} \simeq 3.22$.
\end{itemize}
For comparison recall that the analytic diagonal bounds were $|g_{112}^\text{max}|=|g^\text{max}_{222}|= 4.56$. 

We will now slowly build up towards those conclusions. The first observation, reviewed in appendix \ref{solvable}, is that the full numerical optimization problem can actually be diagonalized and solved exactly for $\alpha = \pm 1$, that is when the two physical couplings are the same up to a sign. For $\alpha=-1$ the result is the S-matrix of the massive deformation of the three-state Potts model \cite{Potts}\footnote{Here we rotated the one particle basis from \cite{Potts} as $|A\> = e^{i \pi/4} \frac{|2\> - i |1\>}{\sqrt{2}} $, $|A^\dagger\> = e^{- i \pi/4} \frac{i |2\> - |1\>}{\sqrt{2}} $, so that the charge conjugation operator is diagonalized. This operator is to be interpreted as the $\mathbb{Z}_2$ symmetry generator. In the $|A\>, |A^\dagger\>$ basis the S-matrix is diagonal and that is why we can solve this point exactly, see Appendix \ref{solvable}. }
\begin{align}
M_{11\to11} = M_{22\to22} = M_{12 \to12}^\text{Forward} =&3m_1^4/(\sqrt{3}m^2_1-\sqrt{(4m^2_1-s) s})-\sqrt{3}m_1^2 -3 \sqrt{(4m_1^2-s) s}\nonumber \\
 M_{11\to22} = -  M_{12\to12}^\text{Backward} = & \sqrt{3} (s-2m_1^2) \sqrt{(4m_1^2-s) s}/({\sqrt{3}m_1^2-\sqrt{(4m_1^2-s) s}}) \,.
   \label{MPotts}
   \end{align}
From this solution we read off $  g_{112}=-g_{222}= \sqrt{3\sqrt{3}} m_1^2$ which matches perfectly with point $B$ in figures \ref{gvgm1m2} and \ref{equalmass}. In appendix \ref{Potts}, we briefly review the 3 state Potts field theory.

As also explained in appendix \ref{solvable}, the point $\alpha=+1$ is the other point where we can find a clever change of basis to diagonalize our problem and compute the maximal couplings analytically to find $  g_{112}=+g_{222}=  \sqrt{6\sqrt{3}} m_1^2$ which again matches perfectly with point $C$ in figures \ref{gvgm1m2} and \ref{equalmass}. What we also observe in the process of deriving that analytic solution is that the S-matrix saturating this bound is not unique; there are zero modes. This is probably the explanation of the shower in figure \ref{YB}. These zero modes are probably only present for $\alpha=+1$ but in the vicinity of this point there is probably still some small numerical remnant thereof.  We thus expect the shower in figure \ref{YB} to be nothing but a zero-mode related numerical artifact; the true solution to the optimization problem probably obeys Yang-Baxter for any $\alpha>1$. Yet, since this seems to be a zero mode issue, we expect the coupling as predicted by the numerics to still be correct. We will soon provide very strong evidence for these claims. 

Point $D$ for $\alpha=\sqrt{3}$ is a potentially interesting point if we interpret the $\mathbb{Z}_2$ symmetry as fermion number and think of particles $1$ and $2$ a Majorana fermion and a boson respectively. Then the condition $g_{222}/g_{112} =\sqrt{3}$ would follow for theories where these two particles are part of a $\mathcal{N}=1$ supersymmetry multiplet, see also \cite{ToAppearSUSY}. 
Inspired by this -- and by \cite{ToAppearSUSY} -- we tried to compare the optimal S-matrices at $g_{222}/g_{112} =\sqrt{3}$ to those of the lightest breathers of the super-symmetric sine-Gordon theory.\footnote{Strictly speaking we are comparing with an analytic continuation of that S-matrix since our bound-states have mass equal to the external particles while the next-to-lighests breathers of super-symmetric sine-Gordon have mass bigger than $\sqrt{2}$ times that of the external particles. In \cite{Paper2} the usual bosonic sine-Gordon S-matrix was identified as the theory with the largest coupling in the S-matrix of the lightest particle with a single bound-state of mass $m_b$. When $m_b >\sqrt{2}$ this is kosher but as $m_b<\sqrt{2}$ (and in particular for $m_b=1$) we also need to extend the definition of the SG S-matrix beyond its original mass range. In that case it amounted to multiplying the S-matrix by $-1$. Here the situation is morally the same but the modification ends up a bit more complicated. {This means that here -- as there -- we do not know a physical theory and we can only write an exact S-matrix that saturates the bound. }} 
Beautifully, although we only impose the SUSY condition at the level of the couplings, we see that SUSY emerges at the level of the full S-matrix elements and indeed the optimal S-matrix saturating our bounds at point $D$ is an analytical continuation of the lightest breather supermultiplet of the super-symmetric Sine-Gordon! Unfortunately, while we are able to check this to very convincing numerical accuracy we have no analytic derivation of this statement. For completeness, here are some super SG formulae \cite{SSG}.

The lightest breather supermultiplet SSG S-matrix $S^{(1,1)}_{SSG}\left(\theta\right)$ is equal to\footnote{In the basis $|11\rangle,|12\rangle,|21\rangle,|22\rangle$ so that the second (third) element on the second row is the forward (backward) $12\to 12$ component, for instance.}
\begin{align}
-\frac{\sinh\left(\theta\right) + i \sin\left(\gamma\right)}{\sinh\left(\theta\right) - i \sin\left(\gamma\right)} \,Y(\theta) Y(i \pi - \theta)
\left(
\begin{array}{cccc}
\frac{i \sin\left(\gamma/2\right)}{\sinh\left(\frac{\theta }{2}\right)
   \cosh\left(\frac{\theta }{2}\right)}-1  & 0 & 0 &\frac{\sin\left(\gamma/2\right)}{
   \cosh\left(\frac{\theta }{2}\right)} \\
 0 & 1 & \frac{i \sin\left(\gamma/2\right)}{ \sinh\left(\frac{\theta }{2}\right) }& 0 \\
 0 & \frac{i \sin\left(\gamma/2\right)}{ \sinh\left(\frac{\theta }{2}\right) } & 1 & 0 \\
\frac{\sin\left(\gamma/2\right)}{
   \cosh\left(\frac{\theta }{2}\right)}  & 0 & 0 &  \frac{i \sin\left(\gamma/2\right)}{\sinh\left(\frac{\theta }{2}\right)
   \cosh\left(\frac{\theta }{2}\right)}+1\\
\end{array}
\right)\nonumber
\end{align}
where 
\begin{align}
Y\left(\theta\right) = &\frac{\Gamma\left(-i \theta/2\pi \right)}{\Gamma\left(1/2 -i \theta/2\pi \right)}\\ &\!\!\!\!\!\!\!\!\!\!\!\!\!\!\!\!\!\!\!\!
\times \prod_{n=1}^\infty
\frac{\Gamma\left(\gamma/2\pi  - (i \theta/2\pi) + n \right)\Gamma\left(-\gamma/2\pi  - (i \theta/2\pi) + n -1\right) \Gamma^2\left(-(i\theta/2\pi) + n -1/2 \right)}
{\Gamma\left(\gamma/2\pi  - (i \theta/2\pi) + n + 1/2\right)\Gamma\left(-\gamma/2\pi  - (i \theta/2\pi) + n -1/2\right) \Gamma^2\left(-(i\theta/2\pi) + n -1 \right)}
\nonumber.
\end{align}
and where $\gamma$ is fixed so that the bound state mass is equal to $1$. That is $\gamma = 2\pi/3$. Note that even though the overall scalar factor in the S-matrix is invariant under $\gamma = 2\pi/3 \leftrightarrow \gamma = \pi/3$, the matrix part is not. This is the sense in which our S-matrix is an analytic continuation of SSG. (compare with SG, in which picking $m_b = 1$ instead of $m_b = \sqrt{3}$ only leads to an overall minus sign). More generally $m_b/m_1=2\cos{\gamma/2}$ and the physical mass range for SSG is $2>m_b>\sqrt{2}$.

Two-particle $\mathbb{Z}_2$ symmetric solutions of the Yang-Baxter equations are classified \cite{YBclass}. It is natural that if an extension of SSG exists with an extra parameter, that it is given by an elliptic solution of the Yang-Baxter equations. In fact, examining the classified solutions we see that the only good candidate for being promoted to an S-matrix with all the symmetry properties we have and that reduces to SSG in the trigonometric limit is solution 8VII of \cite{YBclass}, which is equivalent to
\begin{align}
\mathbf{ED}\left(\theta\right) \equiv
& \left(
\begin{array}{cccc}
\epsilon \frac{\text{dn}(\theta  \omega |\kappa ) \text{sn}(\gamma  \omega |\kappa )}{  \text{cn}(\theta 
   \omega |\kappa ) \text{sn}(\theta  \omega |\kappa )}-\text{dn}(\gamma  \omega |\kappa )& 0 & 0 & \epsilon \frac{\text{dn}(\theta 
   \omega |\kappa ) \text{sn}(\gamma  \omega |\kappa )}{  \text{cn}(\theta  \omega |\kappa )} \\
 0 & 1 &\epsilon \frac{\text{sn}(\gamma  \omega |\kappa )}{  \text{sn}(\theta  \omega |\kappa )} & 0 \\
 0 & \epsilon\frac{\text{sn}(\gamma  \omega |\kappa )}{  \text{sn}(\theta  \omega |\kappa )} & 1 & 0 \\
 \epsilon \frac{\text{dn}(\theta  \omega |\kappa ) \text{sn}(\gamma  \omega |\kappa )}{  \text{cn}(\theta  \omega |\kappa
   )} & 0 & 0 & \text{dn}(\gamma  \omega |\kappa )+\epsilon \frac{\text{dn}(\theta  \omega |\kappa ) \text{sn}(\gamma  \omega |\kappa
   )}{  \text{cn}(\theta  \omega |\kappa ) \text{sn}(\theta  \omega |\kappa )}  \\
\end{array}
\right)
\label{ed1},
\end{align}
where we normalised by the $12 \to 12$ forward component so that comparison with SSG is easier. $\epsilon = \pm 1$ from YB. The $s$-channel poles of $12 \to 12$ forward and backward correspond to the flow of a particle of type one, and therefore the residues of this two amplitudes at the $\theta = 2\pi i/3$ pole must coincide. This fixes $\epsilon\, \text{sn}(\gamma  \omega |\kappa ) = \text{sn}(\frac{2 \pi i}{3} \omega |\kappa )$. Crossing symmetry together with the fact that in the trigonometric limit $\kappa \to 0$ we have to recover SSG fix $\omega = -\frac{i}{\pi} K(\kappa)$ where $K$ is the complete elliptic integral of the first kind (more precisely, crossing gives  $\omega = -\frac{i}{\pi} (2n +1) K(\kappa)$ with $n$ an integer. The $\kappa \to 0$ limit fixes n).  This completely fix a crossing symmetric matrix structure up to one free parameter, $\kappa$, which hopefully is unconstrained. There is a miracle going on. For our amplitudes, we have that 
\beq
\underset{\theta = 2 \pi i/3}{\text{res}} M_{22\to22} \underset{\theta = 2 \pi i/3}{\text{res}}  M_{11\to11} = \left(\underset{\theta = 2 \pi i/3}{\text{res}}  M_{11\to22}\right)^2
\eeq
and, moreover,
\beq
\underset{\theta = 2 \pi i/3}{\text{res}} M_{11\to11}= \underset{\theta = 2 \pi i/3}{\text{res}}  M^\text{Forward}_{12\to12}
\eeq
If (\ref{ed1}) is a candidate of matrix structure of the S-matrices saturating the numerical bounds, we must have the same relation between the respective components. It turns out that this holds automatically for any $\kappa$ after all the conditions above are imposed. Otherwise this would fix $\kappa =0$ and we would conclude that there are no Yang-Baxter deformations respecting the symmetries and spectrum of our problem.
So all we need to do now is to unitarize and introduce the poles. Note that
\beq
\mathbf{ED}\left(\theta\right) \mathbf{ED}\left(-\theta\right) = \left(1-\frac{\text{sn}\left(\left.\frac{2 K(\kappa )}{3}\right|\kappa \right)^2}{\text{sn}\left(\left.\frac{i \theta 
   K(\kappa )}{\pi }\right|\kappa \right)^2}\right) \mathbb{I} \equiv g\left(\theta\right) \mathbb{I}
\eeq 
and $g(\theta) \geq 1$ for $\theta \in \mathbb{R}$. Therefore, as follows from \cite{Paper2}, to unitarize $\mathbf{ED}$ we just need to multiply it by
\beq
U(\theta) = - i \sinh\left(\theta\right) \text{exp} \left(-\int_\infty^\infty \frac{d\theta'}{2\pi i}\frac{\log\left(g^{-1}(\theta')/\sinh(\theta')\right)}{\sinh\left(\theta - \theta' + i \epsilon\right)} \right)\,, 
\eeq
while to introduce the poles, we multiply by $\text{CDD}_\text{pole}$ with direct channel pole at $2\pi i/3$, 
\beq
\text{CDD}_\text{pole}\left(\theta\right) = \frac{\sinh\left(\theta\right) + i \sin\left(2\pi/3\right)}{\sinh\left(\theta\right) - i \sin\left(2\pi/3\right)}
\eeq
At the end of the day, a candidate for a unitary, crossing symmetric, integrable deformation of the supersymmetric sine-Gordon reads
\beq
\mathbb{S}_{\mathbf{ED}}\left(\theta\right)  = -\text{CDD}_\text{pole}\left(\theta\right)  U(\theta)  \mathbf{ED}\left(\theta\right) \la{guess}
\eeq
Points $D$ and $C$ with $\alpha=\sqrt{3}$ and $\alpha= 1$ would now correspond to $\kappa\to 0 $ and $\kappa\to 1$ respectively. As $\kappa \to -\infty$, $\alpha \to \infty$. We can now compute the couplings associated to the elliptic deformation (\ref{guess}), cross our fingers and compare those couplings with the numerics of figures \ref{gvgm1m2} and \ref{equalmass}. The elliptic deformation analytic results are the solid {chartreuse}\footnote{Chartreuse, of selcouth beauty, is a colour half-way between yellow and green.} lines in those figures. The agreement could not be better. Note that the agreement goes all the way to point $C$ and since by construction the elliptic solution obeys Yang-Baxter, the shower in figure \ref{YB} should indeed be a simple zero-mode related numerical artifact. To make precise the elliptic notation used here, we present in appendix \ref{Mathematica} a representation of this S-matrix in Mathematica friendly notation, ready to be copy pasted so the reader can more easily explore this exotic solution. 

An obvious question is whether this elliptic deformation corresponds to a nice physical theory.\footnote{We thank Davide Gaiotto for illuminating discussions on related topics.} Since the supersymmetric sine-Gordon we encountered here is not a totally kosher theory but an analytic continuation thereof, it is natural to first extend this analysis to the mass range were the super-sine Gordon breather lives and to study its elliptic deformation for those more physical set of parameters. This is beyond the scope of this paper and is currently being investigated in \cite{ToAppearSUSY}.

The $m_2=m_1$ line was indeed full of surprises. 

\section{QFT in AdS}
\label{sec:qftinads}
\newcommand{\D}{\Delta}
\newcommand{\vev}[1]{{\langle #1 \rangle}}

In the previous section we have numerically explored the space of scattering amplitudes that allow for a Mandelstam representation and we found examples of amplitudes that appear to maximize couplings subject to the unitarity constraints. As explained in footnote \ref{firstF}, these extremal coupling constants are not true `upper bounds': although our numerical results appear to have converged, a numerically more refined ansatz will find slightly larger values.

An orthogonal approach to the extremization of three-point couplings in field theories was developed in \cite{Paper1}. The idea is to consider a field theory in an AdS background and investigate the `boundary' correlation functions that are so familiar from the AdS/CFT correspondence. In our setup gravity is non-dynamical and this translates into the absence of a stress tensor among the set of boundary operators. Nevertheless it is natural to claim \cite{Paper1} that these correlation functions obey all the other axioms of a unitary CFT, including crossing symmetry, making them amenable to a numerical bootstrap analysis as in \cite{CFTNoFlavour}. In this way any general constraints on CFT data directly imply corresponding constraints for QFTs in AdS, and by extrapolating these results to the \emph{flat-space limit} we can get constraints on flat-space QFTs as well. (For a gapped QFT in AdS$_2$ scaling dimensions and masses are related as $m^2 R^2 = \Delta(\Delta-1)$ and therefore the flat-space limit is typefied by sending all scaling dimensions $\Delta\to\infty$ whilst keeping ratios fixed, $\Delta_i/\Delta_j \to m_i/m_j$.)

The QFT in AdS approach uses CFT axioms to provide rigorous upper bounds, at least modulo our extrapolation procedures. It does not assume analyticity or any particular behavior at large complex energies, and the unitarity constraints are phrased in terms of reflection positivity rather than in terms of probabilities. And yet it was shown in \cite{Paper1} that it provides upper bounds on flat-space couplings that are numerically \emph{equal} (up to three significant digits in some cases) to the extremal couplings obtained with the S-matrix bootstrap methods. In this section we demonstrate that this striking equivalence was not just a fluke by employing once more the QFT in AdS approach to reproduce some of the previous S-matrix bootstrap results from conformal crossing equations.

\subsection{Setup}
\label{subsec:setupextrapol}
We will consider four-point functions of operators on the real line, which we think of as the boundary of an AdS$_2$ space with curvature radius $R$. There are two distinguished operators $\phi_1$ and $\phi_2$ of dimensions $\D_1$ and $\D_2$, which correspond to the two single-particle states of the setup described above -- in particular it is understood that $\D_i(\D_i - 1) = m_i^2 R^2$ for $i = 1,2$. Besides the assumed $\mathbb Z_2$ symmetry, under which $\phi_1$ is odd and $\phi_2$ is even, we will also assume that the QFT is parity invariant and that $\phi_1$ and $\phi_2$ are both parity even.\footnote{It is often helpful to think of the parity odd operators as vectors. Indeed, they are equivalent in one dimension because the rotation group is reduced to the parity group $\mathbb Z_2$ and which has only one non-trivial irreducible representation.} The OPEs are, schematically,
\beq
\begin{split}
\phi_1 \times \phi_1 &= \mathbf 1 + \lambda_{112} \phi_2 + (\text{parity and $\mathbb Z_2$ even operators with $\D \geq 2 \text{min}(\D_1,\D_2)$})\\
\phi_2 \times \phi_2 &= \mathbf 1 + \lambda_{222} \phi_2 + (\text{parity and $\mathbb Z_2$ even operators with $\D \geq 2 \text{min}(\D_1,\D_2)$})\\
\phi_1 \times \phi_2 &= \lambda_{112} \phi_1 + (\text{any $\mathbb Z_2$ odd operators with $\D \geq \D_1 + \D_2$})
\end{split}
\eeq
Here the (non-)appearance of $\phi_1$ and $\phi_2$ on the right-hand sides is dictated by $\mathbb Z_2$ symmetry. The other operators are meant to correspond to multi-particle states for the QFT in AdS and their minimal scaling dimension mimicks the beginning of the two-particule cuts in the corresponding scattering amplitudes. The parity properties are dictated by the parity of the operators on the left-hand side. We should add that the OPE coefficients $\lambda_{ijk}$ are related to bulk couplings $g_{ijk}$ via
\beq \label{CtoT}
g_{123}/m_0^2 = \lambda_{123}C(\D_0; \D_1,\D_2,\D_3)
\eeq
with the unsightly relative normalization coefficient \cite{Freedman:1998tz}
\beq
\begin{split}
C(\D_0;\D_1,\D_2,\D_3) &= \frac{\pi 2^{4-\D_1 - \D_2 - \D_3} \sqrt{\Gamma[2\D_1]\Gamma[2\D_2]\Gamma[2\D_3]}}{\D_0^2 \Gamma[\D_{123}/2]\Gamma[\D_{231}/2] \Gamma[\D_{312}/2] \Gamma[(\D_1 + \D_2 + \D_3 - 1)/2]} \\
\end{split}
\eeq
where $\D_{ijk} = \D_i + \D_j - \D_k$. This relation was explained in \cite{Paper1}.

In one dimension conformal transformations preserve operator ordering modulo cyclic permutations. This leads to the following non-equivalent four-point functions
\beq
\< \phi_1 \phi_1 \phi_1 \phi_1 \>, \qquad \< \phi_2 \phi_2 \phi_2 \phi_2 \>, \qquad \< \phi_1 \phi_1 \phi_2 \phi_2 \>, \qquad \< \phi_1 \phi_2 \phi_1 \phi_2 \>,
\eeq
and we will numerically analyze the lot of them.\footnote{\label{footnote1212}An interesting observation is that the $\< \phi_1 \phi_2 \phi_1 \phi_2 \>$ correlator does not feature the identity operator. A conformal bootstrap analysis of this correlator \emph{in itself} therefore does not give any bounds whatsoever because it lacks an overall normalization. This is completely different from the forward $12 \to 12$ amplitude which we have seen can give a meaningful bound on $g_{112}$. However we will shortly see that the ensemble of correlators does give numerical results that mostly agree with the ensemble of amplitudes.} Our recipe follows that of \cite{Paper1} with minor variations. Suppose that we wish to obtain a bound on $g_{112}^2$ (in units of $m_1$) for a given coupling ratio $\alpha = g_{222}/g_{112}$ and mass ratio $\mu = m_2/m_1$. We then proceed as follows:
\begin{enumerate}
  \item Choose a $\Delta_1$. Then set $\Delta_2 = \mu \Delta_1$ and also fix the ratio
  \beq
  \frac{\lambda_{222}}{\lambda_{112}} = \alpha \frac{C(\D_1;\D_1,\D_1,\D_2)}{C(\D_1;\D_2,\D_2,\D_2)}\,.
  \eeq
  \item A single conformal bootstrap analysis of the four correlators listed above now yields a numerical upper bound on $\lambda_{112}^2$. Our multi-correlator bootstrap analysis is very similar to the one introduced in \cite{CFTMixed} where it was successfully applied it to the three-dimensional Ising model. The systematics of our analysis (normalizations, conformal block decompositions, functionals) can be found in appendix \ref{qftadsapp}. The bound so obtained also depends on the \emph{number} of derivatives of the crossing equations that we analyze and this introduces a new parameter $\Lambda$, so we write
  \beq
  (g_{112}^2)^\text{max}[\mu,\alpha,\Delta_1,\Lambda]
  \eeq
  where we use \eqref{CtoT} to pass from $(\lambda_{112}^2)^\text{max}$ to $(g_{112}^2)^\text{max}$.
  \item Upon repeating step 2 for various $\Lambda$ one finds that $(g_{112}^2)^{\text{max}}$ depends significantly on $\Lambda$. To obtain an estimate of the bound that we would obtain if we could analyze all the crossing equations, \emph{i.e.}, if we possessed infinite computational resources, we extrapolate the results for various $\Lambda$ to estimate
  \beq
  \lim_{\Lambda \to \infty}(g_{112}^2)^\text{max}[\mu,\alpha,\Delta_1,\Lambda]
  \eeq
  In practice we do this by fitting a polynomial through data points ranging from $\Lambda = 32$ up to $\Lambda = 140$.\footnote{In \cite{Paper1} we were able to obtain results up to $\Lambda = 200 $ or $\Lambda = 300$ for the different scenarios. The multi-correlator analysis of this paper is numerically more demanding, even more so because the rho series expansion \cite{Hogervorst:2013sma} for conformal blocks with large unequal dimensions converges much more slowly.} Examples of this extrapolation are shown in figure \ref{fig:extrapolgrid} on page \pageref{fig:extrapolgrid}. This limit provides our estimate for the best possible upper bound for a QFT in AdS with two particles with masses determined by $\Delta_1$ and $\mu$ and bulk coupling constant ratio given by $\alpha$.
  \item We view $\Delta_1$ as a proxy for the AdS curvature radius. We therefore repeat steps $1$ to $3$ for a number of different values of $\Delta_1$ and once more extrapolate to infinite $\Delta_1$ to obtain a result on the flat-space coupling:
  \beq
  (g_{112}^2)^\text{max}(\mu,\alpha) = \lim_{\Delta_1 \to \infty} \Big\{ \lim_{\Lambda \to \infty}(g_{112}^2)^\text{max}[\mu,\alpha,\Delta_1,\Lambda]\Big\}
  \eeq
  This is the coupling we can compare with the flat-space S-matrix bootstrap analysis.
\end{enumerate}
Appendix \ref{subapp:extrapol} contains technical details of the extrapolation procedure.

For finite $\Lambda$ and $\Delta$ our results provide rigorous upper bounds on the three-point couplings for any QFT in AdS that obeys the stated assumptions. Once we begin the extrapolations we introduce errors that are hard to quantify and this is an unavoidable drawback of our method. Nevertheless we will soon see, as was the case in \cite{Paper1}, that the extrapolated bounds appear to accurately reproduce the S-matrix bootstrap results in most cases.

\subsection{Results}
The numerical algorithm outlined above is computationally demanding. For a single $\mu$ and $\alpha$ we need about 10 different values of $\Delta_1$ and for each of these we need about $15$ different values of $\Lambda$, implying about 150 multi-correlator bootstrap runs. We have therefore chosen a few representative values of $\mu$ and $\alpha$ to demonstrate both the feasibility of the multi-correlator conformal bootstrap approach to scattering processes and the match with the flat-space S-matrix bootstrap results.

\subsubsection{Results for equal masses}
\begin{figure}[t]
\begin{center}
\includegraphics[width=10cm]{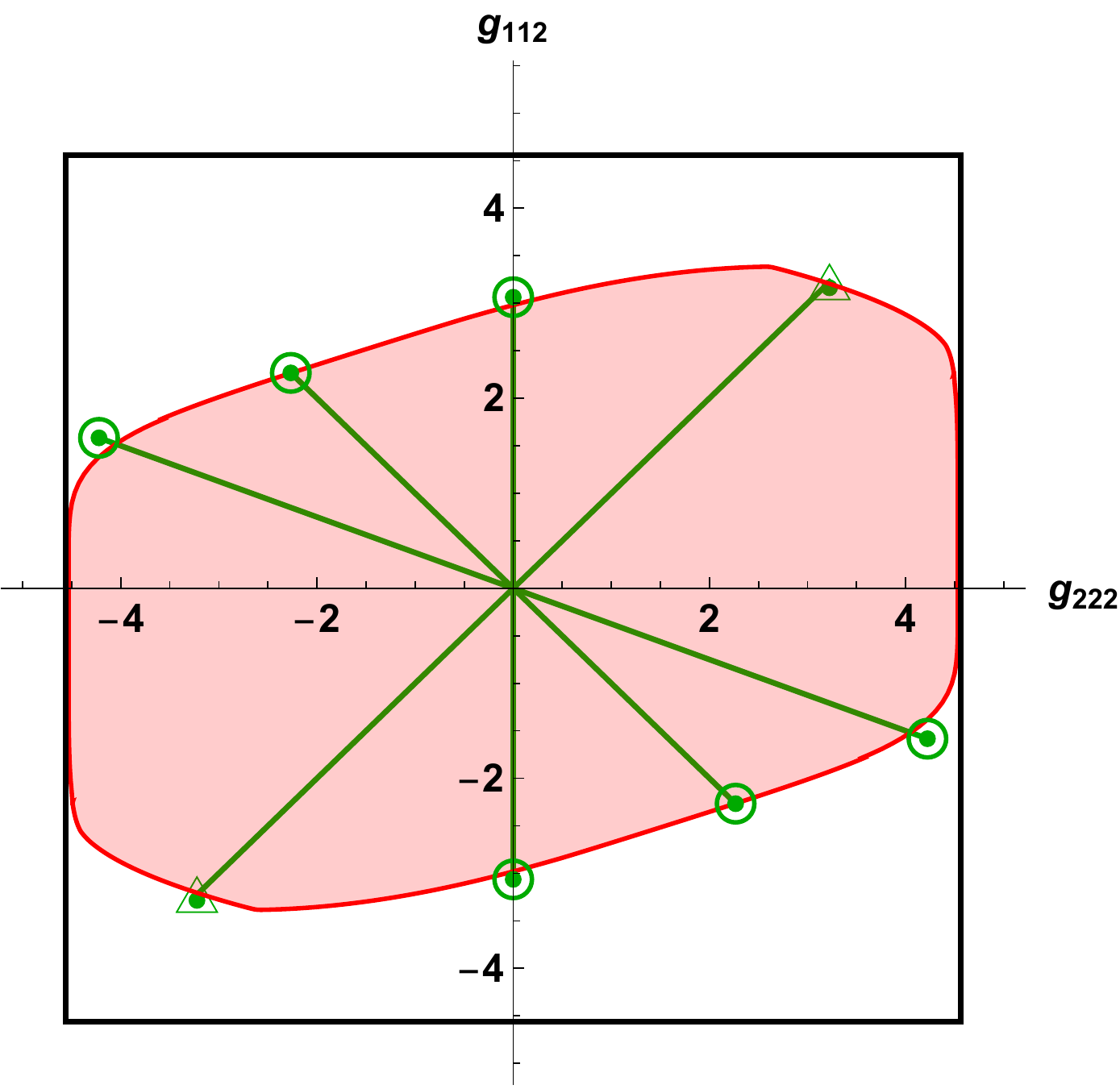}
\caption{\label{equalmassqftinads}Overlaid on a repetition of figure \ref{gvgm1m2}, the green data points show the maximal values of $|g_{112}|$ as derived from the QFT in AdS analysis for fixed values of $\alpha = g_{222}/g_{112}$. (The green straight lines then indicate the allowed range in coupling space.) The conformal bootstrap agrees very well with the S-matrix bootstrap. The point with $\alpha = -1$ is point A in figure \ref{gvgm1m2} which corresponds to the 3-state Potts model, which `emerges' here from the conformal crossing equations in one dimension.}
\end{center}
\end{figure}

Our first plot is for $\mu = 1$ so we have two particles of equal masses. In figure \ref{equalmassqftinads} we overlay the QFT in AdS results (isolated data points) with the S-matrix bootstrap region shown before in figure \ref{gvgm1m2}. The black frame again indicates the single-amplitude bounds, which are in fact equal to the single-\emph{correlator} bounds found in \cite{Paper1}. We have performed a multi correlator QFT in AdS analysis for ratios $\alpha = g_{222}/g_{112}$ equal to $+1$, $0$, $-1$ and $-8/3$ and in all cases we find reasonably good agreement with the multiple amplitude S-matrix bootstrap result. For $\alpha = -8/3$ our bound comes out somewhat higher than the value reached by the S-matrix bootstrap. This might be due to our extrapolation procedure, which also makes it difficult to put error bars on the QFT in AdS points, but it might also be a consequence of the finite truncation level in the S-matrix bootstrap. It is of course reassuring that the S-matrix bootstrap (Yin) always gives lower values than the conformal bootstrap (Yang).

For the \includegraphics[width=0.4cm]{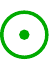} data points in figure \ref{equalmassqftinads} the extrapolation is standard, \emph{i.e.}, as outlined above and elaborated on in appendix \ref{qftadsapp}, but the \includegraphics[width=0.4cm]{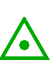} data point requires a comment. In that case we found that the maximal squared coupling $(g_{112}^2)^\text{max}[\mu,\alpha,\Delta_1,\Lambda]$ from the multi-correlator analysis is always numerically equal to one half of the corresponding maximal squared coupling obtained from the single-correlator analysis -- even for finite $\Delta_1$ and $\Lambda$ so before any extrapolations. Therefore, rather than doing a detailed multi-correlator analysis, we just plotted one half the single-correlator result. This factor of one half is understandable: using a change of operator basis similar to the one described in appendix \ref{solvable}, one finds that the multi-correlator problem effectively becomes that of two decoupled single-correlator problems which each feature a squared coupling that is rescaled by a factor 2.\footnote{It is essential here that $\alpha= 1$ so $\lambda_{222} = \lambda_{112}$.}

\subsubsection{Results for $\alpha = -1$}

\begin{figure}[t]
\begin{center}
\includegraphics[width=10cm]{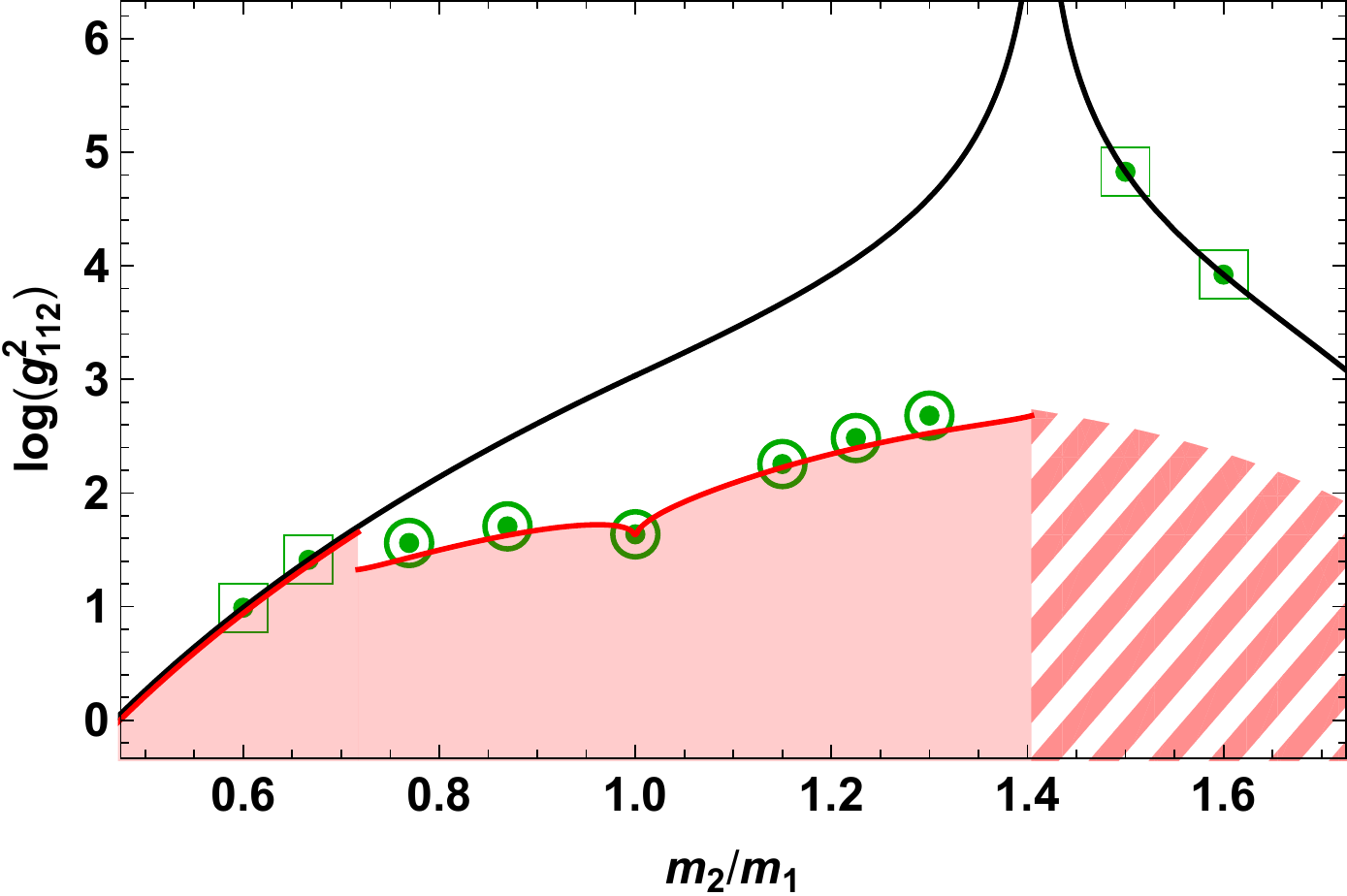}
\caption{\label{ratiominus1qftinads}The green data points show the maximal values of $\log(g_{112}^2)$ provided by the multi-correlator QFT in AdS analysis, now for $\alpha = g_{222}/g_{112} = -1$ and as a function of the mass ratio $m_2/m_1$. For comparison we have also added the single-correlator QFT in AdS bounds from \cite{Paper1}, as solid lines, as well as the S-matrix bootstrap data, in red. The plot naturally splits into three regions. First of all, for $m_2/m_1 < 1/\sqrt{2}$ there is screening and the multi-amplitude bound reduces to the single-amplitude bound. As shown, the correlator bounds nicely follow this behavior. Moving rightward, for $1/\sqrt{2} < m_2/m_2 < \sqrt{2}$ we find a respectable match between the multi-correlator and the multi-amplitude data, in particular we again recover the three-state Potts field theory at $m_2 = m_1$. For $\sqrt{2} < m_2 /m_1$ there are Landau singularities and the multi-amplitude analysis becomes complicated. However we know that the multi-amplitude bound must lie at or below the single-amplitude bound from $S^\text{forward}_{12 \to 12}$, meaning that it must end up somewhere in the striped region. The multi-correlator analysis, on the other hand, appears unable to improve on the weaker $\vev{1111}$ single-correlator bound.}
\end{center}
\end{figure}

Our next result is shown in figure \ref{ratiominus1qftinads}, where we have assumed $g_{222} / g_{112} = -1$. We will discuss in turn the black curve, the red shaded region, and the green (new) data points. 

The black curve corresponds to the best single-correlator bound for the given mass ratio. It is actually made up of two parts: for $m_2 > m_1$ it is the bound obtained from the $\vev{1111}$ four-point function, whereas for $m_2 < m_1$ it is the bound obtained from the $\vev{2222}$ four-point function. These single-correlator bounds were already obtained in \cite{Paper1} and were shown to agree with the single-amplitude analysis of \cite{Paper2}.

In red we show the multi-amplitude results obtained with the methods discussed in section \ref{implementation}. It is again made up of different parts: for $ 1/\sqrt{2} < m_2/ m_1 < \sqrt{2}$ we can use the numerical analysis and we take the $\alpha = -1$ slice from figure \ref{3dplot}. For $m_2 / m_1 < 1/\sqrt{2}$ we have screening and the multiple amplitude analysis does not give stronger results than the analysis of the single amplitude $S_{22 \to 22}$ which, as we stated before, agrees with the $\vev{2222}$ single-correlator bound. For $\sqrt{2} < m_2 /m_1$ there are Landau singularities and a more sophisticated analysis is necessary to obtain multiple amplitude results, but we do know that the maximal coupling from the multiple amplitude analysis can only lie below the single amplitude bounds. In particular, it must lie below the bound obtained from $S^\text{forward}_{12 \to 12}$, which was given as the solid line in figure \ref{g112} in the introduction and here yields the striped region in figure \ref{ratiominus1qftinads}.\footnote{We explain in appendix \ref{Landauappendix} that Landau singularities do not appear in $M^\text{forward}_{12 \to 12}$ for any mass ratio in the range $0< m_2 /m_1 < 2$, so the corresponding single-amplitude bound should be perfectly valid.}

Finally, the new data points obtained from the multi-correlator conformal bootstrap are indicated in green. The \includegraphics[width=0.4cm]{babycircle.pdf} data points are obtained from a standard extrapolation, as before, whereas for the \includegraphics[width=0.4cm]{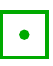} data points the numerical multi-correlator analysis gave identical results to the numerical single-correlator analysis for all $\Delta$ and $\Lambda$. The extrapolation will therefore trivially equal the single-correlator result as is indicated in the plot.

For all points with $m_2 /m_1 < \sqrt{2}$ our extrapolated QFT in AdS results lie at or just above the S-matrix results. As for the previous plot, the small finite difference might be either due to our extrapolation procedure but also due to the S-matrix bootstrap results not yet having converged. The points with $m_2/ m_1> \sqrt{2}$  are more puzzling. The striped domain, as we mentioned, arises from an analysis of $S^\text{forward}_{12 \to 12}$ alone and so a multi-amplitude analysis (with Landau singularities and all) will only be able to land somewhere in that domain. Unfortunately this single-amplitude bound does not seem to be picked up by the multi-correlator analysis at all.\footnote{See also footnote \ref{footnote1212}.} It would be nice to know why this is the case: are we missing constraints to be imposed for the QFT in AdS bootstrap?\footnote{At least the multi-correlator result, which is a hard upper bound, is above or equal to the S-matrix bootstrap results in all cases, so the results are not in direct conflict with each other.} Alternatively, is there maybe a `phase transition' where by pushing to a very high number $\Lambda$ of derivatives the single- and multi-correlator analysis begin to differ? Such a phenomenon would obviously invalidate our large $\Lambda$ extrapolations and might therefore resolve the puzzle. It would be somewhat analogous to the observations discussed in appendix \ref{subapp:functionalslambdas}, see in particular figure \ref{fig:lambda1lamdba2lamdba5lambda6}, where we explain that taking different $\Lambda$'s for different crossing equations leads to non-smooth behavior.

Of course, as we mentioned at the beginning of this section, at a technical level the conformal bootstrap analysis looks completely different from the S-matrix bootstrap. We are confident that both analyses yield valid constraints on three-point couplings, but besides physical intuition there was no a priori guarantee that these constraints had to be exactly the same. From this perspective the aspect most in need of an explanation in figure \ref{ratiominus1qftinads} is the quantitative \emph{match} between the results for $m_2/m_1 < \sqrt{2}$ (and similarly for all points in figure \ref{equalmassqftinads} and the results of \cite{Paper1}) rather than the discrepancy in the other points. Either way, the precise connection between conformal correlators and scattering amplitudes warrants further investigation.

\section{Discussion}
We have demonstrated the feasibility of the multiple-amplitude bootstrap for the lightest two particles, and shown that it gives stronger bounds compared to the simpler S-matrix bootstrap of only the lightest particle. Clearly one expects to get increasingly stronger bounds by considering more and more scattering processes. It is interesting to consider how such results could converge to an `optimal bound' that we would obtain by considering the entire S-matrix, as follows.

In all our numerical experiments it turns out that the unitarity condition at all energies is (numerically) \emph{saturated} in the subspace we work on. To illustrate this, consider for example the various possible outcomes from scattering the lightest particle in our setup -- let us say that it is the $\mathbb{Z}_2$ odd particle -- against itself. Probabilities must add up to $1$ so 
\beq
1=\underbrace{\underbrace{\underbrace{\overbrace{ |S_{11\to 11}|^2}^{{\color{darkgreen} \ge 0}}}_{\color{red} \le 1} +\overbrace{ |S_{11\to 22}|^2}^{{\color{darkgreen}\ge 0}}}_{{\color{blue} \le 1}}+\overbrace{|S_{11\to 112}|^2}^{{\color{darkgreen}\ge 0}}}_{\orange \le 1} +\overbrace{ |S_{11\to 1111}|^2}^{{\color{darkgreen}\ge 0}} +\dots  \la{Unitarity1} \,.
\eeq
where the red and blue inequalities follow trivially from probability positivity as indicated in dark green. In \cite{Paper2} we effectively considered only the weakest red inequality. The theories which lie on the boundary of this space turn out to saturate this inequality for all energies. This would mean that any other process has zero probability, and a theory saturating the bounds of \cite{Paper2} must therefore have zero particle creation or transmutation since the only allowed process is elastic scattering. This is possible for special cases like integrable theories, but generically it cannot be the case. Therefore, by including also the constraints of the other processes we should get better bounds and this is indeed what we have observed in this paper: our improved bounds are due to the stronger constraint given by the lower blue inequality.

However, we once more found that the \textit{optimal} solutions now saturate this \textit{new} condition for all values of the energy, so the remaining processes for the extremal S-matrix are again all zero.\footnote{We do not know why this happens; it stands as an empirical observation.} In other words, theories lying on the boundary of the new space are theories where particles $11$ can continue into $11$ or transmute into $22$ but we still get zero probability for all other processes that have a different final-state particle content. We expect this pattern to continue by including more and more processes in the game, \emph{i.e.} we will continue to observe unitarity saturation within the subspace we consider, no matter how large.
As we increase the size of our truncation we will hopefully asymptotically approach an optimal bound, but we are unlikely to hit a non-integrable theory at our boundary if we consider only a finite number of processes.\footnote{In particular, the beautiful ridge we observe in figure \ref{3dplot} likely does not correspond to a physical theory in itself, but is hopefully close to one that we can uncover by taking into account more processes.} This gives our S-matrix bootstrap approach a more asymptotic flavor than the numerical conformal bootstrap.

Indeed, for our setup with two particles with mass $m_1$ and $m_2$ we find no integrable theories if the masses are different, whereas if they are equal then there are exciting physical theories at our boundary: the three-states Potts model and (an analytic continuation of) the super-symmetric sine-Gordon model. We also find a full segment around the supersymmetric sine-Gordon theory which seems to obey all the necessary factorisation requirements to be an integrable theory; it would be very interesting to see if that is the case. (More on this in \cite{ToAppearSUSY}.) 

We also discussed how the same bootstrap results can be obtained from AdS, using the setup first discussed in \cite{Paper1}. Putting a gapped $\mathbb{Z}_2$ symmetric theory into an AdS box induces a one dimensional $\mathbb{Z}_2$ symmetric conformal theory in its boundary which we can analyze by numerical conformal bootstrap methods. To make contact with flat space, we take this box to be large which corresponds to large scaling dimensions on the boundary.
As it happens the numerical conformal bootstrap results become rather weak at large scaling dimensions and this makes it computationally quite challenging to obtain reliable results. This is the main drawback of the AdS approach. Fortunately there are interesting and potentially very helpful developments on this front: according to \cite{MiguelBernardo}, convergence can be much improved by a smarter choice of functionals (see also \cite{MazacMiguel1,MazacMiguel2}). It would be very interesting to explore this further.

An important advantage of the AdS approach, on the other hand, is that it requires no subtle assumptions about the various analytic properties of scattering amplitudes. The AdS box is thus a literal black box from which we can get beautiful S-matrix bootstrap results even when the analytic properties of such amplitudes might be less obvious. 

A good example of such a subtle assumption is extended unitarity, which we have seen is crucial for our multiple-amplitude bounds. Recall that this is a generalisation of usual unitarity which controls the analytic behaviour of scattering amplitudes for unphysical energies, below physical thresholds. For example, when we scatter the next-to-lightest particle against itself we have a two particle cut associated to the lightest particle starting at $s=(2m_\text{lightest})^2$, before the physical two particle cut at $s=(2m_\text{next-to-lightest})^2$, and extended unitarity governs the discontinuity of scattering amplitudes in the segment between those two values. Extended unitarity is built into perturbation theory \cite{Landau} but it is not straightforwardly justified non-perturbatively. However, the fact that our QFT in AdS approach exactly reproduces the flat space results provides strong evidence for the validity of the extended unitarity assumption. 

\begin{figure}[t]
\center \includegraphics[scale=.45]{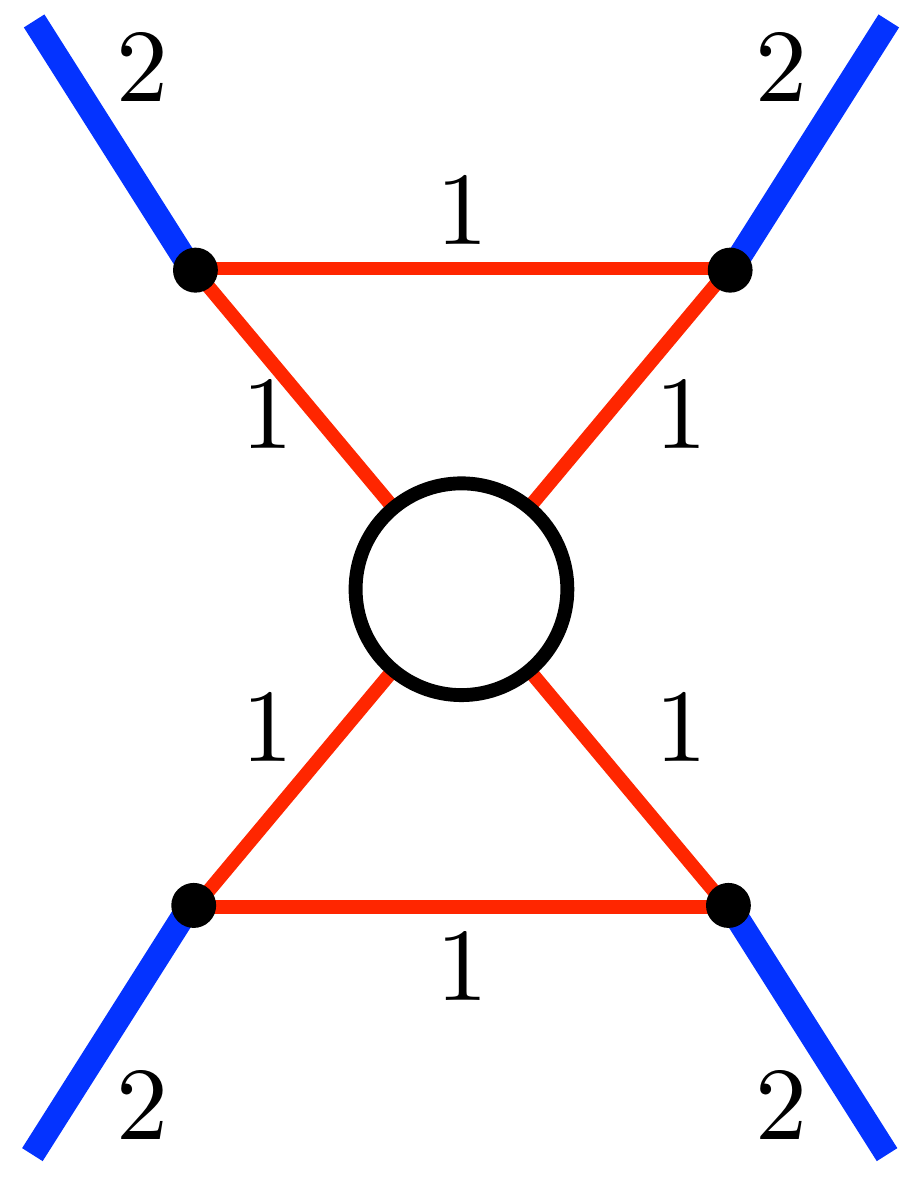}
\vspace{-0cm}
\caption{Landau diagram that gives rise to a double pole in $M_{22\to22}$ if $m_2 > \sqrt{2} m_1$.
} \label{Landau}
\end{figure}

Relatedly, it is puzzling that the AdS bounds for $m_2>\sqrt{2} m_1$ are so much weaker than the $12\to 12$ forward flat space extremal coupling, see figure \ref{ratiominus1qftinads}.
Either the AdS numerics did not converge yet or perhaps there is something deeper to be learned there. It would be very interesting to extend our flat space analysis beyond the mass range (\ref{massRange}) into the~$m_2> \sqrt{2} m_1$ domain. Here we would need to include so-called Coleman-Thun singularities in our setup. An example is shown in figure \ref{Landau}. 

Another very important new ingredient which is not unrelated to the anomalous cuts arising in the extended unitarity region and which appeared in this work is the phenomenon of \textit{screening}. Amplitudes involving heavier particles can sometimes produce discontinuities in some of these anomalous cuts which cancel, i.e. screen, other singularities such as physical poles corresponding to bound-state stable particles. That is, these discontinuities can be tuned so that the amplitudes can often be quite small in the physical energy region where experiments are done. This mechanism is not only possible but it is actually  realised by some amplitudes which lie at the boundary of the truncated $2\to 2$ multiple correlator S-matrix space. It is natural to expect that similar phenomena would also be realised at the boundary of the full S-matrix space. If this boundary is physically significant then screening would be an interesting way for nature to hide strong couplings from the observer.

Let us conclude with some interesting open problems and future directions. 

One open problem for which we now have all tools to explore concerns the tricritical Ising field theory. This theory is obtained by deforming a conformal minimal model with two relevant deformations, see appendix \ref{KinkIsing}, and is integrable if one of the deformations is set to zero. It would be very interesting to bootstrap this theory when both parameters are non-zero. This is a particularly nice case study because the next to lightest particle here is well below $\sqrt{2}$ times the mass of the lightest particle -- see table \ref{tablemasses} for its value at the integrable point -- which means we can readily apply all the methods developed here. The only modification would be to include further poles in the ansatz corresponding to the other stable particles this theory has -- see again table \ref{tablemasses}. So here is a  
homework exercise: consider a line in the mass ratio parameter space which passes by the masses of the integrable theory. Something remarkable should happen: In the anomalous cut of the $22\to 22$ amplitudes we should see a peak developing as we approach the integrable theory. This peak is going to become a new stable particle in the integrable theory with a mass we know. Seeing this peak show up in detail would be great, as it would constitute a ``discovery'' of a new particle through the S-matrix bootstrap. Of course, more interesting still would then be to move away from the masses of the integrable theory and explore the full tricritical Ising field theory, non-integrable and all. The previously discovered sharp peak -- typical of an integrable theory -- would now be smoothened out and correspond to a nearly stable resonance. Because there are so many masses and couplings this would be a challenge numerically, albeit a worthwhile one.

Another open problem which we could now easily address is the problem of multiple amplitudes without $\mathbb{Z}_2$ symmetry. We would now also include amplitudes such as $11\to 12$ which have amusing 2D kinematics by themselves. 
The Ising field theory perturbed by both magnetic field and temperature would be a perfect case study for this case. 

A much more challenging but very interesting open problem would be to extend the multiple amplitude analysis to higher dimensions (as in \cite{Paper3} and \cite{Andrea}). The $\mathbb{Z}_2$ spontaneously broken phase of the $\phi^4$ model in 3D, for instance, seems to have a single stable bound state of mass $m_2 \simeq 1.8 m_1$ \cite{latticephi4}; it would be fascinating to try to bootstrap this S-matrix.

Finally, another frontier in 2D would be to delve into the multiple particle S-matrix bootstrap.  Can we tame scattering of $2$ particles into $3,4,\dots$ final particles? There are two obvious obstacles. One is the analytic structure of these amplitudes. They depend now on more kinematical variables and have a huge plethora of Landau singularities; it is unclear if we can characterise them fully. The other challenge is even more basic: can we close up the system of equations? Suppose we consider a basis of initial and final states with both two and three particles. Then we need to deal with the $3\to 3$ amplitudes. But those, by crossing, are related to $2\to 4$. By unitarity we would then need to include four particles in the final and initial states as well. But then we are forced to consider $4\to 4$ processes which are now related by crossing to $3\to 5$ and $2\to 6$ scattering and so on. It seems we are suddenly obliged to consider any number of final particles at once which of course would be computationally completely infeasible.
Hopefully we can find a suitable truncation scheme. Along these lines, perhaps we could first try to get some inspiration from the AdS side. Some of the necessary higher point conformal blocks are well known in 1D \cite{Rosenhaus}, so can we use this to devise a 1D CFT bootstrap numerical problem dual to the very intimidating flat space multiple particle bootstrap? Even if very challenging numerically, this would prove of extreme conceptual value.

\section*{Acknowledgements}
We thank Gesualdo~Delfino, Carlos~Bercini, Luc\'ia~C\'ordova, Patrick~Dorey, Davide~Gaiotto, Andrea~Guerrieri, Matheus~Fabri, Giuseppe~Mussardo, Miguel~Paulos, Erik~Schnetter and Alexander~Zamolodchikov for numerous enlightening discussions and suggestions. Research at the Perimeter Institute is supported in part by the Government of Canada through NSERC and by the Province of Ontario through MRI. 
This research received funding from the grant CERN/FIS-NUC/0045/2015.
This work was additionally supported by a grant from the Simons Foundation (JP: \#488649, BvR: \#488659, PV: \#488661)
JP is supported  by the Swiss National Science Foundation through the project 200021-169132 and through the National Centre of Competence in Research SwissMAP.

\appendix
\section{Two dimensions and higher dimensional kinematics} \la{triangles}
The Mandelstam plane provides us with a very useful depiction of the (real sections of the) interrelations between the three Mandelstam variables $u,s,t$. The first important object in this plane is the Mandelstam \textit{triangle}. 

\begin{figure}[t]
\center \includegraphics[scale=.45]{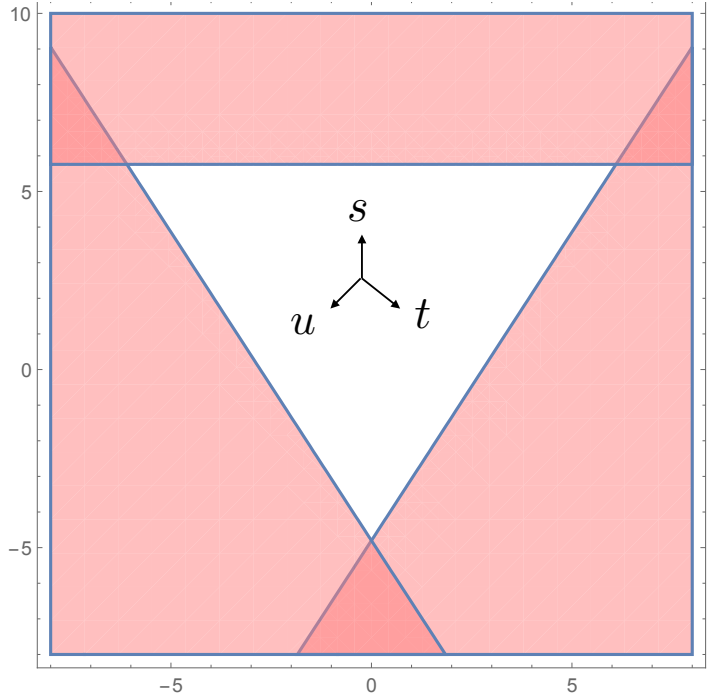}
\caption{The Mandelstam triangle is the region where all Mandelstam variables are below their corresponding two-particle threshold: $s< \text{max}((m_a+m_b)^2,(m_c+m_d)^2) \wedge t< \text{max}((m_a+m_c)^2,(m_b+m_d)^2) \wedge u< \text{max} ((m_a+m_d)^2,(m_b+m_c)^2)$ represented by the white region in this figure. (The $y$ axis is $s$ and the $x$ axis is given by $x=(s+2 t-m_a^2-m_b^2-m_c^2-m_d^2)/\sqrt{3}$ as in the next figures.)} \label{triangle}
\end{figure}

Consider a two-to-two process involving particles with momenta $p_a,p_b,p_c,p_d$ associated to particles of mass $m_a,m_b,m_c,m_d$. We define the three Mandelstam invariants $s=(p_a+p_b)^2$, $t=(p_a+p_c)^2$ and $u=(p_a+p_d)^2$. If particles $p_a,p_b$ are the two incoming particles then $\sqrt{s}$ is the centre of mass energy of the scattering process. The same is true if the incoming particles are particles~$p_c,p_d$. In either of these cases the process can only be physical if we have enough energy to produce both the initial and final state, so for $s\ge \text{max}((m_a+m_b)^2,(m_c+m_d)^2)$. Of course, the same scattering amplitudes can describe other channels.\footnote{To describe other channels we can either swap the masses and always keep $s$ to the be the center of mass energy (as in the main text) or leave the masses untouched but reinterpret which Mandelstam invariant corresponds to the center of mass energy (as in this appendix). It is very simple (and very instructive) to go between these active/passive viewpoints. \label{careful}} If $p_a,p_c$ (or $p_b,p_d$) are the two incoming particles then $\sqrt{t}$ is the centre of mass energy of the scattering process and similarly for $u$ so the physical conditions in those cases would read $t \ge \text{max}((m_a+m_c)^2,(m_b+m_d)^2)$ and $u \ge \text{max} ((m_a+m_d)^2,(m_b+m_c)^2 )$.
The three inequalities are depicted by the shaded pink regions in figure \ref{triangle}. The white region is the Mandelstam triangle.

To be in a physical region we thus need to be in the pink region. This is necessary but not sufficient. We need to have enough energy but we also need to scatter at a real angle.  For instance for incoming particles $p_a,p_b$ we can easily compute the scattering angle to find   
\beq
\cos(\theta_{ab})
= \frac{(s+m_a^2-m_b^2)(s+m_c^2-m_d^2)+2s (t-m_a^2-m_c^2) }{\sqrt{((s-m_a^2-m_b^2)^2-4m_a^2 m_b^2 )((s-m_c^2-m_d^2)^2-4m_c^2 m_d^2 )}} \label{angle} \,.
\eeq
For any left hand side between $-1$ and $+1$ corresponding to a real angle, and for any $s$ in the physical range this equation determines a physical $t$. (Of course, $u=m_a^2+m_b^2+m_c^2+m_d^2-s-t$ is automatically fixed.) The set of physical $s$ and $t$ determined in this way determine the physical region in the $s$-channel. The other channels are treated similarly with 
\beq
\cos(\theta_{ac})= \text{RHS of (\ref{angle})}_{m_{b}\leftrightarrow m_c, \, s \leftrightarrow t} \qquad , \qquad \cos(\theta_{ad})= \text{RHS of (\ref{angle})}_{m_{b}\leftrightarrow m_d, \, s \leftrightarrow u} \la{angle2}
\eeq

\begin{figure}[t]
\center \includegraphics[scale=.6]{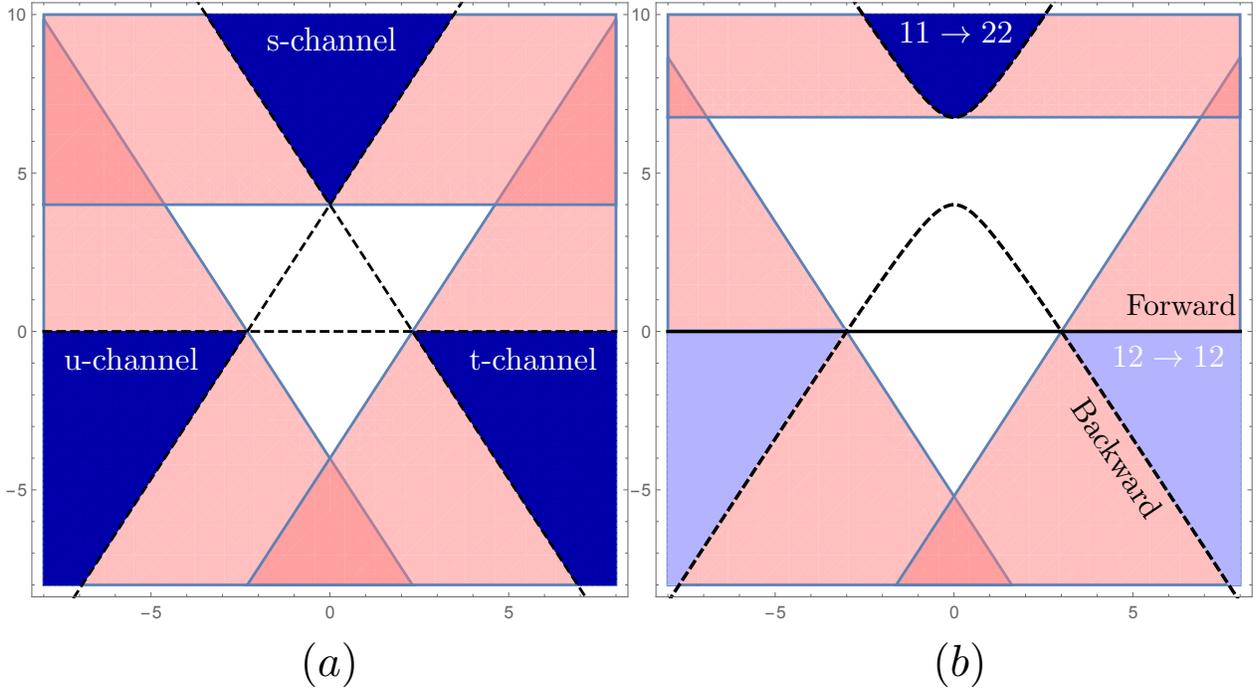}
\vspace{-3.8cm}
\caption{(a) Physical regions (in dark blue) for a two-to-two process with all external particles of identical mass. The boundary of these regions correspond to scattering angle $0$ or $\pi$ and can thus also be identified with the physical scattering `regions" (or better scattering lines) in two dimensions. (b) Physical regions for a process where the masses are equal in pairs. The darker blue region is the process $m_1m_1\to m_2 m_2$ while the two lighter regions (which are equivalent for identical particles) correspond to the $m_1m_2\to m_1m_2$ channels. The boundary of these regions are again the two dimensional physical lines. The boundaries of the lighter blue regions are not identical: one corresponds to forward scattering; the other to backward scattering.  } \label{triangle2}
\end{figure}

If all masses are equal (to $m$) then (\ref{angle}) reduces to the famous relation
\beq
\cos(\theta)=1-\frac{2t}{4m^2-s} \,. \la{simpleAngle}
\eeq
so that the physical region in the s-channel is simply $s>4m^2$ and $4m^2-s \le t \le0$ represented by the top blue region in figure \ref{triangle2}a. In two dimensions the angle ought to be $0$ or $\pi$ so that we have either $t=0$ or $t=4m^2-s$, i.e. $u=0$. These two conditions ($t=0$ and $u=0$) are equivalent if the external particles are indistinguishable so in that case we can pick either; in the main text we took $u=0$. Note that these two conditions are nothing but the boundary of the darker blue region. Similarly we could study all other channels which we can simply obtain by relabelling the Mandelstam variables in (\ref{simpleAngle}). The two extra physical regions are the other two blue regions in the same figure \ref{triangle2}a. Note that the boundary of all these blue regions can be written concisely as $stu=0$ which is nothing but the constraint obtained in the main text from the two dimensional constraint~(\ref{constraint}) and represented by the dashed lines in figure \ref{triangle2}a.

Finally, we come to the more interesting case where the external masses are only pairwise equal: $m_a=m_b=m_1$ and $m_c=m_d=m_2$. This same configuration can describe the $11\to 22$ process (with $\sqrt{s}$ being the centre of mass energy), and the $12\to 12$ processes (with $\sqrt{t}$ being the centre of mass energy). For the first case we use (\ref{angle}) to get 
\beq
\cos(\theta_{11\to 22})=\frac{-2 m_1^2-2 m_2^2+s+2 t}{\sqrt{\left(s-4 m_1^2\right) \left(s-4
   m_2^2\right)}} \,.
\eeq
The physical region corresponding to $-1 \le \cos\theta_{11\to 22} \le +1$ is now a more interesting curved region, represented by the darker blue region in figure \ref{triangle2}b. 
Again, in two dimensions we can only have backward or forward scattering so $t$ must saturate one of these inequalities. If the particles of the same mass are indistinguishable then both solutions are equivalent as before. 
We can thus take $\theta_{11\to 22}=0$ without loss of generality leading to 
\beq
t,u = m_1^2+m_2^2-\frac{s}{2}\pm\frac{1}{2} \sqrt{(s-4m_1^2)(s-4m_2^2)} \,. \la{ts1122}
\eeq
 and reproduce in this way the results below (\ref{1122kin}) in the main text. In the $t$-channel particles $m_a=m_1,m_c=m_2$ are incoming so we are scattering a odd particle against an even particle. Then we use the first relation in (\ref{angle2}) which now reads 
\beq
\cos(\theta_{12\to 12})=1+\frac{2 t s}{-2 m_1^2 \left(m_2^2+t\right)+\left(m_2^2-t\right){}^2+m_1^4} \la{Angle1212}
\eeq
Note that in our convention it is $\sqrt{t}$ (and not $\sqrt{s}$) who is the center of mass energy in this channel.\footnote{Recall footnote \ref{careful} when comparing the results that follow to the main text.} The physical region corresponding to $|\cos(\theta_{12\to 12})|\le 1$ is now represented by the lighter blue region in figure~\ref{triangle2}b. 
The two dimensional conditions that the angle is $0$ or $\pi$ are now quite different. The former corresponds to forward scattering and is obtained by $s=0$ (obtained by equating the RHS of (\ref{Angle1212}) to $+1$) while the later corresponds to backward scattering and yields the more involved relation 
\beq
s=2 \left(m_1^2+m_2^2\right)-t-\frac{\left(m_1^2-m_2^2\right){}^2}{t}
\eeq
(obtained by equating the RHS of (\ref{Angle1212}) to $-1$.) Note that this relation is nothing but (\ref{ts1122}) if we solve for $s$. In other words, these two configurations are simply related by crossing symmetry $s\leftrightarrow t$. 

 To summarize: the boundary of the physical regions are now given by the  black solid line and by the black dashed curve in figure \ref{triangle2}b. Crossing $u \leftrightarrow t$ at $s=0$ relates the left to the right of the straight line -- leading to condition (\ref{C1}). Crossing symmetry also relates the top to the bottom branch of the hyperbolic looking curve -- reflected in equation (\ref{C2}). In two dimensions these two curves (the hyperbola and the straight line) are independent while in higher dimensions they are smoothly connected (by moving in angle space).

\section{Unitarity and final state probabilities} \la{ap:unitarity}

The S-matrix is defined by the expansion of in-states in terms of out-states
\beq
| A \rangle_\text{in} = \sum_B S_{A\to B} |B\rangle_\text{out}\,.
\label{eq:Sdefinition}
\eeq
The in-states and the out-states are both a complete basis of the Hilbert space.
Let us start by discussing the physical meaning of the diagonal unitarity equations \eqref{ex1111} and \eqref{ex2222}.
These follow from the statement that the state $| A \rangle_\text{in}$ above is normalized. However, due to the continuum of states this is a bit subtle. The trick is to contract the state with itself but with different momenta
\beq
\ _\text{in}\langle A' | A \rangle_\text{in} = \sum_B \sum_{B'} S_{A\to B} S_{A'\to B'}^* \ _\text{out}\langle B' |B\rangle_\text{out}\,.
\label{eq:Sunitarity}
\eeq
Here, the state $| A' \rangle_\text{in} $ represents a state with the same particle content but different momenta.
We use the standard normalization for the inner products:
\beq
\ _\text{out}\langle C |B\rangle_\text{out} = 
\ _\text{in}\langle C |B\rangle_\text{in} = \delta_{B,C} \prod_{i \in B} 2E_i 2\pi \delta(p_i^B -p_i^C)\,,
\eeq
where the product runs over each particle in the state $|B\rangle_\text{in}$.
Unitarity then reads
\beq
1= \sum_{B} | S_{A\to B}|^2 {\cal J}_{A,B}\,,
\eeq
where ${\cal J}_{A,B}$ is the jacobian defined by
\beq
  \ _\text{out}\langle B' |B\rangle_\text{out} = {\cal J}_{A,B}
\ _\text{in}\langle A' | A \rangle_\text{in}  \,.
\label{eq:jacobian}
\eeq
The natural physical interpretation is that 
\beq
P_{A\to B} =| S_{A\to B}|^2 {\cal J}_{A,B}
\eeq
is the probability of the in-state $| A \rangle_\text{in}$ end up in the out-state $|B\rangle_\text{out}$.

For two particle states $| A \rangle_\text{in} = |12\rangle_\text{in}$ and  $|B\rangle_\text{out}=|34\rangle_\text{out}$, equation \eqref{eq:jacobian} reduces to
\beq
E_3 E_4 \delta(p_3 - p_3') \delta(p_4-p_4') = {\cal J}_{12,34} 
E_1 E_2 \delta(p_1 - p_1') \delta(p_2-p_2')\,,
\eeq
with $E_i=\sqrt{m_i^2+p_i^2}$ and 
\begin{align}
p_1+p_2&=p_3+p_4\,, \qquad\qquad
E_1+E_2=E_3+E_4 =\sqrt{s}\,,\\
p_1'+p_2'&=p_3'+p_4' \,,\qquad\qquad
E_1'+E_2'=E_3'+E_4' \,.
\end{align}
This gives
\beq
{\cal J}_{12,34} = \frac{\sqrt{s-(m_3-m_4)^2}\sqrt{s-(m_3+m_4)^2}}{
\sqrt{s-(m_1-m_2)^2}\sqrt{s-(m_1+m_2)^2}}=\frac{\rho^2_{12}}{\rho_{34}^2}\,.
\eeq
We conclude that, for two particle states, the transition probabilities are given by
\beq
P_{12\to 34} =| S_{12\to 34}|^2 \frac{\rho^2_{12}}{\rho_{34}^2}\,.
\eeq
For example, for the initial state $|11\rangle_\text{in}$, we can write 
\beq
|S_{11\to 11}|^2 
+
|S_{11\to 22}|^2\frac{\sqrt{s-4m_2^2}}{\sqrt{s-4m_1^2}} 
= 1-P_{11 \to (N\ge 3 \, {\rm particles})}\,.
\label{11prob}
\eeq
This equation is equivalent to \eqref{ex1111} in the energy range 
\beq
\max(2m_1,2m_2) < \sqrt{s} < \min(3m_2, 2m_1+m_2)
\label{2ptrangeeven}
\eeq
where only 2 particle states are available. To show this we contract \eqref{eq:Sdefinition} with a generic out-state $\ _\text{out}\langle C|$, to find
\beq
\ _\text{out}\langle C| A \rangle_\text{in} = \sum_B S_{A\to B}\ _\text{out}\langle C |B\rangle_\text{out}\,,
\eeq
and use the standard definition of the amplitude $M$:
\beq
\ _\text{out}\langle C |A\rangle_\text{in} = 
\ _\text{in}\langle C |A\rangle_\text{in} +  i (2\pi)^2 \delta^{(2)}(P_C -P_A) \,M_{A\to C}\,,
\eeq
where $P_A$ denotes the total momentum of the state $|A\rangle_\text{in}$.
This relates $S$ with $M$. In the particular case of two particle states one finds
\beq
S_{12\to 34}= \delta_{12,34} +\frac{i M_{12\to 34}}{2\sqrt{s-(m_3-m_4)^2}\sqrt{s-(m_3+m_4)^2}}= \delta_{12,34} + i \rho^2_{34} M_{12\to 34} \,,
\label{SandM}
\eeq
where the first term is present (and equal to $1$) if and only if the initial and final states are the same.
This allows us to rewrite \eqref{11prob} in terms of $M$,
\beq
2 \text{Im} M_{11 \to 11}  = \rho^2_{11}  |M_{11 \to 11}|^2 
 + \rho^2_{22} |M_{11 \to 22}|^2  + \frac{P_{11 \to (N\ge 3 \, {\rm particles})}}{\rho^2_{11}}\,,
\eeq 
which should be compared with \eqref{ex1111}. Notice that the physical derivation given here is not valid for $2\min(m_1,m_2)<\sqrt{s}<2\max(m_1,m_2)$ because the two particle state of the heavier particle is not available. This is the regime of extended unitarity where we must use \eqref{ex1111}.

There is also  a more intuitive derivation  of the full matrix form of the unitarity constraints \eqref{matrixextended}.
Consider the $\mathbb{Z}_2$ even sector for simplicity.
The matrix of inner products of the states $\{ |  11\>_\text{in}, |  22\>_\text{in}, | 11\>_\text{out},| 22\>_\text{out} \}$:
\beq
  \bordermatrix{~ & | 11\>_\text{in} & |  22\>_\text{in} & | 11\>_\text{out} & |  22\>_\text{out} \cr
            \  _\text{in}\< 11|  &_\text{in}\< 11 | 11\>_\text{in}  &_\text{in}\< 11 |  22\>_\text{in} & _\text{in}\< 11| 11\>_\text{out}& _\text{in}\< 11|  22\>_\text{out} \cr
            \ _\text{in}\<  22|  & _\text{in}\<  22| 11\>_\text{in}  & _\text{in}\<  22|  22\>_\text{in} & _\text{in}\<  22 | 11\>_\text{out}& _\text{in}\<  22|  22\>_\text{out}  \cr
            _\text{out}\<  11| &_\text{out}\<  11 | 11\>_\text{in} &_\text{out}\<  11 |  22\>_\text{in}&_\text{out}\<  11 | 11\>_\text{out} &_\text{out}\<  11 |  22\>_\text{out} \cr
            _\text{out}\<  22|  &_\text{out}\<  22 | 11\>_\text{in}  & _\text{out}\<  22|  22\>_\text{in} &_\text{out}\<  22 | 11\>_\text{out} & _\text{out}\<  22|  22\>_\text{out} \cr}
\eeq
must be positive semi-definite. In fact, for the range of energies \eqref{2ptrangeeven} 
where there are only two particle states, the rank of this matrix must be $2$ because both the in and the out states are complete basis.
This is a very intuitive way to derive the unitary constraints. However, one must be careful with Jacobian factors that relate different delta-functions.
Factoring out  $
_\text{in}\< 11 | 11\>_\text{in} 
$ and using \eqref{eq:Sdefinition} and \eqref{eq:jacobian}, we can define a positive semi-definite matrix (without delta-functions)
\beq
\bordermatrix{~ & |   11\>_\text{in} & |  22\>_\text{in} & |  11\>_\text{out} & |  22\>_\text{out} \cr
            \<  11|_\text{in} & 1 & 0 & S^*_{11\to11}   & \frac{\rho^2_{11}}{\rho^2_{22}} S^*_{11 \to 22}  \cr
              \<   22|_\text{in} & 0 & \frac{\rho^2_{11}}{\rho^2_{22}} &  S^*_{22 \to 11}   &  \frac{\rho^2_{11}}{\rho^2_{22}}S^*_{22 \to 22}   \cr
              \<  11|_\text{out} & S_{11\to 11}  & S_{22\to 11}  & 1 & 0 \cr
             \<  22|_\text{out} &  \frac{\rho^2_{11}}{\rho^2_{22}} S_{11\to22}  & \frac{\rho^2_{11}}{\rho^2_{22}}S_{22\to22}  & 0 & \frac{\rho^2_{11}}{\rho^2_{22}} \cr}
\eeq
It is convenient to rescale the states $ |  22\> \to \frac{\rho_{22}}{\rho_{11}}  |  22\>$ so that all diagonal entries become 1. This leads to the following positive semi-definite matrix:
\beq
\mathbb{V} = \bordermatrix{~ & |   11\>_\text{in} & \frac{\rho_{22}}{\rho_{11}}|  22\>_\text{in} & |  11\>_\text{out} & \frac{\rho_{22}}{\rho_{11}}|  22\>_\text{out} \cr
          \ \ \ \,   \<  11|_\text{in} & 1 & 0 & S^*_{11\to11}   & \frac{\rho_{11}}{\rho_{22}} S^*_{11 \to 22}  \cr
              \frac{\rho_{22}}{\rho_{11}}\<   22|_\text{in} & 0 & 1 & \frac{\rho_{22}}{\rho_{11}}  S^*_{22 \to 11}   &  S^*_{22 \to 22}   \cr
      \ \ \ \,          \<  11|_\text{out} & S_{11\to 11}  & \frac{\rho_{22}}{\rho_{11}}S_{22\to 11}  & 1 & 0 \cr
            \frac{\rho_{22}}{\rho_{11}} \<  22|_\text{out} &  \frac{\rho_{11}}{\rho_{22}} S_{11\to22}  & S_{22\to22}  & 0 & 1 \cr}
\eeq
Notice that using \eqref{SandM}, the $4\times 4$ matrix $\mathbb{V}$ can be written as
\beq
\mathbb{V} = 
\begin{bmatrix}
\mathbb{I} & \mathbb{S}^\dagger  \\
\mathbb{S} & \mathbb{I} 
\end{bmatrix}\,,\qquad\qquad
\mathbb{S} = 
\mathbb{I} +i \rho \mathbb{M} \rho
\eeq
where $\mathbb{M}$ and $\rho$ are the $2\times 2$ matrices defined in \eqref{matrixextended}. We will now show that the condition $\mathbb{V} \succeq 0$ is equivalent to \eqref{matrixextended} for $\sqrt{s}>2\max(m_1,m_2)$.
First notice that the eigenvalues of the hermitian matrix $\mathbb{I} - \mathbb{V}$ take the form
$(-\lambda_2, -\lambda_1,\lambda_1,\lambda_2)$.\footnote{It is easy to see that the characteristic polynomial $\det(\mathbb{I}- \mathbb{V}-x \mathbb{I})$ is an even function of $x$.}
Then, the condition $\mathbb{V} \succeq 0$  implies that $\lambda_i^2<1$.
On the other hand, if we compute explicitly the square of $\mathbb{I} - \mathbb{V}$, we find
\beq
(\mathbb{I} - \mathbb{V})^2 = 
\begin{bmatrix}
\mathbb{S}^\dagger \mathbb{S} &0 \\
 0 & \mathbb{S} \mathbb{S}^\dagger
\end{bmatrix}\,.
\eeq
Therefore, the eigenvalues of $\mathbb{S}^\dagger \mathbb{S}$ must be less than 1.
Equivalently, we can say that 
\beq
\mathbb{I}-\mathbb{S}^\dagger \mathbb{S} 
= \mathbb{I}-(\mathbb{I} -i \rho \mathbb{M}^\dagger \rho )( \mathbb{I} +i \rho \mathbb{M} \rho)
= \rho \left( 2\text{Im} \mathbb{M} -  \mathbb{M}^\dagger \rho^2 \mathbb{M}
\right)\rho
\succeq 0
\eeq
and  \eqref{matrixextended} follows.\footnote{If a matrix $\rho \mathbb{X} \rho \succeq 0$ then $\mathbb{X} \succeq 0$. This follows from the fact that if $u^\dagger \rho \mathbb{X} \rho u \ge 0$ for any vector $u$ then $v^\dagger   \mathbb{X} v \ge 0$ for any vector $v$ (just choose $u=\rho^{-1} v$).}
In the extended unitarity region $2\min(m_1,m_2)<\sqrt{s}<2\max(m_1,m_2)$ 
this derivation does not apply but we can still use \eqref{matrixextended}.

\subsection{Bounding \texorpdfstring{$\text{Im}M_{11\to 22}$}{Im(M11->22)}}

The discontinuity of the amplitude $M_{11\to 22}$ does not have well defined sign. 
Furthermore, in the region  $2\min(m_1,m_2)<\sqrt{s}<2\max(m_1,m_2)$ below the physical regime, one could worry that the discontinuity could be very large and lead to screening. 
However, the generalized unitarity equations \eqref{ex1111}-\eqref{ex2222} forbid this phenomena in the range of masses we consider.

For $2\min(m_1,m_2)<\sqrt{s}<2\max(m_1,m_2)$, equations (\ref{ex1111}-\ref{ex2222}) reduce to 
\begin{align}
2 \text{Im} M_{11 \to 11} &= \rho_{\ell\ell}^2 \left| M_{11 \to \ell\ell}\right|^2 \\
2 \text{Im} M_{11 \to 22} &= \rho_{\ell\ell}^2 M_{22 \to \ell\ell} M^*_{11 \to \ell\ell} 
\label{extuni1122}\\ 
2 \text{Im} M_{22 \to 22} &= \rho_{\ell\ell}^2 \left|M_{22 \to \ell\ell}\right|^2
\end{align}
where the label $\ell=1$ or $\ell=2$ stands for the lightest particle.
Taking the modulus square of equation \eqref{extuni1122} and using the other two equations, we find
\beq
|\text{Im} M_{11 \to 22}|^2 = \text{Im}  M_{11 \to 11} \text{Im} M_{22 \to 22} 
\eeq
Therefore the size of $\text{Im} M_{11 \to 22}$ is related to the positive discontinuities $\text{Im}  M_{11 \to 11} $ and $\text{Im} M_{22 \to 22} $, which for this reason are bounded.
In fact, in our numerical procedure we impose unitarity as the set of inequalities \eqref{matrixextended}, which in particular implies
\beq
\text{Im} \mathbb{M} 
\succeq 0\,.
\eeq
This leads to
\beq
\det \text{Im} \mathbb{M} \ge 0 \qquad \Leftrightarrow \qquad
|\text{Im} M_{11 \to 22}|^2 \le \text{Im}  M_{11 \to 11} \text{Im} M_{22 \to 22}  \la{SCW}
\eeq
which bounds $\text{Im} M_{11 \to 22}$ in our setup.

\subsection{Phase shifts}
 
In the  $\mathbb{Z}_2$ even sector, it is trivial to define diagonal phase shifts
\beq
S_{11\to 11}(s) = e^{2i \delta_{11}(s)} \,,\qquad\qquad
S_{22\to 22}(s) = e^{2i \delta_{22}(s)} \,.
\eeq
The $\mathbb{Z}_2$ odd sector is slightly more interesting. In this sector, it is convenient to use states of definite parity,
\beq
| 12 \rangle^\pm = \frac{1}{\sqrt{2}} \left( | 12 \rangle \pm | 21 \rangle\right) \,.
\eeq
Then, we define the phase shifts by 
\beq
e^{2i\delta_{12}^\pm}\equiv   \,_\text{out}^{\ \, \pm}\langle 12 | 12 \rangle^\pm_\text{in}  =
S_{12\to12}^\text{Forward} \pm S_{12\to12}^\text{Backward}
=1+i \rho_{12}^2 \left(
M_{12\to12}^\text{Forward} \pm M_{12\to12}^\text{Backward}\right)\,.
\label{oddphaseshifts}
\eeq

In our numerical algorithm, we impose unitarity in the odd sector by the following positive semi-definite condition
\beq
2\text{Im} \mathbb{\tilde{M}} \succeq \rho_{12}^2 \mathbb{\tilde{M}}^\dagger  \mathbb{\tilde{M}}, \qquad \mathbb{\tilde{M}} = \begin{bmatrix}  
  M_{12\to12}^\text{Forward} &  M_{12\to12}^\text{Backward}\\
        M_{12\to12}^\text{Backward}  & M_{12\to12}^\text{Forward}
\end{bmatrix}        \,,\qquad
s>(m_1+m_2)^2\,.
         \eeq
It is instructive to see what this implies for the phase shifts $  \delta_{12}^\pm$.       
Using \eqref{oddphaseshifts}, we conclude that
\beq
\left| e^{2i\delta_{12}^\pm(s)}\right|^2   =
1-\rho_{12}^2\left( a\pm b\right)\,,
\eeq
where
\beq
\begin{bmatrix}  
a &  b\\
b  & a
\end{bmatrix}  =
2\text{Im} \mathbb{\tilde{M}} - \rho_{12}^2 \mathbb{\tilde{M}}^\dagger  \mathbb{\tilde{M}}\succeq0\,.
\eeq
 Thus $|b|\le a$  and we recover the usual unitarity inequality
for the phase shifts 
\beq
\left| e^{2i\delta_{12}^\pm(s)}\right|^2   \le 1\,, \qquad\qquad
s>(m_1+m_2)^2\,.
\eeq

\section{Analytic upper bound on  \texorpdfstring{$g_{222}^2$}{g222}} \la{maxg222}
The goal of this appendix is to prove that \eqref{S2222opt} is the amplitude with maximal coupling $g_{222}^2$ compatible with crossing symmetry and unitarity.
To this end, it is convenient to define
\beq
q(s)= - S_{22\to 22}(s) \frac{h_{22}(s)-h_{22}(m_2^2)}{h_{22}(s)+h_{22}(m_2^2)} \frac{h_{22}(s)-h_{22}(4m_1^2)}{h_{22}(s)+h_{22}(4m_1^2)}
\eeq
such that 
\beq
q(s)=q(4m_2^2-s)\,,
\eeq 
\beq
|q(s)|^2 \le 1 \,,\qquad\qquad s>4m_2^2\,,
\eeq
and 
\beq
q(m^2_2)=\frac{g_{222}^2
   \left(\sqrt{3}
   m_2^2-4 m_1
   \sqrt{   m_2^2-m_1
   ^2}\right)^2}{12 \sqrt{3}
      m_2^4 \left(4
   m_1^2-3
      m_2^2\right) \left(4
m_1^2-   m_2^2\right)}\,.
\eeq
Furthermore, $q(s)$ is analytic in the $s$-plane  minus the s-channel cut  $ \left( 4m_1^2,+\infty \right) $  and the $t$-channel cut $ \left( -\infty, 4m_2^2-4m_1^2 \right)$.
In the extended unitarity region $4m_1^2>s>4m_2^2$, we have
\footnote{Recall that $S_{22\to 22}(s)=1+\frac{  M_{22\to 22}(s) }{2 h_{22}(s)} $.}
\beq
{\rm Im} \,q(s) = - \frac{{\rm Im} M_{22\to 22}(s) }{2 h_{22}(s)}
\frac{h_{22}(s)-h_{22}(m_2^2)}{h_{22}(s)+h_{22}(m_2^2)} \frac{h_{22}(s)-h_{22}(4m_1^2)}{h_{22}(s)+h_{22}(4m_1^2)}
 \le 0\,,
\eeq
where we assumed $m_1^2<m_2^2<\frac{4}{3} m_1^2$.
We conclude that maximizing $g_{222}^2$ is equivalent to maximizing $q(m^2_2)$ subject to $q(s)=q(4m_2^2-s)$, ${\rm Im} \,q(s) \le 0$ for $4m_1^2>s>4m_2^2$ and $|q(s)|^2 \le 1$ for 
$s>4m_2^2$.

To prove that the optimal solution is given by $q(s)=1$ it is useful to change to the coordinate 
\beq
z(s)= \frac{h_{22}(m_2^2)-h_{22}(s)}{h_{22}(m_2^2)+h_{22}(s)} \,,
\eeq
such that the unit disk $|z|<1$ covers (the half ${\rm Re} \, s>2m^2_2$ of) the physical sheet (see figure 1 of \cite{Paper3} for more details).
In these coordinates, the optimization problem translates to maximizing $q(z=0)$ subject to $|q(z)|\le 1$ for $|z|=1$ and ${\rm Im} \,q(z) \le 0$ for $0<z(4m_1^2)\equiv z_0<z<1$, with $q(z)$ analytic on the unit disk minus the cut from $z_0$ to 1.
Then Cauchy's theorem leads to
\beq
q(0)= \frac{1}{2\pi i} \oint_0 \frac{dz}{z} q(z) = \frac{1}{\pi} \int_{z_0}^1 \frac{dz}{z} {\rm Im}\, q(z) + \frac{1}{2\pi}\int_{0}^{2\pi} d\theta \,q(z=e^{i\theta})\,.
\eeq
Clearly, the optimal solution is given by $q(z)=1$ (and ${\rm Im}\, q(z)=0$ inside the unit disk).
\section{3D Plots} \la{3dplotAppendix}

\begin{figure}[t!]
\begin{subfigure}{0.5\textwidth}
\center \includegraphics[width=\textwidth]{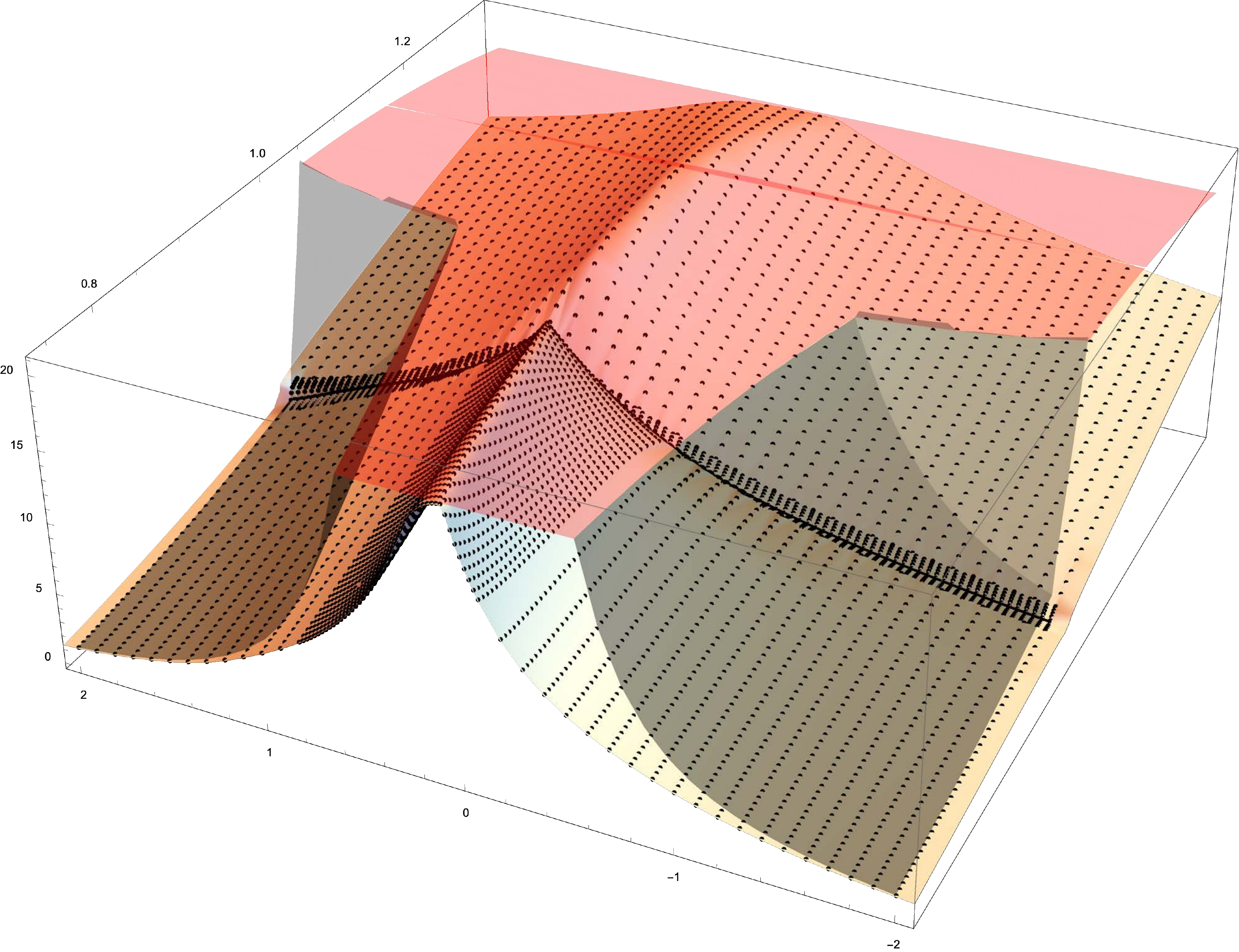} \label{photo1}
\caption{} 
\end{subfigure}\hspace*{\fill}
\begin{subfigure}{0.5\textwidth}
\center \includegraphics[width=\textwidth]{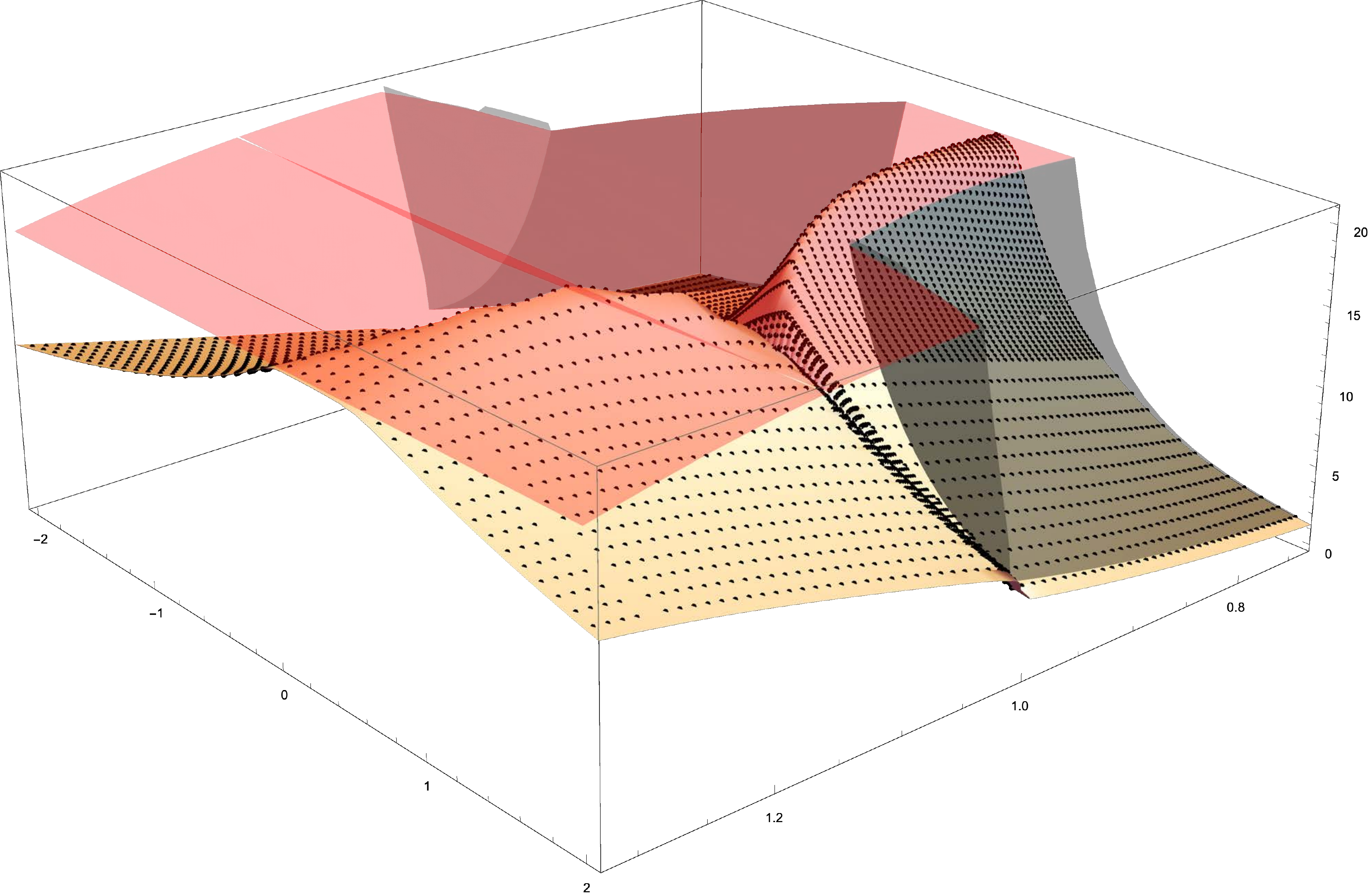} \label{photo2}
\caption{} 
\end{subfigure}
\medskip
\begin{subfigure}{0.5\textwidth}
\center \includegraphics[width=\textwidth]{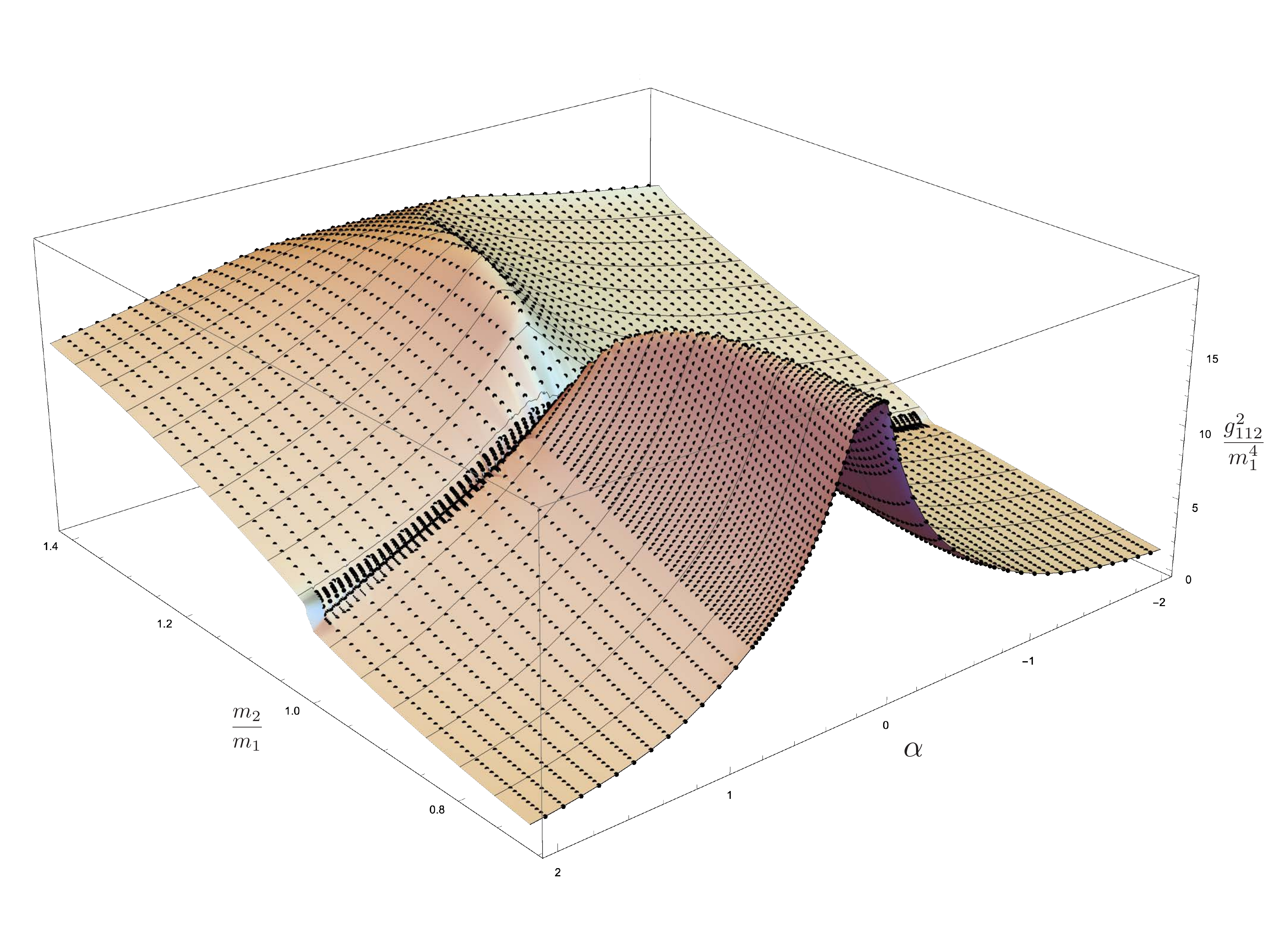} \label{photo3}
\caption{} 
\end{subfigure}\hspace*{\fill}
\begin{subfigure}{0.5\textwidth}
\center \includegraphics[width=\textwidth]{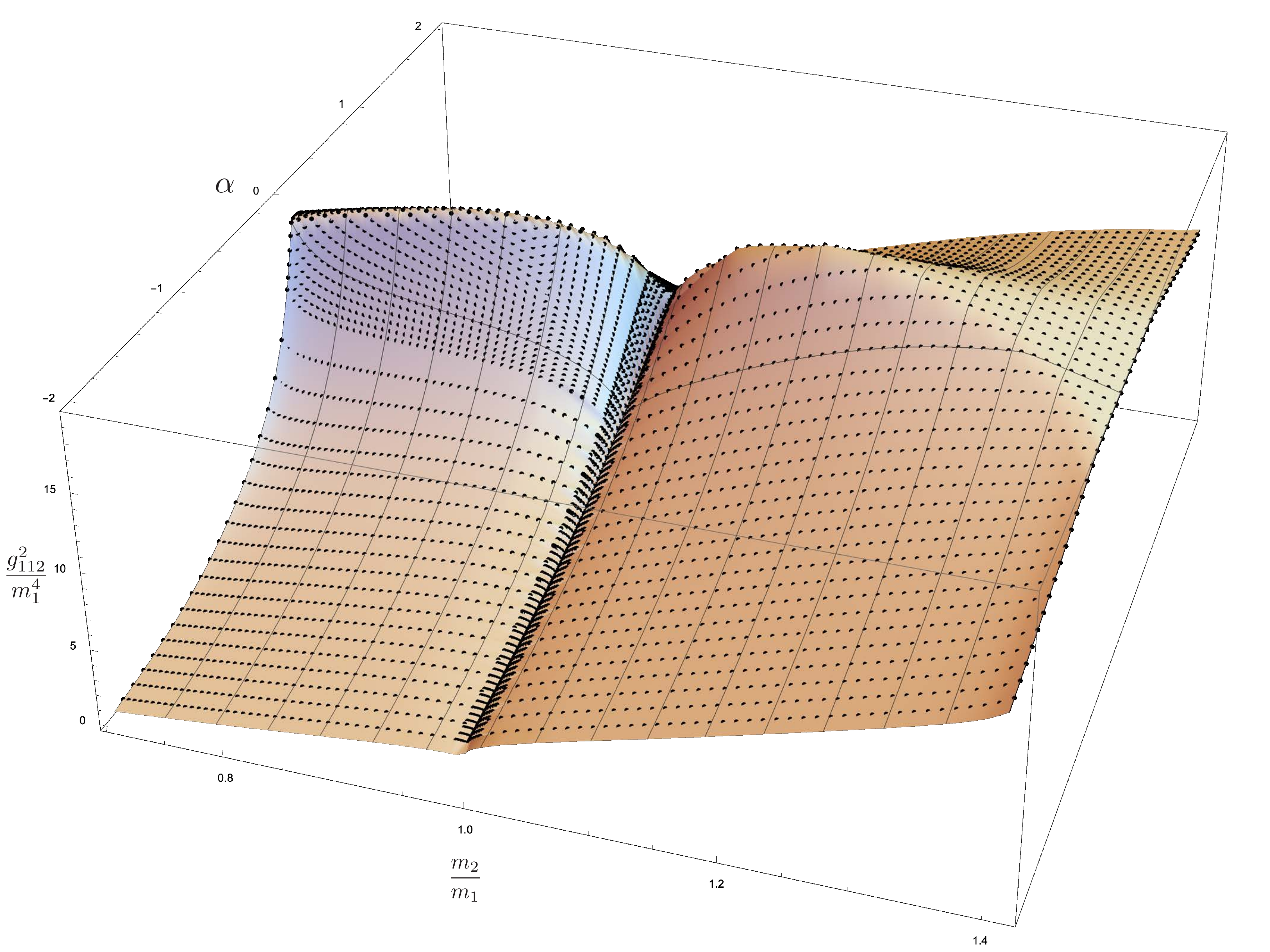} \label{photo4}
\caption{} 
\end{subfigure}
\caption{Bounds on  $g_{112}^2$ following from the multiple amplitudes analysis, as a function of $\alpha$ and $m_2/m_1$. These bounds should hold for any $\mathbb{Z}_2$ symmetric quantum field theory with two particles in the spectrum, $1$ being odd and $2$ being even. They improve the bounds derived from individual amplitudes, corresponding to the red and black surfaces. As can be seen from the various angles (a)-(d), the bound surface has many interesting features that are described in the main text.} \label{3dplot}
\end{figure}

Figure \ref{3dplot} presents  the numerical results for the maximum value of $g_{112}$ for each mass ratio $m_2/m_1$ and for each coupling ratio $\alpha=g_{222}/g_{112}$. By changing axes and looking at different sections we obtain figures \ref{Panels} and \ref{gvgm1m2} from the main text. Many of its features were discussed in section \ref{results}. The black and red surface correspond to the bounds coming from diagonal processes and are, respectively, the translated versions of the horizontal and vertical solid lines in figure \ref{Panels}.

Figure \ref{3dplot} has a ridge where the coupling $g_{112}$ is maximal for each mass ratio. That maximal value set an upper bound for the question: how big can the coupling $g_{112}$ be in a $\mathbb{Z}_2$ symmetric theory with only two stable particles? In figure \ref{g112} we depict this maximum $g_{112}^\text{max}(m_2/m_1)$ (a similar analysis can be done for $g_{222}$, leading to figure \ref{g222}). We see that this maximal coupling approaches the analytic bound derived from the diagonal $12\to 12$ component as the mass approaches the boundary of the mass range (\ref{massRange}) and is otherwise significantly stronger, specially when the particles are mass degenerate where we observe a nice kink feature in figure \ref{g112}. 

This kink has a cute geometrical interpretation in the full 3d plot in figure \ref{3dplot}: The top of the ridge meets a valley at $m_1=m_2$. The valley is a kink for any $\alpha$. For equal masses there is no extended unitarity region and that renders the numerics way more manageable. This is why we can afford to have so many points along the valley as clearly seen in the figure. 

There is one more motivation for resolving this valley region very finely: It is the natural place to look for interesting physical theories. Indeed, each optimal S-matrix in the surface of figure \ref{3dplot} saturates the extended unitarity equations (\ref{ex1111}-\ref{ex12h}). This means that the scattering of two particles of type $1$ or $2$ can never lead to multiparticle production. Processes such as $11 \to 222$ are forbidden. When dealing with 2D S-matrices, in particular extremal examples saturating unitarity such as the ones stemming from this numerical computation, we are commanded to look for integrable field theories. For $m_2 \neq m_1$, these are only possible if $M_{11\to22} = M^\text{Backward}_{12\to12} =  0$. It turns out that no point in the surface (\ref{3dplot}) satisfies this condition.\footnote{This is not an accident, we knew this to be the case a priori since this could only happen if the bound state poles in these amplitudes collided an cancelled or if some extra Landau poles were present. This is not a possibility in the mass range (\ref{massRange}).} 
This leaves the possibility of having physical theories along the equal mass line $m_2 = m_1$. This line is an one-dimensional kink in the maximal coupling surface described in detail in section \ref{Theories}.

\section{Screening}\label{Screening}

\subsection{Invisibility Cloak Toy Model} \la{toyExample}

In this appendix we highlight the importance of not leaving any densities unconstrained as they can lead to very efficient screening thus invalidating any possible bounds. To this purpose consider as a toy model the function 
\beq
f(x) = \frac{1}{x} + \int_{a}^b \frac{\rho(y)}{x-y} 
\eeq
where $0<a<b$. We think of the first term, the pole at the origin, as a target which we would like to screen. The region $[a,b]$ where the density term is defined is denoted as the screening region and that second term is denoted as the screening term. We can think of it as an invisibility cloak whose role is to make the full function small in pre-defined regions. To make it concrete suppose we want the target to be screened in a region to the right of the screening region (as we would usually associate to an invisibility cloak \textit{a la Harry Potter}) but also -- thus making it much more challenging -- in a region between the target and the screening region and even in another region to the left of the target! The point we want to make here is that this screening is trivial to achieve if we put no bounds on the density $\rho$. For that purpose it suffices to consider a discretized version of the problem $f(x) = \frac{1}{x} + \sum_\text{grid} \frac{c_i}{x-x_i}$ and show that by tuning the $c_i$'s we can indeed screen the function remarkably well. Here is an example in Mathematica:  
\begin{figure}[t]
\center \includegraphics[scale=.45]{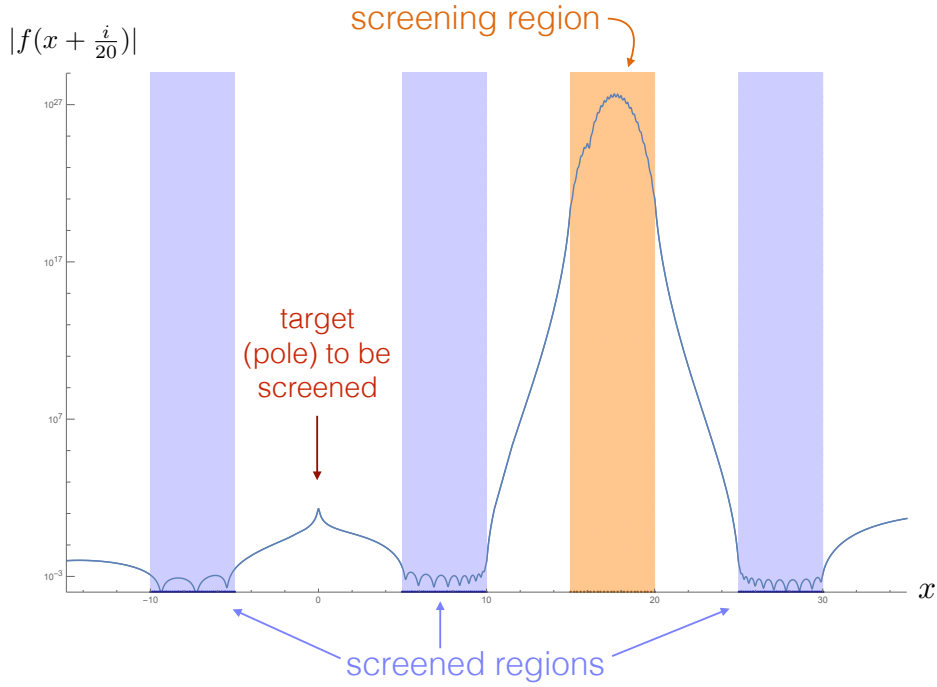}
\caption{Absolute value of the function $f(x)$ defined in the main text. The pole at the origin can be screened anywhere to any desired accuracy by a simple density in the orange region. The price to pay is that the density there is pretty extreme, fluctuating wildly and of huge magnitude. If we have regions where amplitude discontinuities are unbounded, we might expect such screening phenomena to produce such strange behaviour at unphysical regions leading to no bounds.} \label{screening}
\end{figure}
\\ \\ 
\verb"grid=Range[15, 20, 1/5];"\\
\verb"screening=Range[25,30,1/10]~Join~Range[-10,-5,1/10]~Join~Range[5,10,1/10];"\\
\verb"f[x_]= 1/x + Sum[c[y]/(x - y), {y, grid}];"\\
\verb"Total[f[screening]^2]//FindMinimum[#,Variables[#],WorkingPrecision->500]&"\\ \\
leading to the plot in figure \ref{screening}. Of course this only works because the density is unconstrained here otherwise the amount of possible screening is limited. In the screening region between $a$ and $b$ the function we get is pretty huge and wild. In other words, the invisibility cloak is working hard so that spectators in the blue screened regions see nothing. 

\subsection{Screening in our setup} \label{screeningAp2}
The phenomenon of screening happens in our setup for $m_2< m_1/ \sqrt{2}$. In this case, the optimal bounds on the couplings $g_{112}$ and $g_{222}$ are just the same as the ones obtained from the single amplitude analysis. In fact, the off-diagonal amplitudes $M_{11\to 22}$ that saturate these bounds vanish in the physical region. To understand how this is possible it is convenient to first understand why the bounds improve in the region $m_1/ \sqrt{2}<m_2<m_1$.

\begin{figure}[t]
\center \includegraphics[scale=.45]{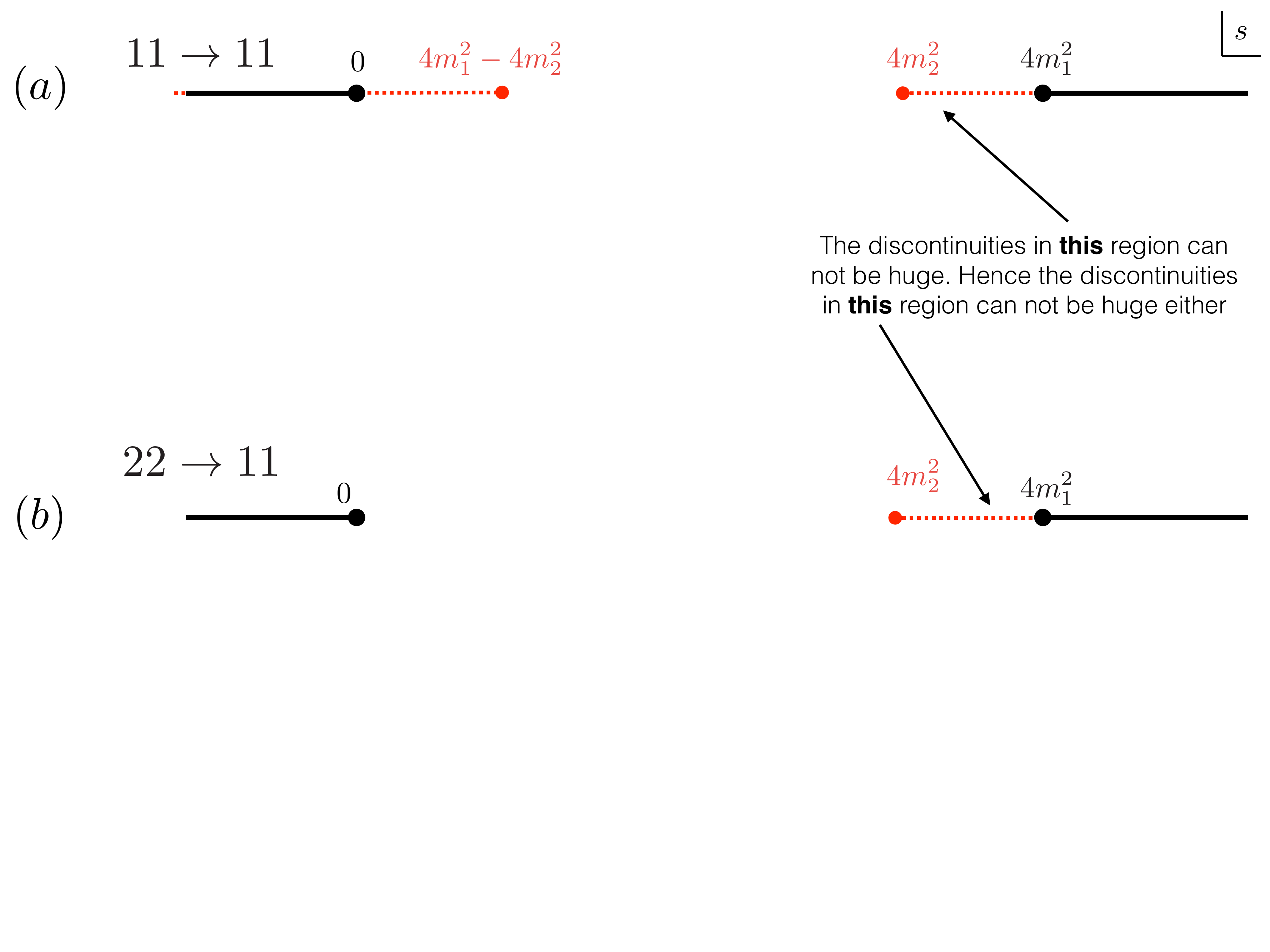}\
\vspace{-3.5cm}
\caption{If $m_2<m_1$ then the  amplitude $M_{22\to 11}=M_{11\to 22}$ contains an extended unitarity region which is bounded by the extended unitarity region in $M_{11\to 11}$. That, in turn, can not be too large or unitarity will be violated in the physical  regions (in solid black). } \label{screening1}
\end{figure}

For $m_1/ \sqrt{2}<m_2<m_1$, the amplitude $M_{11\to 22}$ can  not vanish generically so the diagonal amplitude $M_{22\to 22}$ can not saturate unitarity since some production of $11$ is inevitable.
To see this more precisely note that $M_{11\to 22}$ has poles and  an extended unitarity region where the discontinuity is bounded by the diagonal $M_{11 \to 11}$ and $M_{22 \to 22}$ components as in (\ref{SCW}). On the one hand, $M_{22 \to 22}$ is bounded by unitarity because this amplitude has no extended unitarity region. On the other hand, the discontinuity of the $11\to 11$ amplitude is positive in the extended unitarity region and therefore is bounded by unitarity in the physical region. 
This is depicted in figure \ref{screening1}. 
To summarize: we see that some screening is possible but it can not lead to a vanishing $11\to 22$ amplitude in the physical region and thus to unitarity saturation for $11\to 11$. This is why the multiple amplitude analysis \textit{had} to improve the bound obtained from a purely diagonal analysis. 

\begin{figure}[t]
\center \includegraphics[scale=.45]{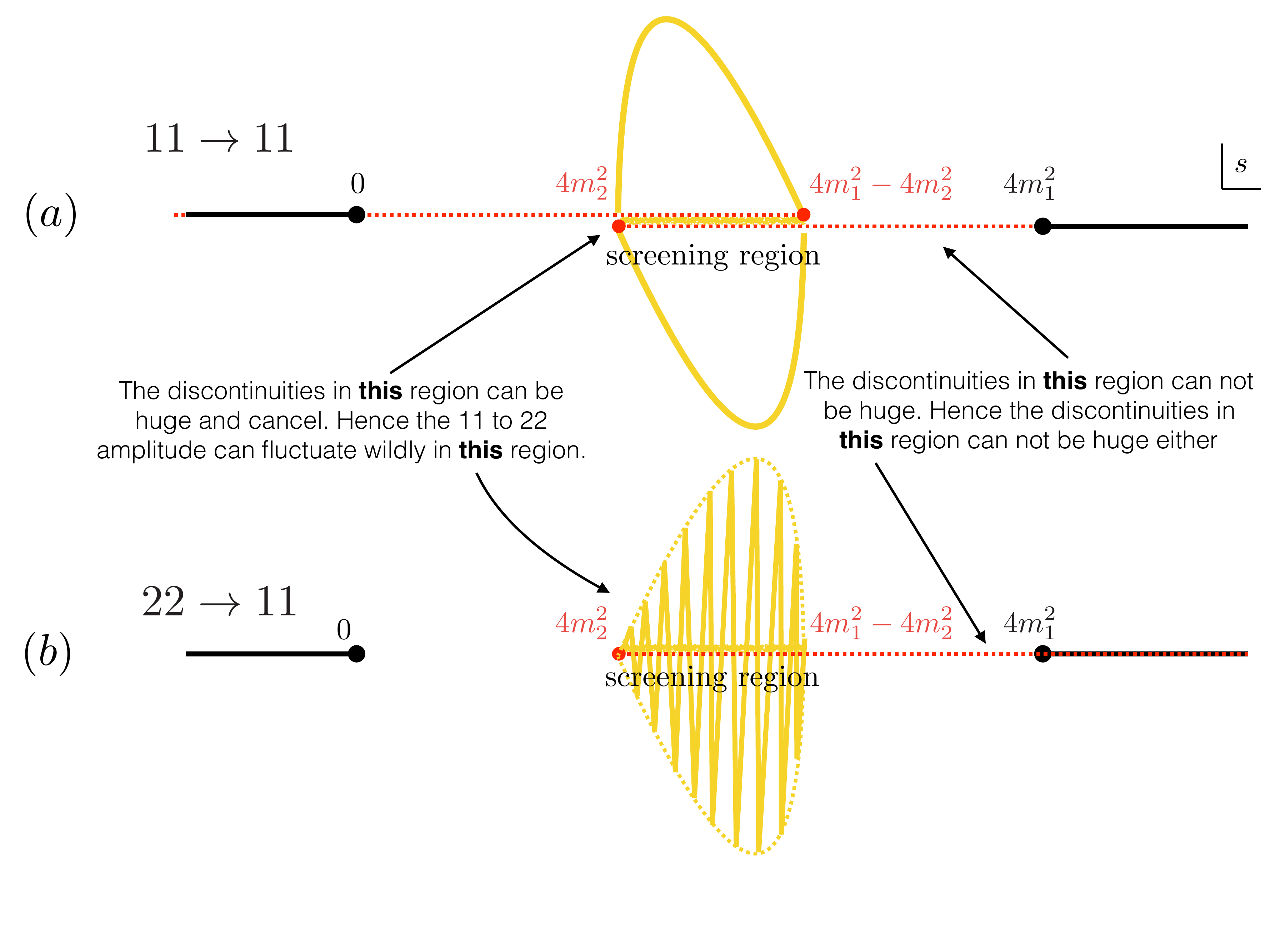}
\vspace{-1cm}
\caption{The s- and t- channel discontinuities come with opposite signs. As such, when there is an overlapping region as in the $11\to 11$ amplitude represented at the top, they can both be very large as long as their sum remains bounded and does not lead to violation of unitarity in the physical region represented by the black solid lines at the top. Furthermore the discontinuity of $M_{11\to 22}$  (at the bottom) in that same kinematical region is bounded by the ${\rm Im} M_{11\to11}$ and has no definite sign. 
Therefore, in the region where ${\rm Im} M_{11\to11}$ is unbounded, the amplitude $M_{11\to 22}$  can take advantage of screenning.} \label{screening2}
\end{figure}

This same analysis also explains why for $m_2<\frac{m_1}{\sqrt{2}}$ the diagonal bound is optimal in our setup. This is because in this range 
we have a collision between the $s$-channel and $t$-channel cuts corresponding to intermediate production of two of the lightest particles  in the scattering of the heaviest particle $11\to 11$, see figure \ref{screening2}a. The s-- and t-- channel discontinuities can now be huge as long as they cancel each other and do not lead to a violation of unitarity for the $11\to 11$ component in the physical region. If they are unbounded, then there is a region in the $11\to 22$ non-diagonal component where this amplitude can also be unbounded. (Note that if the cuts overlap then the imaginary part in the right hand side of (\ref{ex1111}) should be understood as the $s$-channel discontinuity.) If the amplitude can be huge with both signs in a finite segment then it can screen as illustrated in the previous section and can thus kill $11\to 22$ in the physical region (the backward $12\to 12$ amplitude, related by analytic continuation to the $11\to 22$ process can also be killed of course). In other words, for $m_2<\frac{m_1}{\sqrt{2}}$ we can set to zero all non-diagonal processes without violating any of our physical constraints! Therefore the full numerical plots are expected to coincide with the analytical diagonal bounds. This is indeed what we observed in our numerics. 

Would be great to find a way to improve our bounds for $m_2<\frac{m_1}{\sqrt{2}}$. The following section contains a toy model of this screening phenomena which might help elucidate what kind of physics could produce it.
 
For $m_2> \sqrt{2}m_1$ there is a similar screening phenomena that occurs.
Furthermore, in this mass range there are other Landau singularities known in two dimensions as higher pole Coleman-Thun singularities \cite{CT}. Indeed, if the mass $m_2 > \sqrt{2} m_1$ an on-shell diagram as in  figure \ref{Landau}  will produce a double pole. Its residue, as seen in the figure, is governed by the coupling (to the fourth power) and by the $2 \to 2$ on-shell S-matrix of the lightest particle. These are all objects which  we are already manipulating and it should thus be possible to tame these singularities if we properly understand how to deal with the inherent non-linearities. We look forward to reporting on this interesting problem in the future.

\subsection{Multiple resonance toy model} \label{screeningAp3}
One might wonder if the screening mechanism is a numerical artifact or could actually be realized in a reasonable QFT. Here we provide an example pointing towards the later provided we accept some fine tuning.

\begin{figure}[t]
\center \includegraphics[scale=.45]{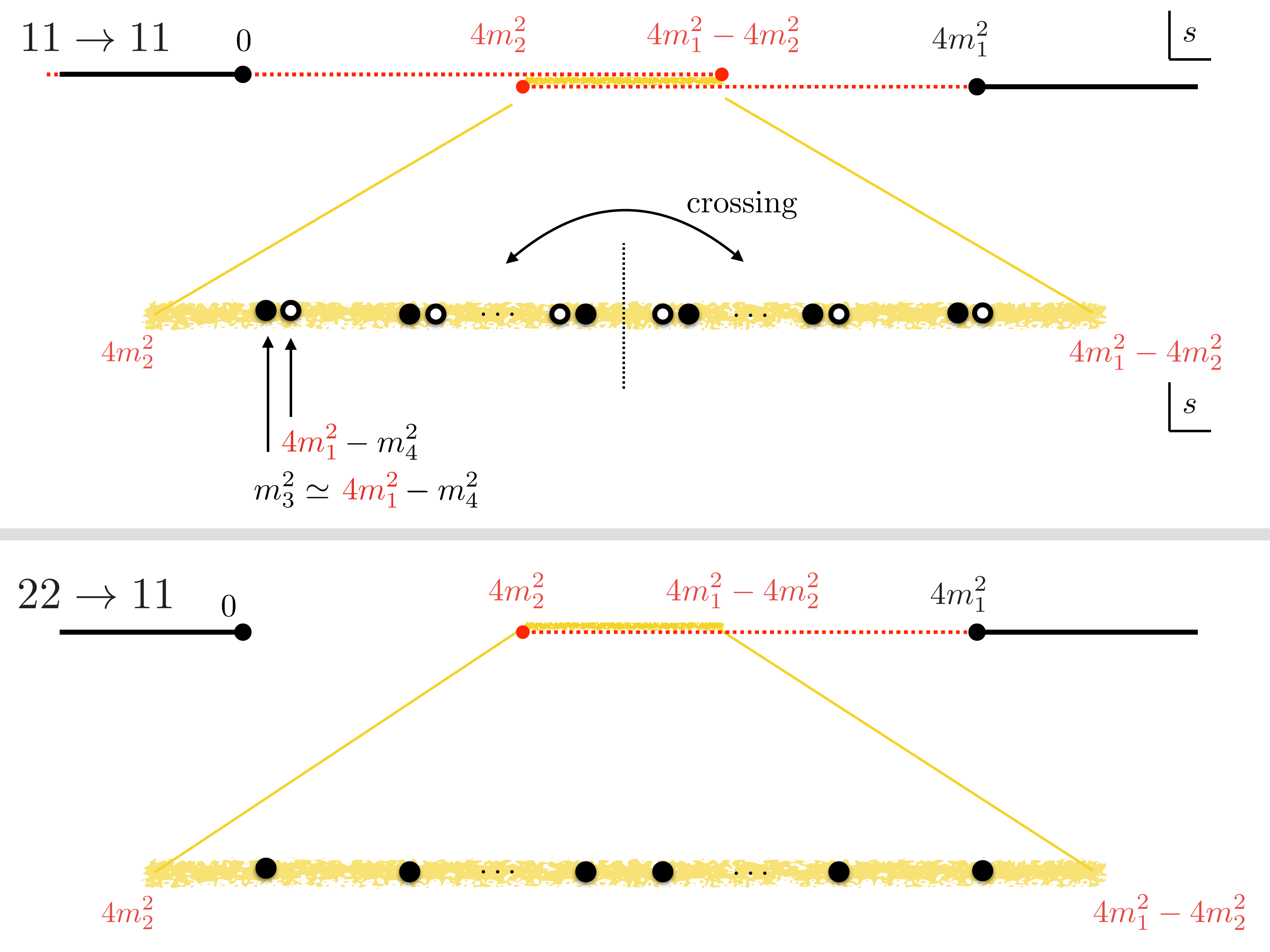}
\vspace{-0cm}
\caption{We can realize the screening mechanism with a large number of resonances with masses $m_A^2 \in [4m_2^2,4m_1^2-4m_2^2]$ in crossing related pairs so that their contribution is moderate in the diagonal channels due to $s$- and $t$-channel pair-wise cancelations while, at the same time, having the potential to screen the non-diagonal processes where no $t$-channel poles show up (since the resonances are taken to be $\mathbb{Z}_2$ even) and where the $s$-channel poles can have arbitrary sign as this is not a reflection symmetric process.} \label{screening3}
\end{figure}

Consider a theory where $m_1 > \sqrt{2} m_2$. The dangerous screening region is the region where the $s$ and $t$ channel cuts overlap in the $11\to 11$ amplitude, i.e. for~$s \in~[4m_2^2,4m_1^2-4m_2^2]$ as described in the previous section. Suppose we have many extra $\mathbb{Z}_2$ even particles $m_3,m_4,\dots,m_{2N}$ with $m_A^2$ in that screening region range\footnote{Strictly speaking these can not be stable particles since their mass is above $2m_1$ so we should think of them as long lived resonances. In other words, we should think of the corresponding poles as coming with a small imaginary part. Our cavalier analysis ignores these subtleties; the conclusions should remain the same since these small imaginary parts are important mostly for the $11\to 11$ amplitude and for this amplitude we will see that these particles do not show up. } and suppose further that for each particle with mass squared$m_A^2$ in that region there is a particle with mass squared
\beq
m_{A+1}^2  = 4m_1^2- m_A^2 + O(\epsilon^2) \,, \la{massCondition}
\eeq
where $\epsilon$ is a small parameter. 
Assume further that the couplings scale with this small parameter as 
\beq
g_{22A} =O( \epsilon) \qquad , \qquad g_{11A} =O( 1/\epsilon) \,, 
\eeq
while $g_{12A}=0$ since the extra particles are even. Finally, assume that the couplings $g_{11A}$ for two particles related as in (\ref{massCondition}) are the same up to small $\epsilon$ corrections. 
In this scenario, several interesting things might happen, including screening:
\begin{itemize}
\item The particles would not appear in the $22\to 22$ channel since they would come as poles with residues of order $\epsilon^2 \to 0$ or in the $12\to 12$ channel since they are $\mathbb{Z}_2$ even.
\item The particles would appear in the $11\to 11$ amplitude. Each particle contributes a huge amount since each coupling is of order $1/\epsilon^2$. However, because of the condition (\ref{massCondition}), for each $s$-channel pole there is a corresponding nearby $t$-channel pole which nearly cancels it,  leading to a finite $O(\epsilon^0)$ result, see figure \ref{screening3}a. If the mass degeneracy if very tiny the couplings could be huge and still lead to a very good cancelation, compatible with unitarity for this $11\to 11$ amplitude. 
\item The particles would also appear as s-channel poles in the $11 \to 22$ amplitude since $g_{11A} g_{22A}=O(\epsilon^0)$, see figure \ref{screening3}b. Note that this product could be very large and can take any sign. Hence it could lead to screening if $N$ is large as explained in the toy example in section \ref{toyExample}. This screening could then lead to $11\to 22$ being very small in the physical region. 
\end{itemize}
Of course we could ask: ``who ordered these extra particles?" No one did but since it is a priori consistent to add them, the numerics will take advantage of them and add them whenever useful for the optimization goal.

\section{Solvable Points at \texorpdfstring{$m_1 = m_2$}{m1=m2}}\label{solvable}

For $m_1 = m_2$ it is sometimes possible to rotate the one-particle basis as
\beq
\label{1particlebasis}
|1'\> = \delta |1\> + \beta |2\> ,\qquad |2'\> =  \bar{\beta} |1\> - \bar{\delta}|2\> ,\qquad \text{with } |\delta|^2 + |\beta|^2 = 1
\eeq
so that the off-diagonal amplitudes have no poles.  If that is the case, we can consistently set these amplitudes to zero and allow for the diagonal processes to saturate unitarity. For the spectrum considered in this paper, the poles terms in the $\mathbb{Z}_2$ basis are
\beq
\text{Poles}_{\mathbb{Z}_2} = \left(
\begin{array}{cccc}
\frac{-g_{112}^2}{s-m_2^2} + \frac{-g_{112}^2}{t-m_2^2} & 0 & 0 & \frac{-g_{112}g_{222}}{s-m_2^2} + \frac{-g_{112}^2}{t-m_1^2} \\
 0 & \frac{-g_{112}^2}{s-m_1^2} + \frac{-g_{112}^2}{t-m_1^2}  & \frac{-g_{112}^2}{s-m_1^2} + \frac{-g_{112}g_{222}}{t-m_2^2} & 0 \\
 0 &  \frac{-g_{112}^2}{s-m_1^2} + \frac{-g_{112}g_{222}}{t-m_2^2}  &  \frac{-g_{112}^2}{s-m_1^2} + \frac{-g_{112}^2}{t-m_1^2} & 0 \\
\frac{-g_{112}g_{222}}{s-m_2^2} + \frac{-g_{112}^2}{t-m_1^2}  & 0 & 0 & \frac{-g_{222}^2}{s-{m_2}^2} + \frac{-g_{222}^2}{t-{m_2}^2}\\
\end{array}
\right).
\label{z2basis}
\eeq
 A straightforward brute force analysis in Mathematica shows that a change of basis as in equation (\ref{1particlebasis}) can diagonalise (\ref{z2basis}) only if
 
\beq
\frac{g_{222}}{g_{112}} = -1 \qquad \implies \qquad \delta = 1/\sqrt{2}, \qquad \beta = -i/\sqrt{2},
\label{pottsc}
\eeq
in which case the rotated poles terms become of the form $\textbf{Diag}\left(\frac{-2g_{112}^2}{s-m_1^2},\frac{-2g_{112}^2}{t-m_1^2}, \frac{-2g_{112}^2}{t-m_1^2}, \frac{-2g_{112}^2}{s-m_1^2}\right)$, or 
\beq
\frac{g_{222}}{g_{112}} =  1 \qquad \implies \qquad \delta = 1/\sqrt{2}, \qquad \beta = -1/\sqrt{2},
\label{edc}
\eeq
where now the poles terms reduce to $\textbf{Diag} \left( \frac{-2g_{112}^2}{s-m_1^2} + \frac{-2g_{112}^2}{t-m_1^2}, 0,0,\frac{-2g_{112}^2}{s-m_1^2} +\frac{-2g_{112}^2}{s-m_1^2}\right)$. In both cases, the S-matrix in the rotated basis 
(\ref{1particlebasis}) is schematically of the form
\beq
\left(
\begin{array}{cccc}
a & c & c & b \\
 c & e & d & c\\
 c & d & e &  c\\
 b  & c & c & a\\
\end{array}
\right).
\eeq
It is then straightforward to apply the maximum modulus principle as in section \ref{intro} to obtain bounds the optimal couplings. For the (\ref{pottsc}) scenario, we conclude that $a$ and $e$ are fixed while $b = c = d = 0$, since the diagonal processes must saturate unitarity. This solution corresponds to the S-matrix of the 3-states Potts model at $T \neq T_c$. Note that in the rotated basis we no longer have $s \leftrightarrow t$ symmetry for diagonal processes in this case, since the external particles no longer diagonalise the charge conjugation operator, see appendix \ref{Potts}. On the other hand, for (\ref{edc}) we conclude only that $a$ is fixed and $b = c = 0$ (note that after changing basis $b$ is no longer related to $d$ by crossing), while $d$ and $e$ correspond to a zero modes that do not affect the optimal coupling. A particular choice of $d$ and $e$ lead to the hyperbolic limit of the elliptic deformation of the supersymmetric Sine-Gordon model, discussed in section \ref{Theories}.

\section{3-state Potts field theory}
\label{Potts}

The 3-state Potts model in two dimensions has a continuous phase transition described by a non-diagonal minimal model with central charge $c=4/5$ and a global permutation symmetry $S_3$.
This conformal field theory (CFT) contains 3 relevant scalar operators invariant under $\mathbb{Z}_2 \subset S_3$ (see \cite{Mussardo:2010mgq} for a nice introduction to this topic).
 This allows us to define a  family of $\mathbb{Z}_2$ symmetric QFTs with action
\beq
A_{QFT} = A_{CFT} + \tau \int d^2x \,\epsilon(x) + h  \int d^2x\, \sigma_+(x)+ h'  \int d^2x\, \Omega_+(x)\,, \label{Pottsaction}
\eeq
where $\tau$, $h$ and $h'$ are relevant couplings and we used the notation of \cite{Lepori:2009ip}.
The scaling dimensions of the relevant operators are $\Delta_\epsilon=\frac{4}{5}$, $\Delta_{\sigma}=\frac{2}{15}$ and $\Delta_{\Omega}=\frac{4}{3}$.
The purely thermal deformation ($\tau\neq 0$ and $h=h'=0$) preserves the $S^3$ symmetry and leads to an integrable QFT. This theory has only two stable particles with the same mass $m$ transforming as a doublet of $S^3$. 
These particles are usually described in a basis $|A\rangle, |A^\dagger \rangle$ where the $\mathbb{Z}_2$ acts as charge conjugation $C|A\rangle = |A^\dagger \rangle$ with $C^2=1$  \cite{Potts, hep-th/0511168}. In this basis, the S-matrix is diagonal, \emph{i.e.} $S_{AA \to A^\dagger A^\dagger}=S_{AA^\dagger \to A A^\dagger}^\text{Backward}=0$ and 
\begin{align}
S_{AA \to AA}   = S_{A^\dagger A^\dagger \to A^\dagger A^\dagger}  = 
\frac{\sinh\left( \frac{\theta}{2} + \frac{i\pi}{3} \right)}
{\sinh\left( \frac{\theta}{2} - \frac{i\pi}{3} \right)}\,,\qquad
S_{AA^\dagger \to AA^\dagger}^\text{Forward}   = -
\frac{\sinh\left( \frac{\theta}{2} + \frac{i\pi}{6} \right)}
{\sinh\left( \frac{\theta}{2} - \frac{i\pi}{6} \right)} \,,
\end{align}
where we used the rapidity $\theta$ to parametrize the Mandelstam invariant $s= 4 m^2 \cosh^2\frac{\theta}{2}$. Notice that, in this basis, the Yang-Baxter equations are trivially satisfied.

In this paper, we work in the eigenbasis of the $\mathbb{Z}_2$ global symmetry generated by charge conjugation,
\beq
|1\rangle =e^{i\pi/4} \frac{ |A\rangle -  |A^\dagger \rangle}{\sqrt{2}}  \,,\qquad \qquad
|2\rangle =e^{-i\pi/4} \frac{ |A\rangle +  |A^\dagger \rangle }{\sqrt{2}} \,.
\eeq
In this basis, we find
\begin{align}
S_{11\to 11}&=S_{22\to 22}=S_{12\to 12}^\text{Forward}=
-\frac{i\sinh\theta}{2i\sinh\theta+\sqrt{3}} \,,\\
S_{11\to 22}&=-S_{12\to 12}^\text{Backward}=
\frac{\sqrt{3} \cosh\theta}{2i\sinh\theta+\sqrt{3}}\,.
\end{align}
Using equation \eqref{SandM} we can obtain the expressions \eqref{MPotts} for the connected scattering amplitudes that maximize $g_{112}^2$ for $m_1=m_2$ and  $g_{222}=-g_{112}$ (point A in figure \ref{gvgm1m2}). 

The magnetic deformations $h$ and $h'$ in \eqref{Pottsaction} preserve $\mathbb{Z}_2$ and therefore must be compatible with our bounds, at least for small magnetic deformations $h \tau^{-\frac{14}{9} }\ll 1$  and $h' \tau^{-\frac{10}{9} }\ll 1$ that do not give rise to more stable particles. In fact, the mass spectrum of these theories (with $h\neq 0$ and $h'=0$) has been studied in \cite{Lepori:2009ip} using the Truncated Conformal Space Approach.
The authors observed that the degeneracy between the two particles is lifted for $h\neq 0$. It would be interesting to study the cubic couplings and the S-matrices of this 2-parameter family of $\mathbb{Z}_2$ symmetric QFTs and compare them to our bounds.

\section{Tricritical Ising (cusp)}
 \label{KinkIsing}
 
The tricritical Ising model in two dimensions has a continuous phase transition described by a diagonal minimal model with central charge $c=7/10$ and a  $\mathbb{Z}_2$ (spin flip) symmetry. 
This conformal field theory (CFT) contains 2 relevant scalar operators invariant under $\mathbb{Z}_2$ (see \cite{Mussardo:2010mgq} for a nice introduction to this topic).
 This allows us to define a  family of $\mathbb{Z}_2$ symmetric QFTs with action
\beq
A_{QFT} = A_{CFT} + \tau \int d^2x \,\epsilon(x) + \tau'  \int d^2x\, \epsilon'(x)\,, \label{Tricriticalaction}
\eeq
where $\tau$, $\tau'$ are relevant couplings.
The scaling dimensions of the relevant operators are $\Delta_\epsilon=\frac{1}{5}$ and
$\Delta_{\epsilon'}=\frac{6}{5}$.
The purely thermal deformation ($\tau\neq 0$ and $\tau'=0$)  leads to an integrable QFT. This theory has seven  particles but only four of them have masses below the continuum of multi-particle states.
The masses of these particles are given in table \ref{tablemasses}.

In this paper, we constrained the space of 2D S-matrices with $\mathbb{Z}_2$ symmetry by requiring unitarity and analyticity for the two-to-two S-matrix elements involving only the two lightest particles $\{m_1, m_2\}$ as external states. We restrained ourselves to study the subset of theories for which only $m_1$ and $m_2$ themselves appeared as bound states in these matrix elements.  This excludes the tricritical Ising field theory from our analysis in the many body of this paper.
However, one can easily relax this restriction. We shall not do a full numerical study of the multiple amplitude bootstrap in this more general setup. We will just derive the analytic bounds that follow from the amplitudes $S_{11\to 11}$ and $S_{12\to 12}^{\text{Forward}}$.
For concreteness, we consider  a theory with $\mathbb{Z}_2$ symmetry and a mass spectrum as in the table \ref{tablemasses}.\footnote{
Any masses within the range $0<4m^2_1-m^2_4<m^2_2<2m^2_1$ and $m^2_1<2m^2_1 + 2m^2_2 - m^2_3< m^2_1 + m^2_2$ would lead to qualitatively the same conclusions regarding single amplitude bounds as below but the precise locations of cusps and edges on the bounds does depend on the masses. If we deviate away from these mass constraints then we we would have to redo the analysis. This is the same discontinuous nature of the bounds already observed in \cite{Paper2}, see e.g. figures 10 and 11 therein. }
\begin{table}[ht]
\centering
 \begin{tabular}{||c c c||} 
 \hline
 Particle & Mass & $\mathbb{Z}_2$ charge \\ [0.5ex] 
 \hline\hline
 $m_1$ &  $m$ & \text{odd} \\ 
 \hline
 $m_2$ & $2m\cos{5\pi/18}\approx 1.29 \,m$ & \text{even} \\
 \hline
 $m_3$ & $2m\cos{\pi/9}\approx 1.88 \,m$ & \text{odd} \\
 \hline
 $m_4$ & $2m\cos{\pi/18}\approx 1.97 \,m$ & \text{even} \\
 \hline
 & + possibly extra particles with masses bigger than $m_1 + m_2$. &\\
 \hline
 \end{tabular}
 \caption{Spectrum assumed for the analysis in this appendix.}
  \label{tablemasses}
\end{table}

With this setup there is a richer structure of couplings to play with. For example, in the ${11 \rightarrow 11}$ amplitude  we have bound state poles corresponding to particles $m_2$ and $m_4$:
\begin{equation}
S_{11 \rightarrow 11}(s) = -\mathcal{J}(m_2^2)\frac{g_{112}^2}{s-m_2^2} - \mathcal{J}(m_4^2)\frac{g_{114}^2}{s - m_4^2} + \text{t-channel poles} + \text{cuts}
\end{equation}
where $\mathcal{J}(s)=\frac{1}{2\sqrt{s(4m_1^2-s)}}$. One question that could be asked is: what values for the pair $(g_{112}^2, g_{114}^2)$ are allowed by the unitarity constraint $|S_{11 \rightarrow 11}| \leq 1$?

\begin{figure}[t]
\center \includegraphics[scale=.45]{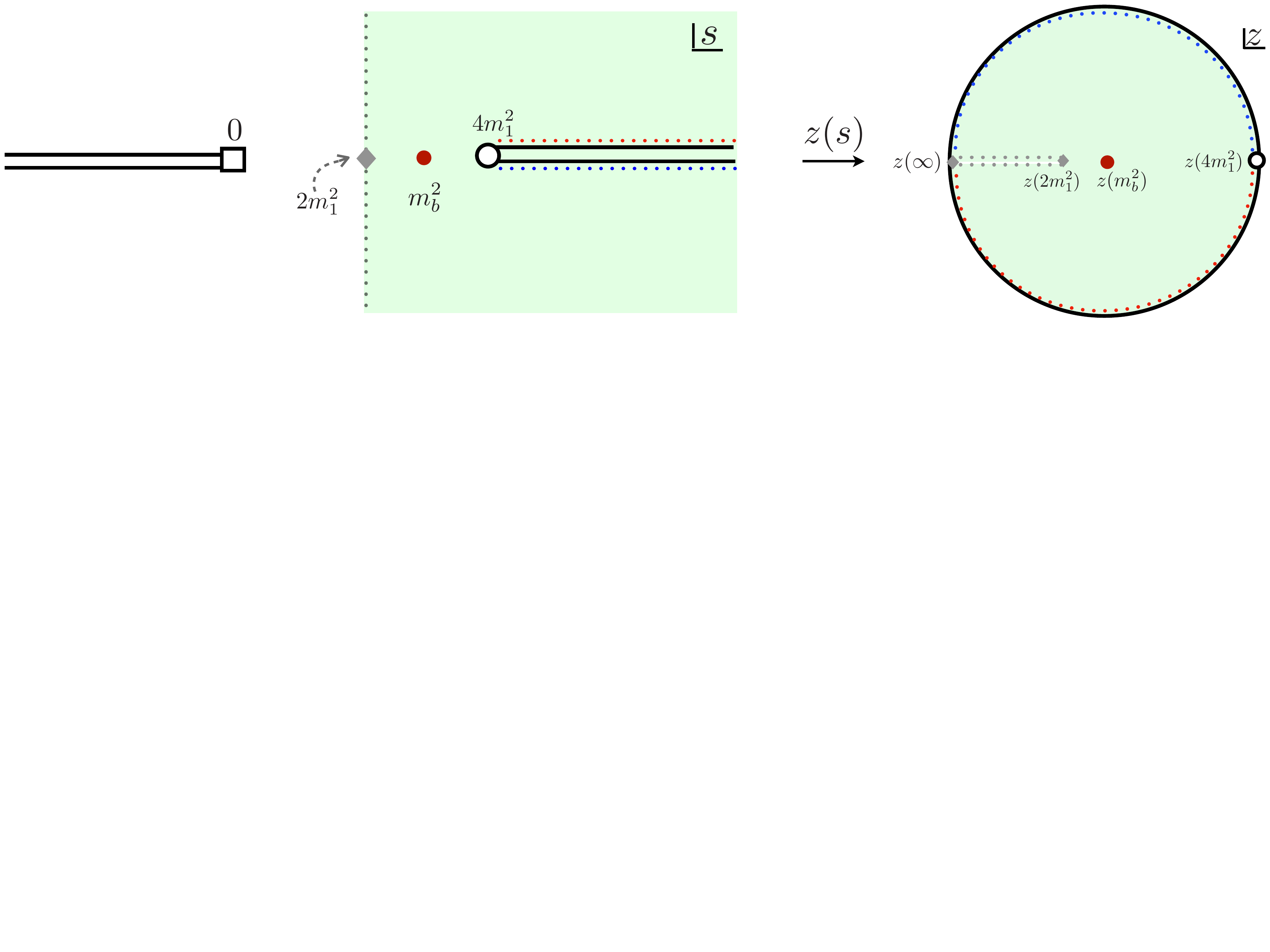}
\vspace{-8cm}
\caption{ The change of variables from $s$ to $z$ trivialises the crossing symmetry of the  S-matrix and maps the cut half-plane $s>2m_z^2$ to the unit disk by ``opening" the cut and mapping it to the boundary of the unit disk. In doing so, it maps $s=m_b^2$ to $z=0$, $s=\infty$ to $z=-1$ and the imaginary axis $s \in [2m^2_1 -i \infty, 2m^2_1 + i \infty]$ to a segment in the real $z$ axis.} 
\label{zmap}
\end{figure}

\begin{figure}[t]
\center \includegraphics[scale=.35]{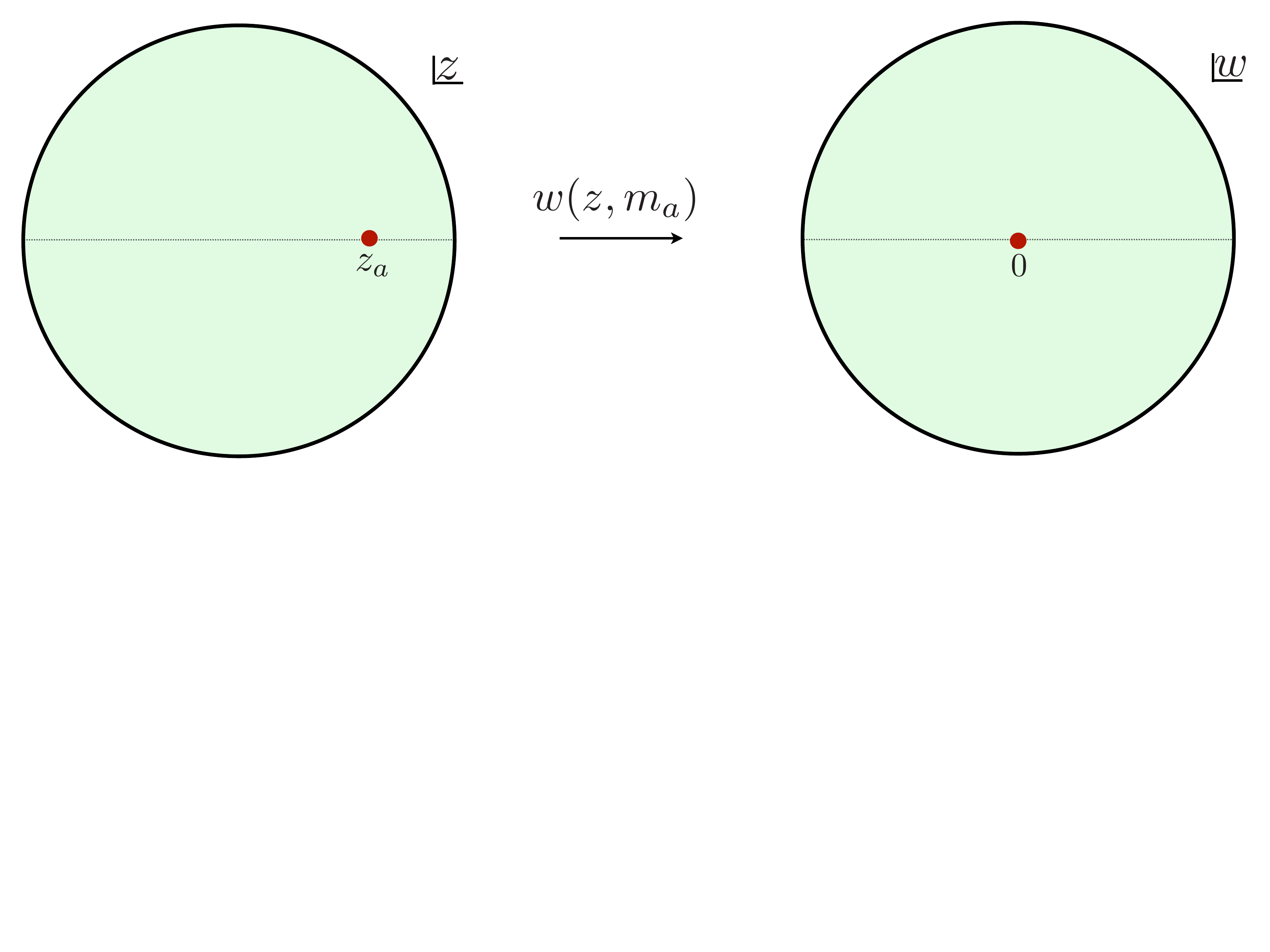}
\vspace{-4cm}
\caption{The map $w_a$ is an automorphism of the unit disk. It maps $z_a$ to the origin. When $z_a$ is real, it preserves the real segment $[-1,1]$.} \label{wmap}
\end{figure}

 To answer this question, 
it is useful to introduce the variables
\begin{align}
z(s) &= \frac{m_b \sqrt{4m_1^2 -m_b^2} - \sqrt{s} \sqrt{4m_1^2-s} }{m_b \sqrt{4m_1^2 -m_b^2} + \sqrt{s} \sqrt{4m_1^2-s}},\\
z_a &\equiv z(m_a^2)\\
w_a &\equiv w(z,m_a) = \frac{z-z_a}{1- \bar{z_a} z}, \hspace{2cm}|z_0|\leq1 \label{w}
\end{align}
where we take $m_b^2 = 2m_1^2$  for convenience. This choice of $m_b$ maps $s \in [2m^2_1,4m^2_1]$ to $z \in [0,1]$ and the imaginary axis $s \in [2m^2_1 -i \infty, 2m^2_1 + i \infty]$ to $z \in [-1,0]$.  See figures \ref{zmap} and \ref{wmap} for an illustration of such maps.

 As a function of the $z$ variable, $S_{11\rightarrow 11}$ has poles at $z(m_2^2)$ and $z(m_4^2)$. Therefore the function
 \begin{equation}
  f(z) = -S(z)w_2 w_4
  \end{equation} 
 is a holomorphic function on the unit disk which satisfies, as a consequence of unitarity, $|f(z)|\leq1$. Moreover, from the fact that $S(s)$ has a negative residue at $s = m_2^2$ and a positive residue at $s= m_4^2$, we find that $f(z)$ is positive at $z_2$ and $z_4$.

 Suppose we want to maximise $g_{112}^2$. This is equivalent to maximising $f(z^*=z(m_2^2))$. But by the maximum modulus principle, $|f(z)| \leq 1$ everywhere inside the unit disc and, moreover, the optimal value $|f(z^*)| =1$ is only achieved when $|f(z)| =1$. From the fact that $f$ is positive at $z_2$ and $z_4$ we conclude that a maximal $g_{112}^2$ is obtained for $f(z) = 1$ and
 \begin{equation}
 (g^{max}_{112})^2 = \text{res}_{s=m_2^2} \frac{1}{\mathcal{J}(s) w_2 w_4}.
 \end{equation}
 Note that this solution also maximises $g_{114}^2$ so that
 \begin{equation}
 (g^{max}_{112})^2=\text{res}_{s=m_4^2} \frac{1}{\mathcal{J}(s) w_2 w_4}.
\end{equation}

 Now we maximise $g_{112}^2$ under the extra constraint that $g_{114}^2 = \alpha (g_{114}^{max})^2$ with $\alpha \in [0,1]$. This maximisation problem (together with the equivalent problem of maximising $g_{114}^2$ with $g_{112}^2$  fixed) completely determines the subspace of $(g_{112}^2, g_{114}^2)$ compatible with $|S_{11\rightarrow 11}| \leq 1$, since this space is convex.
 
 Under this extra constraint, we know that $f(z_4) =\alpha$ and so the solution $f(z) = 1$ is no longer possible. Consider, however, the function
 \begin{equation}
 g(w_4) =w(f(z(w_4),\alpha) = \frac{f(z(w_4)) - \alpha}{1-\alpha f(z(w_4))},
 \end{equation}
where we now think of $f$ as a function of $w_4$ by inverting equation \ref{w}. Since $w(f,\alpha)$ is an increasing function of $f$, to maximise $f(w_4(z_2)) = \frac{g_{112}^2}{(g_{112}^{max})^2}$ is equivalent to maximise $g(w_4(z_2))$. Moreover, since $g$ is an automorphism of the disk, unitarity implies $|g(w_4)|\leq1$ for $w_4$ on the unit disc. Finally, $g(0) =0 $. Now recall Schwartz Lemma:
\begin{lemm}
\textbf{Schwartz Lemma}: Let $\mathbf{D}$ be the unit disk and $g:\mathbf{D}\rightarrow\mathbf{D}$ be a holomorphic map such that $g(0)=0$ and $|g(w)|\leq1$ on $\mathbf{D}$. Then $|g(w)| \leq |w|$. Moreover, if the inequality is saturated for any non-zero point in $\mathbf{D}$, then $g(w) = a w$ with $|a| = 1$.
\end{lemm}

We conclude that under the extra constraint $g_{114}^2 = \alpha (g_{114}^{max})^2$, the maximal value for $g_{112}^2$ is given by the solution of the following algebraic equation on $S_{11 \rightarrow 11}$:
\begin{equation}
w_4 =  \frac{f(z(w_4)) - \alpha}{1-\alpha f(z(w_4))}.
\end{equation}

\begin{figure}[t]
\centering
\includegraphics[scale=0.38]{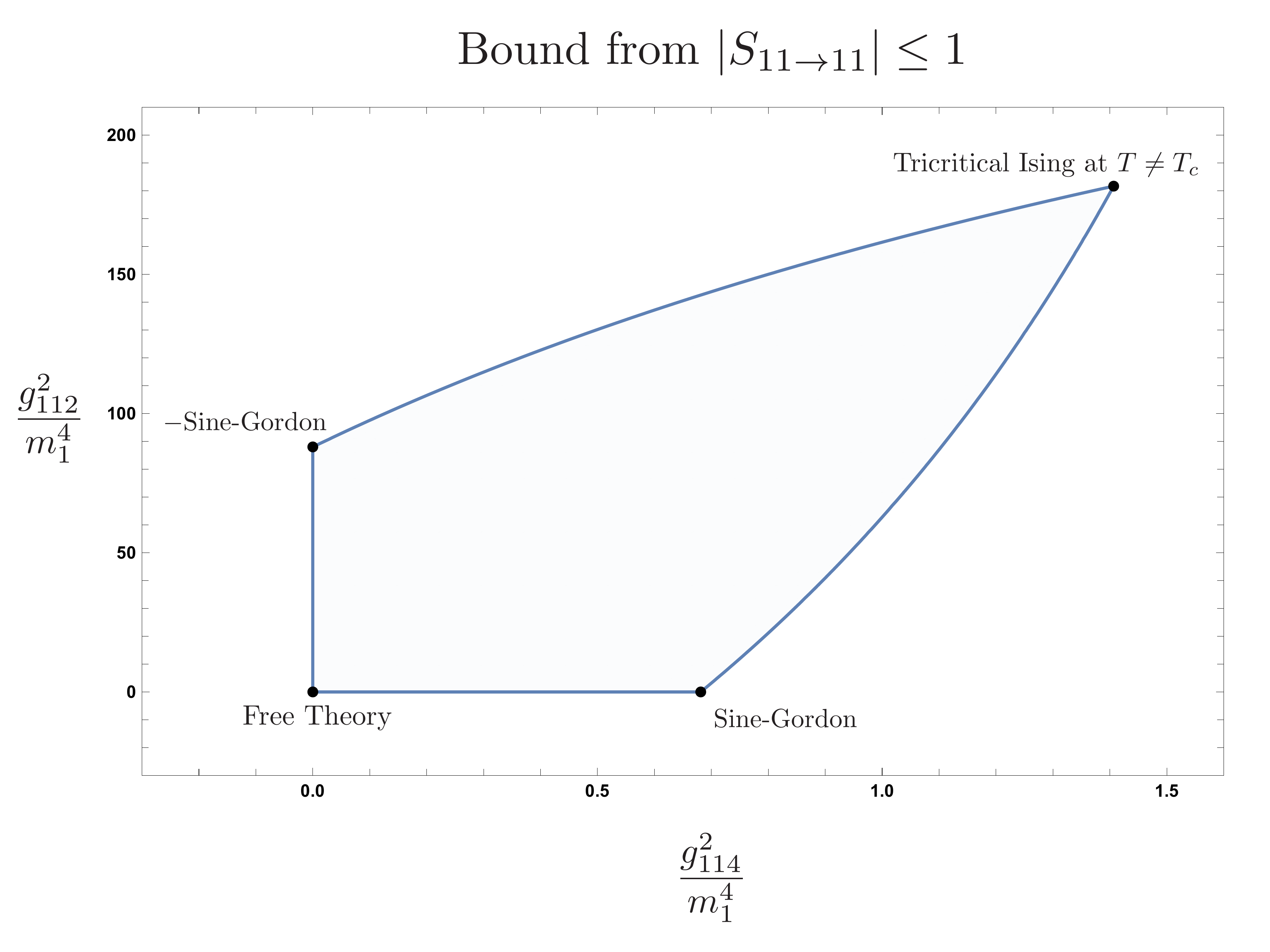}
\caption{Space of allowed couplings compatible with $|S_{11 \rightarrow 11}| \leq 1$ and spectrum according to table (\ref{tablemasses}). This space has kinks that are related to integrable theories.}
\label{S1111}
\end{figure}

After performing the equivalent exercise for the maximisation of $g_{114}^2$ under $g_{112}^2 = \beta (g_{112}^{max})^2$, where $\beta \in [0,1]$, we obtain the allowed space of $(g_{112}^2,g_{114}^2)$ as depicted in the figure \ref{S1111}.

\begin{figure}[ht]
\centering
\includegraphics[scale=0.45]{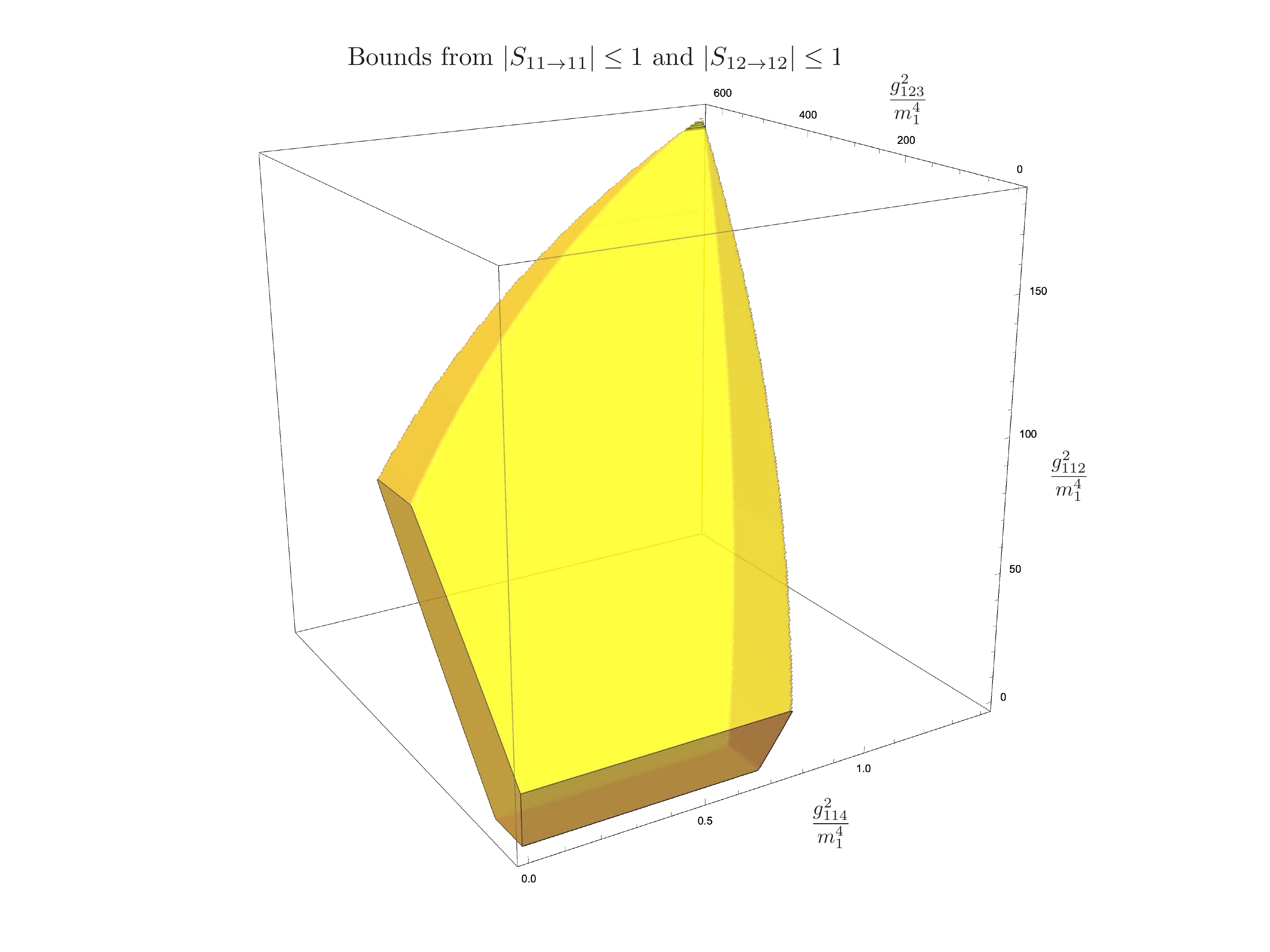}
\caption{Space of allowed couplings compatible with $|S_{11 \rightarrow 11}| \leq 1$ and $|S_{12 \rightarrow 12}| \leq 1$ and spectrum as in table (\ref{tablemasses}). The tip of this spear is the thermal deformation of the Tricritical Ising Model.}
\label{3dplottc}
\end{figure}

One can play the same game for the $S_{12 \rightarrow 12}^{\text{Forward}}$ matrix element, which contains bound state poles corresponding to particles $m_1$ and $m_3$ (and therefore constrain the space $(g_{112}^2,g_{123}^2)$. One can then combine both results into a $3D$ plot of allowed triplets $(g_{112}^2, g_{114}^2,g_{123}^2)$ compatible with $|S_{11 \rightarrow 11}| \leq 1$ and  $|S_{12 \rightarrow 12}| \leq 1$. This is figure \ref{3dplottc}.

It would be interesting to perform a multiple amplitudes analysis for this setup and explore the space of masses $(m_1, m_2, m_3, m_4)$ in the vicinity of the values of table \ref{tablemasses}.
Notice that in such an analysis, the values in  table \ref{tablemasses} would be single out by the condition that the multiple amplitude bounds saturate the single amplitude bounds 
of figure \ref{3dplottc} at the tip of the spear. That is because only for those particular masses can the multiple poles in the off-diagonal channels coincide and cancel (\`a la integrable bootstrap), allowing for the off-diagonal amplitudes to vanish and for the diagonal processes to saturate unitarity, as in the boundary of the yellow surface in figure \ref{3dplottc}.  
It would also be interesting to see if the sub-leading (non-integrable) deformation $\tau'$ of the tricritical Ising model leads to any feature of the bounds.

\section{Numerical optimization as a SDP}\label{semidefiniteequivalence}

As discussed in section \ref{implementation}, once we fix $\alpha = \frac{g_{222}}{g_{112}}$, our discretised ansatz for the amplitudes depends only on the variables $\vec{\eta} =\{g^2_{112}, C_{a \to b}, \sigma_{a \to b}(x_i)\}$. To reduce the maximisation problem to an SDP, all we need to do is to write the extended unitarity constraint (\ref{matrixextended}) as a semidefinite constraint linear on $\vec{\eta}$, as in (\ref{matrixextendedlinear}). The purpose of this appendix is to prove the equivalence of (\ref{matrixextended}) and  (\ref{matrixextendedlinear}) or, explicitly,
\beq
 \bordermatrix{~ & ~ & \cr
         ~ & \mathbb{I} & \rho  \mathbb{M} \cr
         ~ & (\rho \mathbb{M})^\dagger & 2\text{Im } \mathbb{M} \cr} \succeq 0 \iff 2\text{Im} \mathbb{M} \succeq \mathbb{M}^\dagger \rho^2 \mathbb{M} .\label{toprove}
\eeq 

\textbf{Proof of} $\Rightarrow$:
\beq
 \bordermatrix{~ & ~ & \cr
         ~ & \mathbb{I} & \rho  \mathbb{M} \cr
         ~ & (\rho \mathbb{M})^\dagger & 2\text{Im } \mathbb{M} \cr} \succeq 0 \Rightarrow 
          \left(\begin{array}{c}
          \vec{v}\\
          \vec{w}\\
           \end{array} \right)^\dagger\left(\begin{array}{c c}
         \mathbb{I} & \rho  \mathbb{M} \\
         (\rho \mathbb{M})^\dagger & 2\text{Im } \mathbb{M} \\
         \end{array}\right)\left(\begin{array}{c}
          \vec{v}\\
          \vec{w}\\
           \end{array} \right) \geq 0, \qquad \forall \vec{v}, \vec{w} \in \mathbb{C}^2
\eeq
This becomes the r.h.s\ of (\ref{toprove}) if we pick $\vec{v} = - \rho \mathbb{M} \vec{w}$.

 \textbf{Proof of} $\Leftarrow$:
\begin{align}
0 &\leq  (\vec{v} + \rho \mathbb{M} \vec{w})^\dagger  (\vec{v} + \rho \mathbb{M} \vec{w}) =  \vec{v}^\dagger \vec{v}  + \vec{v}^\dagger \rho \mathbb{M} \vec{w} + \vec{w}^\dagger \mathbb{M}^\dagger \rho \vec{v}  + \vec{w}^\dagger \mathbb{M}^\dagger \rho^2 \mathbb{M} \vec{w} \\
&\leq  \vec{v}^\dagger \vec{v}  + \vec{v}^\dagger \rho \mathbb{M} \vec{w} + \vec{w}^\dagger \mathbb{M}^\dagger \rho \vec{v}  + \vec{w}^\dagger\text{Im} \mathbb{M} \vec{w} \iff \left(\begin{array}{c c}
         \mathbb{I} & \rho  \mathbb{M} \\
         (\rho \mathbb{M})^\dagger & 2\text{Im } \mathbb{M} \\
         \end{array}\right)\succeq 0 \nonumber
\end{align}
where we used the r.h.s.\ of (\ref{toprove}) in the second inequality.

\section{Elliptic Deformation}\label{Mathematica}

Bellow one can find the definition of the supersymmetric sine-Gordon elliptic deformation S-matrix in \verb"Mathematica" friendly notation.

\noindent \\
\verb"w = EllipticK[k]/\[Pi];"\\
\verb"e = 1;"\\
\verb"g = 2 \[Pi]/3;"\\ \\
\verb"ED = {{JacobiDN[g w, k] - (JacobiDN[I q w, k] JacobiSN[g w, k])/(e"\\
\verb"JacobiCN[I q w, k] JacobiSN[I q w, k]), 0, 0,"\\
\verb"(JacobiDN[I q w, k] JacobiSN[g w, k])/(e JacobiCN[I q w, k])},"\\
\verb"{0, 1, -(JacobiSN[g w, k]/(e JacobiSN[I q w, k])), 0},"\\
\verb"{0, -(JacobiSN[g w, k]/(e JacobiSN[I q w, k])), 1, 0},"\\
\verb"{(JacobiDN[I q w, k] JacobiSN[g w, k])/(e JacobiCN[I q w, k]), 0, 0, -JacobiDN[g w, k] "\\
\verb"- (JacobiDN[I q w, k] JacobiSN[g w, k])/(e JacobiCN[I q w, k] JacobiSN[I q w, k])}};"\\ \\
\verb"IntR[x_, k_] := 1/(2 \[Pi] I)Block[{w = (EllipticK[k]/\[Pi])},"\\
\verb" (NIntegrate[Log[((1 - JacobiSN[g w, k]^2/JacobiSN[I q w, k]^2)^ -1)"\\
\verb"/Sinh[q]^2]/Sinh[q - x], {q, -Infinity, Infinity}])];"\\ \\
\verb"U[x_, k_] := -I Sinh[x] Exp[IntR[x, k]];"\\
\verb"CDDPole[q_] := (Sinh[q] + I Sin [g])/(Sinh[q] - I Sin[g])"\\\\ 
\verb"SED[Q_, K_] := U[Q, K] CDDPole[Q] ((ED /. w -> EllipticK[k]/\[Pi])"\\
\verb" /. {k -> K, q -> Q});"\\ \\
\noindent
Note that to compute the S-matrix for physical $\theta \in \mathbb{R}$, one must be careful and take the appropriate principal value around the singularity at $\theta=x$ in the integrand of \verb"IntR".

\section{Conformal computations} \label{qftadsapp}
\newcommand{\bxx}[1]{\begin{#1}}
\newcommand{\be}{\bxx{equation}}
\newcommand{\ee}{\end{equation}}
\newcommand{\du}[2]{_{ #1 }^{\phantom{ #1 } #2 }}
\renewcommand{\ud}[2]{^{ #1 }_{\phantom{ #1 } #2 }}
\renewcommand{\AA}{\mathcal A}
\newcommand{\CC}{\mathcal C}
\newcommand{\FF}{\mathcal F}
\newcommand{\GG}{\mathcal G}
\newcommand{\LL}{\mathcal L}
\newcommand{\OO}{\mathcal O}
\newcommand{\TT}{\mathcal T}
\newcommand{\VV}{\mathcal V}

\subsection{Crossing symmetry in one dimension}
We consider correlation functions of primary operators $\phi_i(x_i)$ in a one-dimensional boundary field theory. The weight of the operator $\phi_i$ will be denoted as $\D_i$. We will work in the Euclidean theory on $\mathbb R \cup \{\infty\}$. The conformal group is a two-fold cover of $PSL(2,\mathbb R)$; its elements act by the usual fractional linear transformations on the positions,
\be
x \to x' = \frac{a x + b}{c x + d}
\ee
with $a d - b c = \pm 1$. The elements with negative determinant involve the parity transformation $x \to - x$. The Jacobian of this transformation is $(ad - bc)(c x + d)^{-2}$ and the fields transform as
\be
\phi(x) \to \phi'(x') = (ad - bc)^{P_\phi} |c x + d|^{2 h} \phi(x)
\ee
with $P_\phi \in \{0,1\}$ dictated by the parity of $\phi$.\footnote{The parity operator is unitary and its square is an internal symmetry transformation. Up to a well-known caveat \cite[section 3.3]{Weinberg:1995mt} we can always redefine the parity operator so it squares to $1$ and its eigenvalues are then $\pm 1$ as we assumed.} In a correlation function we should remember that the parity operation also reverses the operator ordering.

In a suitable basis the two-point functions take the familiar form
\be
\vev{\phi(x) \phi(0)} = \frac{(-1)^{P_\phi}}{|x|^{2\D_\phi}}
\ee
The two-point function of a parity odd operator is negative, so the associated norm $\vev{\phi|\phi}$ is positive.\footnote{It may help the reader that parity odd operators are the same as one-dimensional vectors $\phi_\mu(x)$. The  reflection-positive two-point function is then $\vev{\phi_\mu(x) \phi_{\nu}(0)} =|x|^{-2 \D_\phi} \left(  \delta_{\mu \nu} -2\frac{x_\mu x_\nu}{x^2}  \right) =- |x|^{-2 \D_\phi} $, in one dimension.}

The operator product expansion reads
\be
\phi_1(x) \phi_2(0) = \frac{C\du{12}{k}}{|x|^{\D_1 + \D_2- \D_k}}  \phi_k(0) + \ldots
\ee
where we assume that $ x < 0$. If we act with the parity operator on both sides then we find
\be
(-1)^{P_1 + P_2} \phi_2(0) \phi_1(-x) = (-1)^{P_k} \frac{C\du{12}{k}}{|x|^{\D_1 + \D_2- \D_k}}  \phi_k(0) + \ldots
\ee
and therefore the reflected OPE coefficients between primaries are
\be
C\du{21}{k} = (-1)^{\s_{12k}} C\du{12}{k}
\ee
with
\be
\s_{12k} \colonequals P_1 + P_2 + P_k \mod 2\,.
\ee
The previous relation in particular implies that parity odd operators can never appear in the OPE of two identical operators. We will work in a basis where two-point functions are diagonal. The structure of three-point functions then dictates that
\be
C_{12k} = (-1)^{\s_{12k}} C_{k21}
\ee
and so the OPE coefficients transform either in the trivial representation (if $\s_{12k} = 0$) or the sign representation (if $\s_{12k} = 1$) of the permutation group $S_3$. Notice that, even if parity is broken, the cyclic symmetry is always preserved. This for example means that $C_{12k} = C_{k12}$, whereas $C_{12k}$ and $C_{21k}$ are not always the same.

We will be specifically interested in four-point functions,
\be \label{fourptgeneralform}
\vev{\phi_1(x_1) \phi_2(x_2) \phi_3(x_3)\phi_4(x_4)} = \left|\frac{x_{14}}{x_{24}}\right|^{\D_{21}} \left|\frac{x_{14}}{x_{13}}\right|^{\D_{34}} \frac{\GG_{1234}(x)}{|x_{12}|^{\D_1 + \D_2} |x_{34}|^{\D_3 + \D_4}}
\ee
with, as usual, $(\cdot)_{ij} = (\cdot)_i - (\cdot)_j$ and
\be
x \colonequals \frac{x_{12} x_{34}}{x_{13} x_{24}}\,.
\ee
If $x_ i < x_{i+1}$ for $i = 1,2,3$ then $0 < x < 1$. The $s$-channel conformal block decomposition reads
\be
\GG_{1234}(x) = \sum_k C\du{12}{k} C_{34k} \, g(\D_{21},\D_{34};\D_k;x)
\ee
with the conformal blocks
\be
\label{generalblock}
g(a,b;\D;z) \colonequals |z|^\D {}_2 F_1(\D+a, \D+b; 2\D; z)\,.
\ee
where we added an absolute value sign so the expression is unambiguous also for negative values of its argument.

Let us briefly discuss operator ordering. Correlation functions with operator orderings that are cyclic permutations of each other are directly related, as follows from covariance under the (orientation-preserving) inversion $x_i \to - 1/x_i$. Furthermore, parity symmetry dictates that
\be
\GG_{1234}(x) = \GG_{4321}(x) (-1)^{\s_{1234}}
\ee
To see the complications that arise if we just swap two adjacent operators, let us swap operators 1 and 2. A simple relabeling leads to the block decomposition
\be
\GG_{2134}(x) = \sum_k C\du{21}{k} C_{34k} g(\D_{12},\D_{34};\D_k;x)
\ee
and assuming a parity invariant theory this is equal to
\be
\GG_{2134}(x) =(1-x)^{\D_{34}} \sum_k C\du{12}{k} C_{34k} (-1)^{\s_{12k}}\, g\left(\D_{21},\D_{34};\D_k;\frac{x}{x-1}\right)
\ee
where we used a standard hypergeometric transformation formula, valid for $x < 1$. The factor $(-1)^{P_k}$ and the absolute value sign in the definition \eqref{generalblock} imply that $\GG_{2134}$ and $\GG_{1234}$ are not, in general, related in an obvious manner. The symmetries and non-symmetries altogether leave us with three independent four-point functions from which the others follow. We can take these to be $\vev{2134}$, $\vev{1234}$, and $\vev{1324}$, which respectively correspond to $x < 0$, $0 < x < 1$, and $1 < x$. We will here be interested in just the second of these correlators.

With the ordering fixed there are only two OPE channels. For the $\vev{1234}$ ordering we gave the first one above; the crossed channel OPE reads
\be
\GG_{2341}(y) = \sum_p C\du{23}{p} C\du{p41}\, g(\D_{32},\D_{41};h;y)
\ee
with $y = x_{23} x_{41} / x_{24} x_{31} = 1 - x$. Crossing symmetry then takes the form\footnote{One foolproof way to obtain this expression is to relabel the operators in the original expression \eqref{fourptgeneralform} and then use a conformal transformation to relate $\vev{2341}$ to $\vev{1234}$. To verify that directly fusing operators 2 and 3 together in \eqref{fourptgeneralform} gives the same OPE limit requires that $C_{1 p 4} = C_{p41}$ which we proved previously.}
\be 
\GG_{1234}(x) = \frac{|x|^{\D_3 + \D_4}}{|1-x|^{\D_2 + \D_3}} \GG_{2341}(1-x)
\ee

Let us consider all correlation functions of two parity-even operators $\phi_1$ and $\phi_2$. We then find, in a diagonal operator basis, the following set of non-trivial crossing equations:
\be
\label{allcrossingeqns}
\begin{split}
0 &= \sum_k C_{11k}^2 \, g(0,0;\D_k;x) - \frac{|x|^{2\D_1}}{|1-x|^{2 \D_1}} \sum_k C_{11k}^2 g(0,0;\D_k;1-x)\\
0 &= \sum_k C_{22k}^2 \, g(0,0;\D_k;x) - \frac{|x|^{2\D_2}}{|1-x|^{2 \D_2}} \sum_k C_{22k}^2 g(0,0;\D_k;1-x)\\
0 &= \sum_k C_{22k} C_{12k}\, g(0,\D_{21};\D_k;x) - \frac{|x|^{2\D_2}}{|1-x|^{2\D_2}} \sum_k C_{22k} C_{12k}\, g(0,\D_{12};\D_k,1-x)\\
0 &= \sum_k C_{11k} C_{12k}\, g(0,\D_{12};\D_k;x) - \frac{|x|^{2\D_1}}{|1-x|^{2\D_1}} \sum_k C_{11k} C_{12k}\, g(0,\D_{21};\D_k,1-x)\\
0 &= \sum_p (-1)^{P_p} C_{12p}^2 \, g(\D_{21},\D_{12};\D_p;x) - \frac{|x|^{\D_1 + \D_2}}{|1-x|^{\D_1 + \D_2}} \sum_p (-1)^{P_p} C_{12p}^2\, g(\D_{12},\D_{21};\D_p;1-x)\\
0 &= \sum_k C_{11k} C_{22k} \, g(0,0;\D_k;x) - \frac{|x|^{2\D_2}}{|1-x|^{\D_1 + \D_2}} \sum_{p} C_{12p}^2\, g(\D_{21},\D_{21};\D_p;1-x)
\end{split}
\ee
where the operators labeled $k$ are parity even, whereas the operators labeled $p$ can be either parity even or parity odd.

\subsection{Transition to convex optimization}
In matrix form, the $I$'th crossing symmetry equation can be written as:
\be
\sum_{p, P_p = 1} C_p^t M^I_p C_p + \sum_{p, P_p = -1} C_p^t N^I_p C_p = 0
\ee
with
\be
C_p^t = 
\begin{pmatrix}
C_{11p} & C_{12p} & C_{22p}
\end{pmatrix}^t
\ee
and with
\begin{align}
M^1_p &=
\begin{pmatrix}
F^1_p(x) & 0 & 0 \\
0 & 0 & 0 \\
0 & 0 & 0
\end{pmatrix}
&
N^1_p &= 0 \nonumber \\
M^2_p &=
\begin{pmatrix}
0 & 0 & 0 \\
0 & 0 & 0 \\
0 & 0 & F^2_p(x)
\end{pmatrix}
&
N^2_p &= 0 \nonumber
\\
M^3_p &=
\begin{pmatrix}
0 & 0 & 0 \\
0 & 0 & F^3_p(x) \\
0 & F^3_p(x) & 0
\end{pmatrix}
& N^3_p &= 0\\
M^4_p &=
\begin{pmatrix}
0 & F^4_p(x) & 0 \\
F^4_p(x) & 0 & 0 \\
0 & 0 & 0
\end{pmatrix}
& N^4_p &= 0  \nonumber\\
M^5_p &=
\begin{pmatrix}
0 & 0 & 0 \\
0 & F^5_p(x) & 0 \\
0 & 0 & 0
\end{pmatrix}
& N^5_p &=  - M^5_p  \nonumber
\\
M^6_p &=
\begin{pmatrix}
0 & 0 & F^6_p(x) \\
0 & G^6_p(x) & 0 \\
F^6_p(x) & 0 & 0
\end{pmatrix}
& N^6_p &=
\begin{pmatrix}
0 & 0 & 0 \\
0 & G^6_p(x) & 0 \\
0 & 0 & 0
\end{pmatrix}  \nonumber
\end{align}
where
\be
\label{FsandGs}
\begin{split}
F^1_p(x) &= |1-x|^{2 \D_1} g(0,0;\D_p;x) - |x|^{2 \D_1} g(0,0;\D_p;1-x)\\
F^2_p(x) &= |1-x|^{2 \D_2} g(0,0;\D_p;x) - |x|^{2 \D_2} g(0,0;\D_p;1-x)\\
F^3_p(x) &= |1-x|^{2\D_2} g(0,\D_{21};\D_p;x) - |x|^{2\D_2} g(0,\D_{12};\D_p,1-x)\\
F^4_p(x) &= |1-x|^{2\D_1} g(0,\D_{12};\D_p;x) - |x|^{2\D_1} g(0,\D_{21};\D_p,1-x)\\
F^5_p(x) &= |1-x|^{\D_1 + \D_2} g(\D_{12},\D_{21};\D_p;x) - |x|^{\D_1 + \D_2} g(\D_{12},\D_{21};\D_p;1-x)\\
F^6_p(x) &= \frac{1}{2} |1-x|^{\D_1 + \D_2} g(0,0;\D_p;x)\\
G^6_p(x) &= - |x|^{2 \D_2}  g(\D_{21},\D_{21};\D_p;1-x)
\end{split}
\ee
We can act with a functional on each equation, and then add all of them. This yields
\be
\sum_{p,P_p = 1} C_p^t \begin{pmatrix}
F^1_p & F^4_p & F_p^6\\
F^4_p & G^6_p + F^5_p & F^3_p \\
F_p^6 & F^3_p & F^2_p
\end{pmatrix} C_p
+ 
\sum_{p,P_p = -1} C_p^t \begin{pmatrix}
0 & 0 & 0 \\
0 & G^6_p-F^5_p & 0 \\
0 & 0 & 0
\end{pmatrix} C_p = 0
\ee
where $F_p^I$ (without an argument) is shorthand for the function of $\D_p$ obtained by acting with the corresponding component of the functional.

If we single out the identity (with $C_{110} = C_{220} = 1$ and $C_{120} = 0$) and the external operators from the sums then we obtain
\be
\begin{split}
0 &= F_0^1 + 2 F_0^6 + F_0^2\\
&\phantom{=}+
\begin{pmatrix}
C_{111} & C_{112} & C_{122} & C_{222}
\end{pmatrix} 
\begin{pmatrix}
F^1_1 & F^4_1 & F_1^6 & 0\\
F^4_1 & G^6_1 + F^5_1 + F^1_2 & F^3_1 + F^4_2 & F_2^6 \\
F_1^6 & F^3_1 + F^4_2 & F^2_1 + G^6_2 + F^5_2 & F^3_2 \\
0 & F_2^6 & F^3_2 & F^2_2
\end{pmatrix}
\begin{pmatrix}
C_{111} \\ C_{112} \\ C_{122} \\ C_{222}
\end{pmatrix}\\
&\phantom{=}+
\sum_{p,P_p = 1} C_p^t \begin{pmatrix}
F^1_p & F^4_p & F_p^6\\
F^4_p & G^6_p + F^5_p & F^3_p \\
F_p^6 & F^3_p & F^2_p
\end{pmatrix} C_p\\
&\phantom{=}+ 
\sum_{p,P_p = -1} C_p^t \begin{pmatrix}
0 & 0 & 0 \\
0 & G^6_p-F^5_p & 0 \\
0 & 0 & 0
\end{pmatrix} C_p
\end{split}
\ee
For a feasibility study we can normalize the functionals on the unit operators, giving
\be
F^1_{0} + 2 F^6_{0} + F^2_{0} = 1
\ee
and furthermore demand positive semidefiniteness of the three square matrices listed above. We can also get bounds on products of OPE coefficients. For example, if we set
\be
F^1_1 = 1\qquad F^4_1 = F^6_1 = 0
\ee
and then maximize/minimize $F^1_{0} + 2 F^6_{0} + F^2_{0}$, we obtain a lower/upper bound on $C_{111}^2$. More precisely, if we extremize and the result is positive then crossing cannot be solved. If the result is negative then the absolute value of the result is our upper (for minimization) or lower (for maximization) bound.

\subsection{Setup with \texorpdfstring{$\mathbb Z_2$}{Z2} symmetry}
In the previous section we discussed the general one-dimensional conformal bootstrap analysis for two operators. Let us now specialize to the case discussed in the main text, so we consider a QFT in AdS$_2$ with a $\mathbb Z_2$ symmetry and only two stable parity even particles; a $\mathbb Z_2$ odd one created by an operator $\phi_1$ and a $\mathbb Z_2$ even one created by an operator $\phi_2$.

With this additional symmetry $F^3 = F^4 = 0$ automatically which leaves four non-trivial correlation functions of $\phi_1$ and $\phi_2$. In these correlators we should also label the internal operators with an even/odd quantum number depending on the OPE channel in which they appear. Our notation above is already adapted to this situation: in equation \eqref{allcrossingeqns} the operators labeled $k$ are necessarily parity and $\mathbb Z_2$ even, whereas the operators labeled $p$ are parity even or odd but always $\mathbb Z_2$ odd. (Operators that are $\mathbb Z_2$ even but parity odd do not feature in this set of correlation functions.) With the exception of $\phi_1$ and $\phi_2$ themselves we take their dimensions to lie above the following `gap' values:
\begin{center}
\begin{tabular}{ccll}
$\mathbb Z_2$ & $P$ & gap & index in \eqref{Z2matrixcrossingeqns}\\
\hline
even & even & $\min(2 \D_1, 2\D_2)$ & $k$ \\
odd & even & $\D_1 + \D_2$ & $p, P_p = 1$\\
odd & odd & $\D_1 + \D_2$ & $p, P_p = -1$ 
\end{tabular}
\end{center}
These gaps are precisely the two-particle values for a QFT in AdS, reflecting our assumption that there are no further stable single-particle states.

Going then through the same logic as before we write the crossing equations in matrix form as
\be \label{Z2matrixcrossingeqns}
\sum_{p, P_p = 1} C_p^2 M^I_p + \sum_{p, P_p = -1} C_p^2 N^I_p + \sum_k C^t_k Q^I_k C_k = 0\,,
\ee
with (note some redefinitions with respect to the previous formulae)
\be
C_k = 
\begin{pmatrix}
C_{11k} & C_{22k}
\end{pmatrix}^t
\qquad
C_p = C_{12p}
\ee
and with
\be
\begin{split}
M^1_p = 0
\qquad
N^1_p &= 0
\qquad
Q^1_k =
\begin{pmatrix}
F^1_k(x) & 0 \\
0 & 0
\end{pmatrix}
\\
M^2_p = 0
\qquad
N^2_p &= 0
\qquad
Q^2_k =
\begin{pmatrix}
0 & 0 \\
0 & F^2_k(x)
\end{pmatrix}
\\
M^5_p = F^5_p(x)
\qquad
N^5_p &=  - F^5_p(x)
\qquad
Q^5_p = 0
\\
M^6_p = G_p^6(x)
\qquad
N^6_p &= G_p^6(x)
\qquad
Q^6_k =
\begin{pmatrix}
0 & F^6_k(x) \\
F^6_k(x) & 0
\end{pmatrix}
\end{split}
\ee
After acting with the functional and singling out the important operators again we find
\be
\begin{split}
0 &= 
F^1_0 + 2 F^6_0 + F^2_0 
+
\begin{pmatrix}
C_{112} & C_{222}
\end{pmatrix}
\begin{pmatrix}
G^6_1 + F^5_1 + F^1_2 & F^6_2 \\
F^6_2 & F^2_2
\end{pmatrix}
\begin{pmatrix}
C_{112} \\ C_{222}
\end{pmatrix}
\\
&+
\sum_{p, P_p = 1} C_p^2 (G^6_p + F^5_p) + \sum_{p, P_p = -1} C_p^2 (G^6_p - F^5_p) + 
\sum_k  
\begin{pmatrix}
C_{11k} & C_{22k}
\end{pmatrix}
\begin{pmatrix}
F^1_k & F^6_k \\
F^6_k & F^2_k
\end{pmatrix}
\begin{pmatrix}
C_{11k} \\ C_{22k}
\end{pmatrix}
\end{split}
\ee
If we want to find bounds in the $(C_{112}, C_{222})$ plane we can set $C_{222} = \beta C_{112}$ and set a normalization
\be
1 = G^6_1 + F^5_1 + F^1_2 + 2 \beta F^6_2 + \beta^2 F^2_2
\ee
With this normalization our problem takes the form of a semidefinite program and we can proceed using the well-known numerical bootstrap methods of \cite{CFTNoFlavour,CFTMixed} and the specialized solver of \cite{SDPB}. Further technical details are available upon request from the authors.

\subsection{Functionals and derivative combinations}
\label{subapp:functionalslambdas}
As in previous works, our choice of functionals are linear combinations of derivatives of the crossing equations at $z = 1/2$, so our functionals $\alpha$ can be written as
\be \label{functional}
\alpha[F(z)] = \sum_{n = 0}^{\Lambda} \a_n \frac{d^nF}{dz^n}(1/2).
\ee
The coefficients $\a_n$ should be thought of as the main degrees of freedom to be fixed by the optimization procedure. Furthermore, in a multi-correlator study we can act with a different functional on each of the crossing equations in \eqref{allcrossingeqns}, so the degrees of freedom are
\be
\alpha^I_n, \qquad I \in \{1, 2, 5, 6\}, \qquad n \in \{0, 1, 2, \ldots \Lambda\}\,.
\ee
where, as before, $I$ denotes the different crossing equations and we used that the third and fourth equation are identically zero by our $\mathbb Z_2$ symmetry assumptions.

Of course the functions that are odd around $z = 1/2$, like the first and second crossing equation in \eqref{allcrossingeqns}, contribute only about $\Lambda/2$ non-trivial degrees of freedom since the $\a_n$ for odd $n$ are meaningless (and in fact should be set to zero to prevent numerical instabilities). This reduces the scaling of the number of components to $3 \Lambda$ rather than $4 \Lambda$.

\begin{figure}
\centering
\includegraphics[width=10cm]{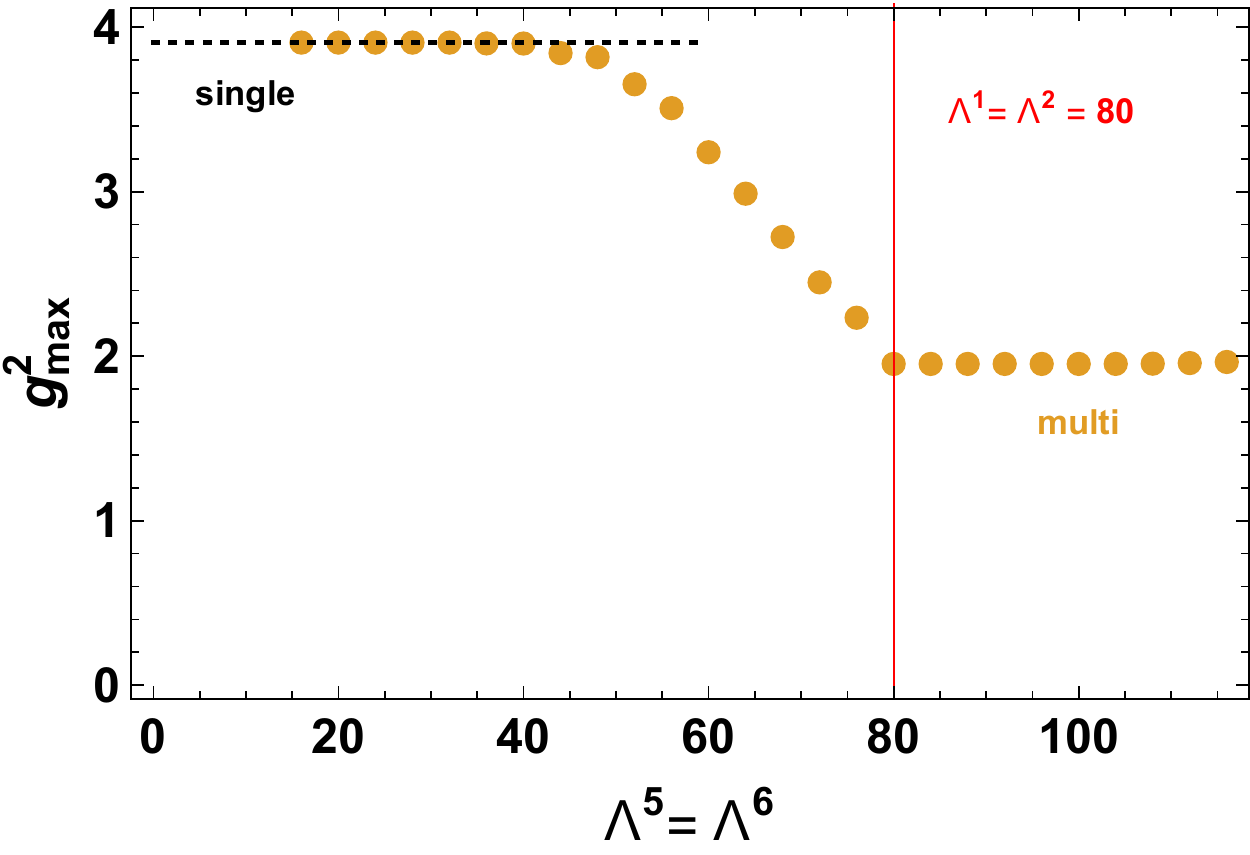}
\caption{\label{fig:lambda1lamdba2lamdba5lambda6}Bound on the maximal coupling (without any extrapolations) as a function of $\Lambda^5 = \Lambda^6$ for fixed $\Lambda^1 = \Lambda^2 = 80$. We see that the multi-correlator bound does not improve over the single-correlator bound (in black) for a range of values of $\Lambda^5 = \Lambda^6$, and this may lead one to believe that no improvement is possible whatsoever. On the other hand, increasing $\Lambda^5 = \Lambda^6$ above the natural value given by $\Lambda^1 = \Lambda^2$ does not seem to lead to further improvements for the range of values that we tested. The case shown here has $\Delta_1 = \Delta_2 = 6.58895$ and $g_{222}/g_{112} = 1$, but other cases look similar: we start with a plateau, then a kink (not necessarily at $\Lambda^1/2$) marks the start of a downward trajectory (which is not necessarily this linear), and then another kink at $\Lambda^5 = \Lambda^1$ leads to a second plateau where the bound is constant.}
\end{figure}

In our earliest attempts we thought it unfair to first two crossing equations that they could only contribute half as many functional components as the fifth and sixth equation. Therefore we decided to cut off the sum in \eqref{functional} at $\Lambda/2$ for the fifth and sixth equation but at $\Lambda$ for the first and second equation. Such an egalitarian approach, however, would have led to completely different and incorrect results. To illustrate this we plot in figure \ref{fig:lambda1lamdba2lamdba5lambda6} the upper bound on the coupling as a function of $\Lambda^5=\Lambda^6$, i.e. the cutoffs for the fifth and sixth equation, whilst holding fixed every other parameter, including the cutoffs $\Lambda^1 = \Lambda^2$ for the first and second equation. Clearly for $\Lambda^5 = \Lambda^1/2$ the multi-correlator bound offers exactly no improvement over the single-correlator bound. So, if we had continued to work with the egalitarian cutoffs then we would erroneously conclude that no improvement would have been possible over the single-correlator bound! Only when increasing $\Lambda^5$ beyond $\Lambda^1/2$ do we begin to see a gradual improvement, which stops as abruptly as it started at $\Lambda^5 = \Lambda^1$.

Notice that the plot is drawn for the \includegraphics[width=0.4cm]{babytriangle.pdf} data point in figure \ref{equalmassqftinads}, which has the special property that the multi-correlator bound turns out to be exactly half that of the single-correlator bound. We however observed very similar behavior, including the kinks and stabilization, also for other data points where the final multi-correlator bound is completely non-trivial.

\subsection{Extrapolations}
\label{subapp:extrapol}
An example of the extrapolation procedure outlined in section \ref{subsec:setupextrapol} is shown in figure \ref{fig:extrapolgrid}. One important subtlety not mentioned in the main text is that we extrapolate the log-ratio of the multi-correlator and the single-correlator result. That is, for every multi-correlator optimization run we also ran a single-correlator optimization run (for either $\vev{1111}$ or $\vev{2222}$) and so we get raw data that we can denote $(g_{112}^2)^\text{max,multi}[\mu,\alpha,\Delta_1,\Lambda]$ and $(g_{112}^2)^\text{max,single}[\mu,\alpha,\Delta_1,\Lambda]$. We found that direct extrapolation of the multi-correlator bounds led to a relatively large dependence on our fitting procedure, whereas extrapolation of the log-ratio
\be
\log(g_{112}^2)^\text{max,multi}[\mu,\alpha,\Delta_1,\Lambda]- \log(g_{112}^2)^\text{max,single}[\mu,\alpha,\Delta_1,\Lambda]
\ee
could, as shown, be done with relatively low-degree fits. Since we know that the single-correlator bounds match the analytic single-amplitude bounds with large accuracy \cite{Paper1}, we can add (the logarithm of) this known answer to our extrapolated log-ratio to obtain a better estimate of the flat-space multi-correlator bound.

\begin{figure}
\centering
\includegraphics[width=17cm]{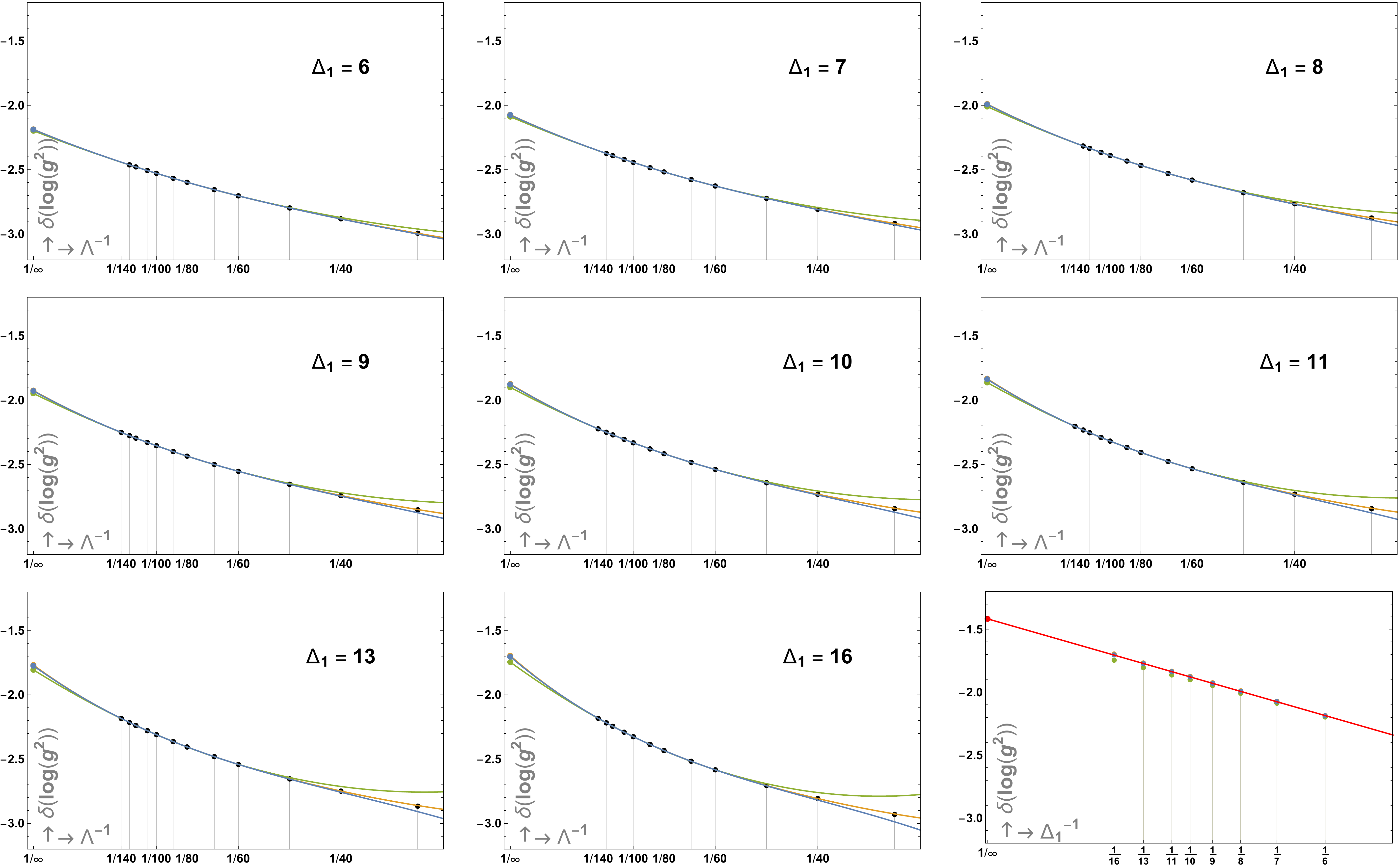}
\caption{\label{fig:extrapolgrid}The double extrapolation for the data point with $m_2/m_1 = 1.15$ in figure \ref{ratiominus1qftinads}. Vertically we plot the logarithm of the maximal coupling, but with the logarithm of the single-correlator bound subtracted, so $\delta(\log(g^2_\text{max})) = \log(g^2_{\text{max,multi}}) - \log(g^2_{\text{max,single}})$. In the first 8 plots we show our raw data in black, and the curves correspond to three different extrapolations to infinite $\Lambda$. In the final plot we collected the $\Lambda \to \infty$ extrapolations and the single red line represents our $\Delta_1 \to \infty$ extrapolation. Our final answer gives $\delta(\log(g^2_\text{max})) \approx -1.446$ in the flat-space limit, and adding the single-correlator bound $3.673$ gives the $2.226$ plotted in figure \ref{ratiominus1qftinads}.}
\end{figure}

The first 8 plots show the raw data and subsequent extrapolations to infinite $\Lambda$. The three curves correspond to fits with a polynomial in $\Lambda^{-1}$ of degree 3 (in blue), degree 4 (in orange, mostly coinciding with blue), and degree 2 (in green). In the fits we did not include the (three) data points with $\Lambda < 60$. We observe a rather small difference between the different extrapolations, and we have checked that these fits give good predictions for the high $\Lambda$ raw data points if we exclude one or more of them by hand.

The final plot in figure \ref{fig:extrapolgrid} collects all the extrapolated points to infinite $\Lambda$. The data points line up nicely, providing evidence for a small non-systematic error in our first extrapolation. We have fitted a linear function in $\Delta^{-1}$ to the degree 3 (in blue) points, leading to the extrapolated value of $-1.446$ for the log-ratio of this data point. From this plot it is clear that there is no meaningful difference if we had extrapolated the degree 4 (in orange) points instead.

We employed exactly the same extrapolation procedure, including the choice of the degree of the fitting functions, for all the other data points shown in the main text.

\newpage

\section{Landau Singularities in \texorpdfstring{$12\to 12$}{12->12} Scattering}
\label{Landauappendix}
Landau singularities are associated to diagrams representing particle interactions with all lines on-shell and momentum conservation at each vertex \cite{Landau, CT}. We claimed in the text that for the forward $12\to12$ scattering there ought to be no new such singularities (to be added to the bound state poles already there) whereas for the $12\to 12$ backward component we claimed that when $m_2 > \sqrt{2} m_1$ we do find such new on-shell processes. It is easy to convince oneself of that with a few pictures. For that purpose, we adapt here some beautiful discussion in chapter 18 of the book \cite{BD} by Bjorken and Drell, translating it to our two dimensional case of interest. 

\begin{figure}[t!]
\center \includegraphics[scale=.6]{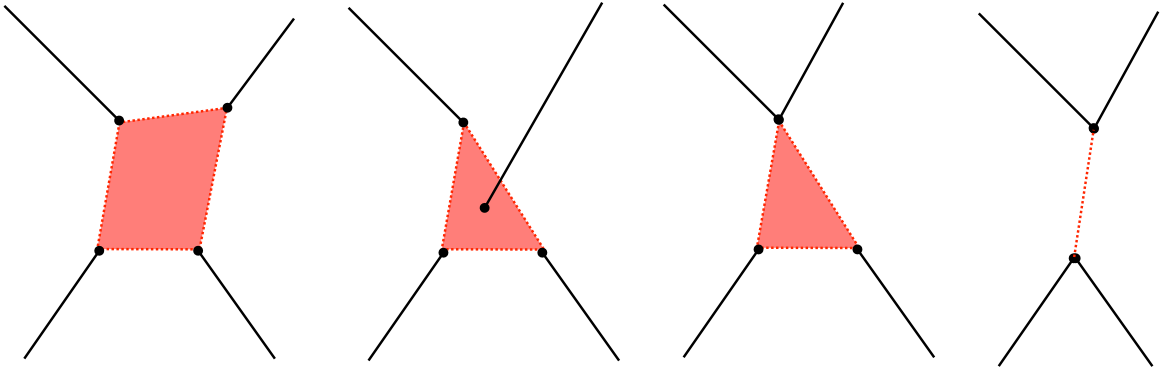}
\vspace{-7cm}
\caption{An on-shell scattering process must lie inside the convex hull defined by the vertices attached to external lines. This is due to momentum conservation: an on-shell internal particle that wandered outside the convex hull would never be able to move back inside, since there would be no external momentum available to kick it back in. There are 3 possible convex hulls for $2 \to 2$ amplitudes: a quadrilateral, a triangle, or a line.}\label{convexHull}
\end{figure}

Start with a diagram representing a two-to-two on-shell process. Each of the four external particles eventually encounter a vertex (several can meet at the same vertex). Consider those vertices containing an external line. They define a convex hull which can be a quadrilateral, a triangle or a line as shown in figure~\ref{convexHull}. Next we draw all other internal lines to complete the Landau diagram. The first important claim is that all those lines must lie inside the convex hulls just defined. That is simply because of momentum conservation: if they got out of the convex hull there would be no external particle to kick them back in (and the diagram would never close)! With this in mind we can go on to analyse the three possible convex hulls in turn. 

\begin{figure}[t]
\center \includegraphics[scale=.6]{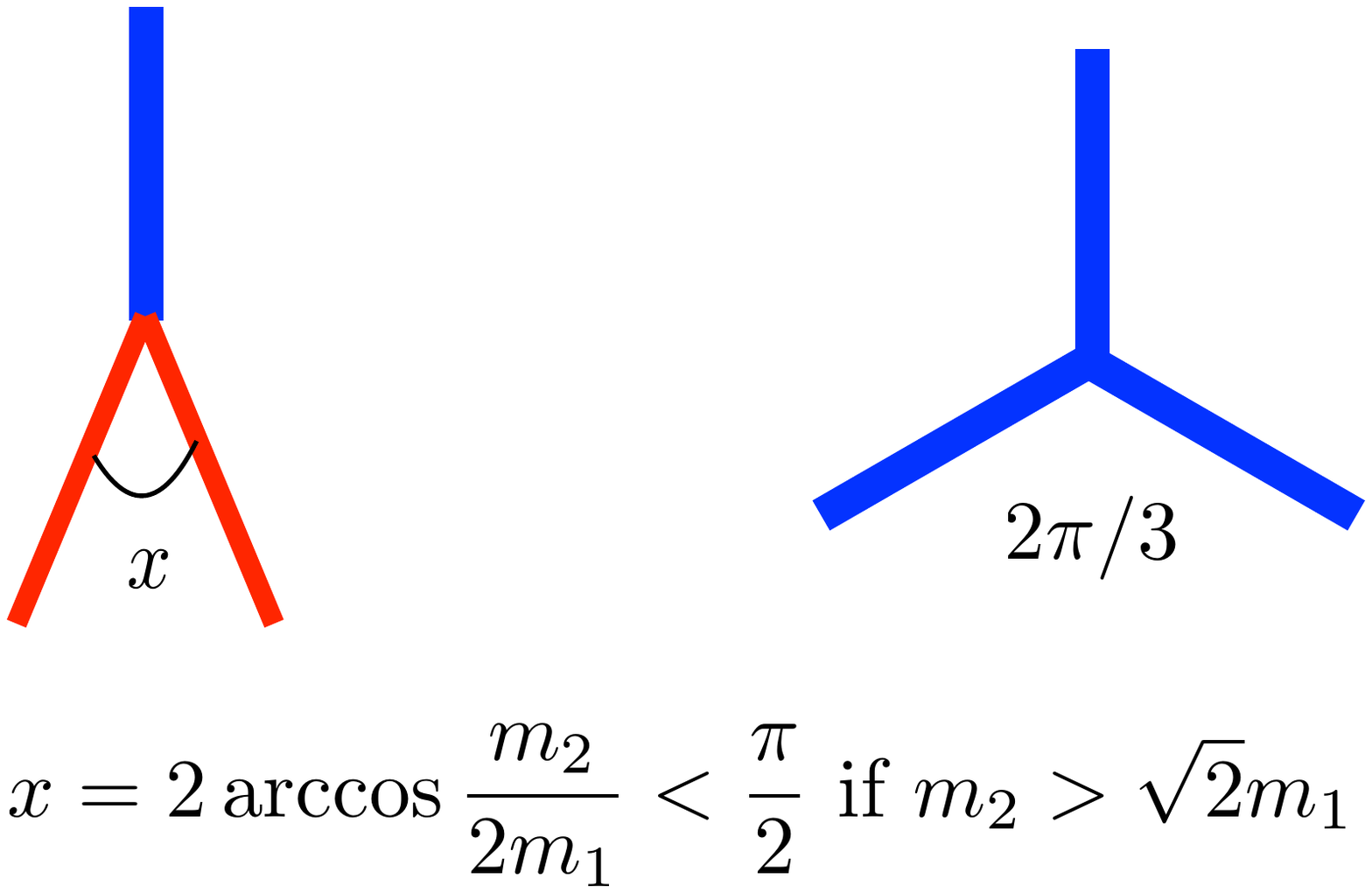}
\vspace{-6.5cm}
\caption{On-shellness and energy-momentum conservation fix angles in cubic vertices. Importantly, when $m_2> m_1 \sqrt{2}$ there are acute angles in cubic couplings and the analysis of Landau singularities becomes more intricate. }\label{cubic}
\end{figure}

Consider first the quadrilateral case and assume first that at each of the four external vertices we have a cubic vertex. Then, momentum conservation and on-shellness constrains the angles at those cubic vertices\footnote{In this appendix we assume that the diagrams and four-momenta are euclidean. This is the case in-between two-particle cuts, since then all spatial momenta can be chosen to be purely imaginary, making the lorentzian metric effectively euclidean. The absence of Landau diagrams in the full physical sheet follows from their absence in the euclidean region after a causality argument, as detailed in \cite{BD}.}. For example, for the case at hand with two particles of different masses we have the angles in figure~\ref{cubic}. For  $m_2 < \sqrt{2} m_1$ we realize right away that the angles are obtuse so it is generically impossible to form a quadrilateral! What about using other vertices when particles first interact? Well, that would be even worse as the total opening angles would be even greater in that case. If $m_2>\sqrt{2} m_1$ then one of the angles is smaller than $\pi/2$ so we have a better chance of finding extra singularities and, indeed, we can sometimes form such diagrams, as the one in figure \ref{Landau}, but not if two external particles are of type $2$ and two particles are of type $1$ as illustrated in figure \ref{explainQuadri}(a). Hence, there are no ``quadrilateral convex hull" Landau diagrams for $12 \to 12$ scattering. 

What about ``triangular convex hull'' singularities? When $m_2 > \sqrt{2} m_1$, as explained in figure \ref{explainQuadri}(b), those are indeed  present in the backward component, but not in the forward one. So the backward amplitude should have extra Landau poles but the forward amplitude should not. As such, the bound on $g_{112}^2$ derived from the forward component should hold even for $m_2 > \sqrt{2} m_1$. As discussed in the main text, this bound is not captured by the QFT in AdS bootstrap, see figure \ref{ratiominus1qftinads}.
 
We did not discuss the case where the convex hull is the line: those are just the usual singularities such as the bound state poles and all the production cuts which open for multi-particles at rest, and thus moving parallel to each other, along the convex hull line.

\begin{figure}[t!]
\center \includegraphics[scale=.6]{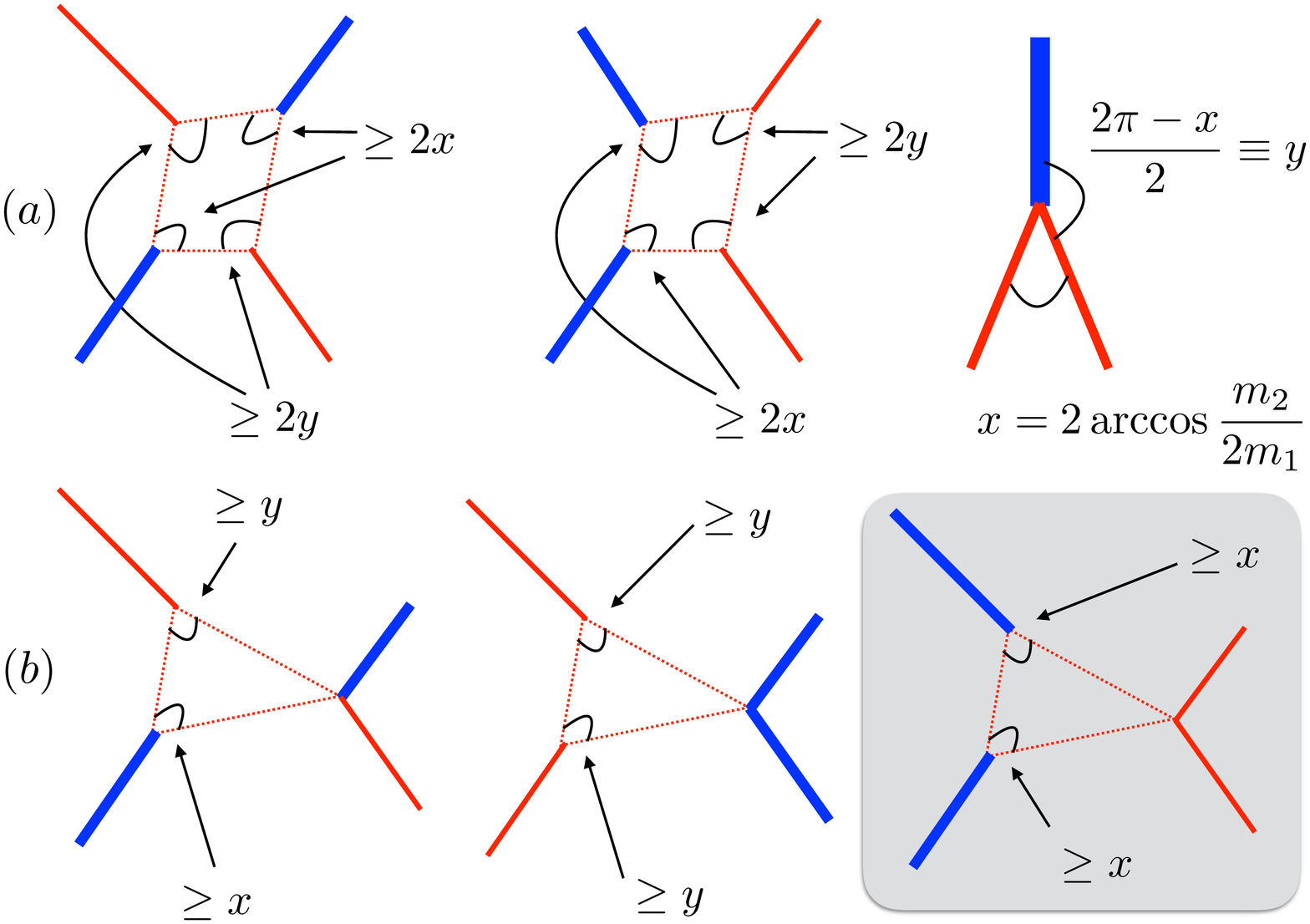}
\caption{(a) There are no possible ``quadrilateral convex hull" Landau singularities in the $12 \to 12$ amplitude even when $m_2>\sqrt{2}m_1$. To understand this, consider a vertex with an external leg attached. The total opening between internal edges on such vertex  must be greater than $x$ ($y$) for an external even (odd) leg. Hence, for the $12 \to 12$ process, the total internal angles at the four external vertices must be greater than $2(x + y) \geq 2 \pi$, so that it is inconsistent to have a closed singular diagram inside the convex hull. (b) The same is true for ``triangular convex hull" diagrams unless two odd particles meet at the same external vertex, as on the bottom right diagram - as in other cases the total opening at external vertices  would be greater than $\pi$. For the forward component there would be no momentum transfer through such vertex, so that by on-shellness and energy-momentum conservation the internal opening at this vertex would be at least $\pi$, once again making it impossible to have a singular Landau diagram. For the backward component, on the other hand, the momentum transfer through the vertex, and hence the opening angle, can be arbitrary. In turn, this amplitude will have extra Landau singularities for $m_2>\sqrt{2}m_1$.}\label{explainQuadri}
\end{figure}

\pagebreak

\end{document}